%% file: He_sdo_revised_accepted_arxiv.tex
\documentclass[]{aa}

\usepackage{rotating, graphicx}
\usepackage{txfonts}
\usepackage{pdflscape} 
\usepackage{isotope}
\usepackage{color}
\usepackage{setspace}
\usepackage{multirow}

\begin{document} 
\newcommand{\CD}{\object{CD--31$^{\circ}$4800}}
\newcommand{\lsIV}{\object{LS\,IV\,+10$^{\circ}$\,9}}   
\newcommand{\lss}{\object{LSS\,1274}}
\newcommand{\uv}{\object{[CW83]\,0904$-$02}}
\newcommand{\pg}{PG\,1544$+$488}
\newcommand{\jl}{JL\,87}
\newcommand{\lb}{LB\,1766}

\title{A quantitative NLTE analysis of visual and ultraviolet spectra of four helium-rich subdwarf O stars}

\author{M. Schindewolf\inst{1}
\and    P. N\'{e}meth\inst{1,2,3}
\and	U. Heber\inst{1}
\and	T. Battich\inst{4,5}
\and	M. M. Miller Bertolami\inst{4}
\and    A. Irrgang\inst{1}
\and	M. Latour\inst{1,6}
}

\institute{Dr.\,Karl Remeis Observatory, University of Erlangen-Nuremberg, Sternwartstr. 7, D96049 Bamberg\\
\email{markus.schindewolf@fau.de}
\and Astronomical Institute of the Czech Academy of Sciences, CZ-251\,65, Ond\v{r}ejov, Czech Republic 
\and Astroserver.org, 8533 Malomsok, Hungary
\and Instituto de Astrofísica de La Plata, CONICET–UNLP, Argentina
\and Facultad de Ciencias Astronómicas y Geofísicas, UNLP, Argentina
\and Institut für Astrophysik, Georg-August-Universit\"at, G\"ottingen, Germany
}

\date{Received: 20 October 2017 / Accepted: 20 August 2018}
\abstract{}{Hot subdwarf stars represent a poorly understood late phase of stellar evolution. While binary evolution plays an important role for the formation of B-type subdwarfs (sdB), the origin of the helium dominated subclass of O-type subdwarfs (He-sdO) is still unknown. We search for chemical signatures of their genesis by means of quantitative spectral analyses of high-quality visual and ultraviolet spectra.}
{Four prototypical He-sdO stars, one belonging to the nitrogen-rich and three to the C-rich subclass, were selected for which archival FUV spectra from the FUSE satellite as well as high-resolution visual and UVA spectra taken with the ESO-UVES/FEROS spectrographs are available. Using {\sc Tlusty200/Synspec49} to compute line blanketed-
non-local thermodynamic equilibrium (NLTE) model atmospheres and synthetic spectra, atmospheric parameters and the abundances patterns have been derived. The final models included H, He, C, N, O, Ne, Mg, Al, Si, P, S, Fe, and Ni represented by the most detailed model atoms available. Because of the enrichment of either nitrogen or carbon, it turned out, that models including these elements at the appropriate high abundance provide sufficiently accurate approximations to the temperature stratification of full models.}
{No indications
for binarity were found, neither radial velocity variations nor photometric evidence for the presence of a companion could be detected.
All stars have helium-dominated atmospheres almost free of hydrogen and temperatures between 42\,000\,K and 47\,000\,K while their surface gravities lie between $\log{g}= 5.4$ and 5.7. The abundance pattern of \CD\ displays the signatures of CNO burning, while heavier elements are subsolar by about 0.4 dex, except for Ne and Si which are close to solar. The abundance patterns of the C-rich He-sdOs are more complex. A slightly subsolar metallicity is accompanied by N-enrichment and O-deficiency, less pronounced than in \CD. Neon is mildly to strongly enriched, up to a factor of ten with respect to the sun in \lsIV. The nickel-to-iron ratio is significantly super-solar. Using spectral energy distributions and Gaia parallaxes the masses of the stars were determined. They are found to scatter around the canonical mass for the core helium flash, although the uncertainties are large.}
{The abundance pattern observed for \CD\ is consistent with predictions of models for slow (cold) mergers of pairs of equal mass helium WDs except for the low oxygen abundance observed. Models for composite mergers were considered for the C-rich stars, but predict abundance pattern dissimilar to those determined. \uv, though, may be a candidate for a composite He-WD merger, as it rotates and appears to be more massive than the other program stars. New evolutionary models for the hot flasher scenario predict abundance patterns similar to those determined for the C-rich stars. Hence, C-rich He-sdO may well result from late He flashes with deep-mixing episodes.}

% 5 {} token are mandatory
 
\keywords{stars: abundances, stars: atmospheres, stars:individual, stars: evolution, stars: subdwarfs}

\titlerunning{A quantitative NLTE analysis of four helium-rich subdwarf O stars}

\maketitle
%
%-------------------------------------------------------------------

\section{Introduction}
Hot subluminous stars (sdB and sdO) represent late stages of the evolution of low-mass stars and can be found between the main sequence and the white dwarf (WD) sequence in the Hertzsprung-Russell diagram. 
Most hot subdwarf stars, in particular the sdB stars, are core helium-burning stars with very thin hydrogen envelopes. Hence, the sdB stars populate the extreme horizontal branch (EHB). Unlike normal horizontal branch stars EHB stars are unable to sustain hydrogen shell burning \citep{heber2009} and evolve directly to the WD cooling sequence without an excursion to the AGB.\\
They can be found both in the old disk and the halo of the Galaxy \citep{Ferraro1997,napiwotzki2008} and are sufficiently abundant to dominate the population of faint blue stars at high Galactic latitudes to about eighteens  
magnitude. Hot subdwarf stars are believed to be the dominating source for the UV upturn phenomenon that can be observed in elliptical galaxies and galaxy bulges \citep{Brown1997,han2007}. \\
While sdB stars form a rather homogeneous class of stars, subluminous O stars have more diverse stellar and atmospheric properties. Luminosities range from a few tens of solar to some thousands and effective temperatures of sdO stars span a large range from 40\,000\, K to 100\,000~K \citep{napiwotzki2008,heber2009,heber2016}. The most luminous ones, however, are believed to be post-AGB stars similar to central stars of planetary nebulae. The atmospheric composition of sdO stars is diverse as well, in particular with respect to the abundances of helium, carbon, and nitrogen. In hydrogen dominated sdO atmospheres the helium content may be subsolar while helium is dominating over hydrogen in the He-sdO subclass. Hydrogen is hardly detectable for some of those stars. The origin of these He-sdOs is ambiguous as they are believed to form differently from the hydrogen-dominated sdO stars, which are naturally explained as the progeny of the sdB stars in a post-EHB stage of evolution. The low fraction of close binaries among He-sdO stars \citep{napiwotzki2008} suggests that they may originate from the merger of two helium WDs \citep{webbink1984}. However, this scenario is rivaled by the so called late hot flasher scenario, which does not implicate binary evolution. Instead, the helium, carbon, and/or nitrogen enrichments are attributed to internal flash phenomena occurring when a star has already embarked on the WD cooling sequence \citep{1997fbs..conf....3S}.  %Sweigart 
Interest in the hot flasher scenario revived, when the small group of helium rich sdB stars attracted attention \citep{2003A&A...402..335A}. \citet{2010MNRAS.409..582N} suggested that two subclasses should be distinguished according to their helium content: the intermediate iHe-sdBs with helium-to-hydrogen ratios  between solar and He/H=4 (by number), because He-sdBs with higher helium abundance (extreme He-sdBs) have surface compositions distinctively different from the very peculiar composition of iHe-sdBs. The results of the spectral analyses of hot subdwarf stars from the ESO-SPY project \citep{2005A&A...430..223L,Stroer2007} suggest that such a distinction may be 
useful for the He-sdOs as well \citep[see][for a discussion]{heber2016}. Therefore, we shall restrict the term He-sdO to those with almost pure helium line spectra (hydrogen Balmer components too weak to significantly contribute to the equivalent widths of the He\,{\sc ii}/H\,{\sc i} blends) throughout this paper.  %Lisker et al. Stroeer et al. 2007    
While spectral analyses of many subluminous B stars are available \citep[e.g.][]{2003A&A...400..939E,2005A&A...430..223L,peter2012,2013A&A...549A.110G,2014ASPC..481...83F}, the sdO stars have not been studied extensively.  %Edelmann et al. 2003,  Lisker et al. 2005, Nemeth et al. 2012, Fontaine et al. 2014 
This is partly caused by the need of model atmospheres in non-Local Thermodynamic Equilibrium (NLTE). When such atmospheres became available, first quantitative analyses were carried out from low \citep[e.g.][]{1990A&A...235..234D} as well as high-resolution spectra \citep[e.g.][]{Dreizler1993}. 
\citet{nemeth2012} have presented a NLTE analysis from low resolution optical spectra of a magnitude limited subdwarf sample from the solar neighborhood, including 124 sdB and 42 sdO stars. 
Detailed abundance analyses, however, remained scarce.
  
The detailed abundance pattern of He-sdO stars are important to test the formation scenarios. The latest generation of NLTE model atmospheres is now able to treat NLTE and line blanketing effects by heavy elements consistently and allow abundances of low, intermediate and higher mass elements (iron, nickel) to be determined.
In recent years several quantitative spectral analyses have been carried out, making use of visual spectra of low \citep[e.g.][]{nemeth2012} as well as high resolution \citep{Hirsch2009}. Those analyses, however, were restricted to the chemical elements H, He, C \& N \citep{Hirsch2009} and H, He, C, N \& O \citep{nemeth2012}, only. Accordingly, He-sdO stars may be grouped into three subclasses, according to their carbon and/or nitrogen line strength; the C strong-lined, the N strong-lined and those where both carbon and nitrogen lines being strong \citep[CN,][]{Stroer2007}. Actually, the CN subclass is not well defined, because the classification scheme by \citet{Stroer2007} is based on spectra of different S/N, most being much lower than the spectra presented here. Therefore, we distinguish N strong-lined (N-type) from C strong-lined (C-type) stars, only. This calls for more detailed abundances studies. 

Visual Echelle spectra of excellent quality (high spectral resolution and S/N, as well as broad wavelength range) have become available and the FUSE mission has opened up the FUV spectral window to the Lyman series limit. FUSE spectra of high spectral resolution and S/N are now at hand superseding IUE UV spectra previously available for some He-sdOs both in terms of S/N and spectral resolution and allowing access to a  larger number of elements than from visual spectra.
Hot subdwarf stars have been observed in the far UV with the aim to probe the interstellar medium, in particular by making use of the FUSE satellite \citep[e.g.][]{2002ApJS..140...51S,2002ApJS..140...37F,2013ApJ...764...25J}. Similarly, high-resolution spectra in the UVA (3200 -- 3900\AA) range have been taken to determine elemental abundances in the interstellar medium \citep[e.g. Ti\,{\sc ii},][]{2008A&A...481..381L}. 
    
Most of the available quantitative spectral analyses of the photospheric FUV spectra have targeted the hydrogen-rich sdB and sdO stars \citep{2000ApJ...538L..95O,2008ApJ...678.1329B,2006BaltA..15...99F,2006BaltA..15..131C,2013ApJ...773...84L,2018A&A...609A..89L} and three helium-rich sdB stars \citep{2004ApJ...602..342L}. 
To our knowledge no He-sdO has been analyzed from FUSE or UVA spectra. 
Hence, we considered it worthwhile to revisit some of the brightest He-sdO stars, for which data from modern ground-based Echelle spectrographs as well as from the FUSE satellite are at hand. We judged four stars to be the best suited ones, from both subclasses of He-sdOs, \CD\ (N strong-lined), \lss, and \uv, and \lsIV\ (C strong-lined). All of them have been analyzed before using an early generation of NLTE model atmospheres. 
We consider these stars as benchmarks for the abundance patterns of He-sdO stars to test the predictions of different evolutionary scenarios. In addition, they provide a benchmark for NLTE model atmosphere diagnostics. Hence, we also embarked on a detailed investigation of NLTE model atmospheres of different degree of sophistication and a reanalysis of previously analyzed spectra to identify potential systematic differences.

We begin with a description of the available spectra and discuss radial velocity variability in Sect. \ref{observation}. In Sect. \ref{atmospheres} we present our atmospheric models and spectral synthesis and provide a comparison of models of different degree of sophistication. 
In Sect. \ref{atmos_params} we describe our approach to analyze visual and UVA Echelle and FUSE spectra. Atmospheric parameters are derived from visual spectroscopy in Sect. \ref{tghe} along with stellar masses from spectral energy distributions and Gaia parallaxes in Sect. \ref{sect:mass}. The analysis of the metal abundances follows in Sect. \ref{sect:results} and results are discussed in the context of rivalling scenarios in Sect. \ref{discussion}. The paper ends with a brief summary and conclusions in Section \ref{sect:summary}.

\section{The sample and spectral observations}\label{observation}

\CD\ was discovered by \citet{1973ApJ...179L.117G} % Garrison and Hiltner 1973
to be an extremely helium rich subluminous O star. \citet{1981MNRAS.197..241W} %Walker 1981
found the visual spectrum of \lsIV\ to be similar to that of \CD\ and concluded that it must be of the same spectral type as \CD. 
\citet{1980A&A....85..367B} %Berger \& Fringant 1980
observed low resolution UV spectra from the TD1 satellite and from a comparison with other sdO stars (including \CD) concluded that \uv\ is a sdO star similar to \CD.
\lss\ was first described as a sdO star by \citet{drilling1987}.
The first quantitative spectral analysis of visual spectra of \CD\ was carried out by \citet{giddings1981}. %Giddings 1981
With the advent of the IUE satellite high resolution ultraviolet spectra became available and early Echelle spectrographs for the visual wavelength range, such as CASPEC at the ESO 3.6m telescope, allowed high resolution spectra of sdO stars to be observed. Such visual and UV spectra were analyzed by \citet{Dreizler1993} using the TMAP NLTE code \citep{1999JCoAM.109...65W} and by \citet{bauer1995} using the NLTE code described by \citet{1989A&A...222..150H} 
%Bauer and Husfeld, Husfeld et al.
who derived atmospheric parameters and the abundance of light elements using NLTE model atmospheres and spectral synthesis. \citet{Hirsch2009} derived H, He, C and N  abundances from high-resolution visual spectra using TMAP NLTE model atmospheres and spectrum synthesis \citep{1999JCoAM.109...65W}. % Werner \& Dreizler 1999

%FUSE: ISM \citet{2002ApJS..140...51S} % Sonneborn et al. 2002, BD+28 
%\citet{2002ApJS..140...37F} % Friedman et al. 2002, Feige 110
FUSE spectra were used to determine the deuterium-to-oxygen and the deuterium-to-hydrogen ratios in the interstellar (ISM) medium toward \CD\ and \lss\ \citep{2003ApJ...599..297H,2004ApJ...609..838W}. %D/H toward \lss\ %Wood et al. 2004
%\CD and \lsIV\ and \uv. %Jenkins 44 hot subdwarf stars. %Hebrard et al. 2003
\citet{2013ApJ...764...25J} used FUSE spectra of \CD\ and \lsIV\ and \uv\ to study the respective line-of-sight warm ISM. %Jenkins 44 hot subdwarf stars 
Also the hot ISM has been investigated using FUSE spectroscopy of hot subdwarf stars to search for 
 diffuse O\,{\sc vi} emission \citep{2006ApJ...647..328D,2009ApJ...692..335B}. %Dixon et al. 2006, Blair et al.
Hence, for the ISM studies the
stars just provided background sources. 
Here we turn the analysis around and re-use the observations to 
analyze the stellar photospheric spectra. 

%--------------------------------------------------------------------
\subsection{Spectroscopic Observations}
\label{spectroscopy}
For the four He-sdO stars (\CD, \uv, \lss\ and \lsIV) under investigation both visual and FUV spectra are available. The space-based FUV spectra from FUSE were obtained from the Mikulski Archive for Space Telescopes (MAST). The FUSE spectrograph covers the far UV wavelength range from 905 to 1185\AA\ at a spectral resolution of R=20\,000. The design of the instrument and the spectroscopic data products are described by \citet{2000ApJ...538L...7S},  %Sahnow et al. (2000), 
\citet{2000ApJ...538L...1M}, %Moos et al. (2000),
and \citet{dixon2007}. % Dixon et al. (2007).
 In addition, visual spectra, obtained with the UVES and FEROS spectrographs, were downloaded from the European Southern Observatory (ESO) Data Archive. The FEROS spectrograph covers the spectral range from 3530-9200\AA\ at a resolution of R $\sim$ 48\,000 \citep{1999Msngr..95....8K}, whereas the UVES spectrograph allow spectra with different instrument setups \citep{2000SPIE.4008..534D} to be taken, resulting in different spectral coverages and resolutions. % Kaufer et al. 1999
  
 Table \ref{sample-overview} gives an overview of our sample and the properties of the available spectral observations. Visual spectra taken with the CASPEC spectrograph at the ESO 3.6m telescope were also available
\citep[see][]{Dreizler1993} in reduced and normalized form. Those spectra complement the UVES spectra of \lss\ and \lsIV\ to cover the full visual spectral range. We take the opportunity to reanalyze the CASPEC spectra and compare the results to the published ones in Sect. \ref{sect:caspec} in the appendix. The UVES spectrum of \CD\ has been observed with the image slicer to enhance the spectral resolution to about R=70\,000 in the blue and R=100\,000 in the red channel, respectively. Because this is a non-standard mode of observing, the archive data showed serious data reduction artifacts. S. Moehler kindly re-reduced those spectra. Despite of her extensive efforts some parts of the spectrum show residual fringing in the red part, which, however, does not compromise the spectral analysis.     
In addition to the visual and FUV range, we can make use of the extreme blue coverage of UVES in the wavelength range between 3200\,\AA\ and about 3800\,\AA\ (the UVA spectral range).

\subsection{Line identification and selection}\label{line_id}

Line identification was done by consulting the NIST database \citep{nist18} as well as evaluating synthetic spectra calculated with the SYNSPEC code \citep{synspec94}. 
Especially the UVES spectrograph allows for high resolution spectra in the visual region to the atmospheric cut-off (the UVA range), which is important because it covers the limits of the hydrogen Balmer as well of helium line series. Most importantly numerous lines of light metals, in particular from N\,{\sc iii}, O\,{\sc iii}, and Ne\,{\sc ii} are present. 
For the abundance analysis we had to select spectral lines in the first place that are not blended with lines of other chemical elements nor with interstellar or telluric lines.  Because the metals are added to the models subsequently (see Sect. \ref{atmos_params}), we are also able to treat line blends when the abundance of the blending element has been determined beforehand.

\subsubsection{The visual spectral range}

Spectral lines in the visual selected for the spectral analysis are listed in Tables \ref{list_h_he} and \ref{1xx}. Besides hydrogen and helium, lines from two ions of  carbon (C\,{\sc iii} and C\,{\sc iv})\footnote{For C\,{\sc ii}, 4267\AA\ see Sects. \ref{sect:cno} and \ref{sect:c2}.}, three ions of nitrogen (N\,{\sc ii}, N\,{\sc iii}, and N\,{\,\sc IV}) were considered. Two ions of oxygen (O\,{\sc ii} and O\,{\sc iii}), silicon (Si\,{\sc iii} and Si\,{\sc iv}), and phosphorus (P\,{\sc iv} and P\,{\sc v} line), and two stages of sulfur (S\,{\sc iii}, S\,{\sc iv}) were also used. In addition spectral lines of Ne\,{\sc ii}, Mg\,{\sc ii} and Al\,{\sc iii} were used as well. No spectral lines from iron group elements were identified.

\subsubsection{The FUV spectral range}

Many atomic and molecular interstellar lines are present in the FUV. Most of the atomic lines arise from neutral atoms, such as N\,{\sc i}, O\,{\sc i}, and Ar\,{\sc i} \citep[see Fig. 1 of][for an example, (\uv)]{2013ApJ...764...25J}. We used the table of \citet{2003ApJS..149..205M} %Morton (2003).
to identify the atomic interstellar lines. More difficult are the molecular interstellar lines, most prominently the Lyman and Werner bands of H$_2$. %\citep{1993A&AS..101..273A,1993A&AS..101..323A}. %Abgrall et al. 1993
 We used the list of \citet{1976ApJ...204....1M} %Morton & Dinerstein%
 for identification.
 While the Lyman and Werner lines are weak in \CD\ and \uv, they provide more severe line confusion for \lss\ and \lsIV.

The most obvious difference of the far-ultraviolet spectrum to the visual one is the large amount of lines of iron group elements. We concentrate on iron 
(Fe\,{\sc iii}, Fe\,{\sc iv}, and Fe \,{\sc v}) and nickel (Ni \,{\sc iv} and Ni \,{\sc v}). 
Since the lines are heavily blended, great care has to be taken to chose appropriate spectral lines.  As found for other sdO stars many spectral lines remained unidentified \citep[e.g.][]{2002ApJS..140...37F}. \citet{2014A&A...566A...3R} pointed out, that opacity is missing even when spectral lines from additional iron group elements are included in the calculation. % Rauch et al. 

For the lighter elements (C to S) the same ions as identified in the visual are accessible in the FUSE spectral range as well  with the exception of magnesium. In addition, O\,{\sc iv}, S\,{\sc v} and S\,{\sc vi} lines are  present.  %The FUV continuum forms deeper in the atmosphere where the local temperatures is higher, which causes a shift towards lines that require higher excitation.

\subsubsection{Radial velocities}
The observations were shifted to the rest wavelength before the analysis of atmospheric parameters was started. To this end we determined the radial velocity of the targets for each spectrum. This also allows us to co-add different single spectra and, thus to increase the S/N ratio. The radial velocity was measured by fitting a Voigt line profile to selected sharp metal lines. In the FUV, this was often not possible due to line blending and the lower spectral resolution. In these cases a synthetic spectrum was shifted manually and the best fit was determined by eye. Table \ref{rvs} gives an overview on the measured velocities. Results from FUSE have to be dealt with caution as the wavelength calibration turned out to be non-consistent for some spectra. \citet{dixon2007} found a systematic error of up to 10 km\,s$^{-1}$ for the wavelength calibration of FUSE data, depending on the angle under which light hits the optics. We added this systematic error to the statistical ones.

\begin{table*}
\setstretch{1.3}
\caption{Radial velocities for the four sample stars with time of observation (start). For radial velocities from FUSE data, we added the systematic uncertainty arising from wavelength calibration.\label{rvs}}
\begin{center}
\begin{tabular}{l l l l l}
\hline
\hline
Name&Date (UTC)&Instrument&Spectral range [\AA]&RV [km\,s$^{-1}$]\\
\hline

\CD&2002-06-03;23:31.23&UVES&3300...6690&$30\pm2$\\
%\CD&1978-07-02;22:32&IUE&$25\pm3$\\
%\CD&2004-02-23;04:46:00&FORS&3480...5890&$28\pm4$\\
\CD&2004-04-12;14:22:44&FUSE&905...{1185}&$27\pm4\pm10$\\
%\hline
\uv&2005-02-24;04:24:19.13&FEROS&3530...9200&$22\pm4$\\
\uv&2005-02-24;06:20:23.80&FEROS&3530...9200&$22\pm4$\\
\uv&2004-05-04;05:19&FUSE&905...1185&$30\pm5\pm10$\\
\uv&2004-05-05;04:34&FUSE&905...1185&$30\pm5\pm10$\\
%\uv&1993-12-23;09:48&IUE&$25\pm4$\\
%\uv&1993-12-23;13:45&IUE&$25\pm4$\\
%\uv&1994-03-16;04:22&IUE&$24\pm6$\\
%\hline
\lss&2006-04-10;00:54&UVES&3025...6830&$13\pm1$\\
\lss&2006-04-10;01:29&UVES&3025...6830&$13\pm1$\\
\lss&2002-03-11;12:13&FUSE&905...1185&$17\pm4\pm10$\\
\lss&2002-03-08;06:00&FUSE&905...1185&$16\pm4\pm10$\\
%\hline
\lsIV&2001-07-01;09:42:48&UVES&3280...4560&$-32\pm3$\\
%\lsIV&1983-04-29;10:36&IUE&$-3\pm3$\\
\lsIV&2002-07-03;17:49&FUSE&905...1185&$-35\pm4\pm10$\\
\lsIV&2004-05-10;18:11&FUSE&905...1185&$-6\pm3\pm10$\\
\hline
\end{tabular}
\end{center}
\end{table*}

Few radial velocities are reported in the literature. For \CD\ \citet{giddings1981} and \citet{drilling1987} measure consistently 36 $\pm 2\pm 7$ km\,s$^{-1}$ (statistical and systematic uncertainties) and 37$\pm$ 1 km\,s$^{-1}$, while \citet{kilkenny1989} give 18 $\pm$4 km\,s$^{-1}$. \citet{Hirsch2009} find the radial velocity of \CD\ to be constant at 36 $\pm$ 4\,km\,s$^{-1}$ in four consecutive nights and the same 174 days later. 
It is therefore likely that the \citet{kilkenny1989} as well as our measurements from FUSE spectra are less accurate than estimated, especially when taking the propagated error for FUSE from \citet{dixon2007} into account. We conclude that there is no evidence for RV variations in \CD.

For \uv\ \citet{Hirsch2009} reported radial velocities of 20 $\pm$4 km\,s$^{-1}$ and 22 $\pm$3 km\,s$^{-1}$ measured during a single night, which is consistent with v$_{\rm rad}$=24 $\pm$1 km\,s$^{-1}$ reported by \citet{drilling1987}.
For \lss\ \citet{drilling1987} give v$_{\rm rad}$=24 $\pm$1 km\,s$^{-1}$ somewhat larger than derived here. We are not aware of any published radial velocities for \lsIV\ and, therefore, conclude that there is insufficient observational material to judge whether \lss\ and \lsIV\ are radial velocity variable or not.   
%--------------------------------------------------------------------

%--------------------------------------------------------------------
\section{Model atmospheres \& synthetic spectra}
\label{atmospheres}
Visual spectra are well suited to derive the $T_{\text{eff}}$, $\log{g}$ and the helium abundance for example from the Pickering series of He\,{\sc ii}, 4686\AA, and several He\,{\sc i} lines.
Model atmospheres in NLTE and the corresponding synthetic spectra were calculated with the publicly available {\sc Tlusty 200}\footnote{\url{http://nova.astro.umd.edu/Tlusty2002}} and {\sc Synspec} 49\footnote{\url{http://nova.astro.umd.edu/Synspec49/synspec.html}} codes \citep{Hubeny1995,synspec94}.  The models are calculate in horizontally homogeneous, plane parallel geometry. A detailed description and manual is provided by \citet{2017arXiv170601859H,2017arXiv170601935H,2017arXiv170601937H}. The stellar atmospheres were assumed to be in radiative and hydrostatic equilibrium and include metal line-blanketing and the important broadening mechanisms. The most detailed model atoms were used that are available on the {\sc Tlusty} website. We used the default values of the standard keyword parameters. 

We calculated three different model grids. A grid containing only hydrogen and helium was constructed in order to compare with previous results. In addition, two grids with H/He and additional nitrogen or carbon (at the corresponding abundance for each sample star) were calculated to determine the atmospheric parameters and compare them to those derived by \citet{Hirsch2009}.\\
To fit the observational data, we made use of FITSB2 \citep{2004ASPC..318..402N}
spectral analysis program, and of SPAS, the Spectrum Plotting and Analysis Suite \citep{Hirsch2009}, which are based on the fitting procedure described by \citet{napiwotzki1999}. These programs allow us to measure and correct for the radial velocity shift of each spectrum and derive T$_{\text{eff}}$ and $\log{g}$ along with the abundance of a single element by comparing precalculated synthetic spectra to the observations by means of $\chi^2$-minimization. A spline fit is used to interpolate between the synthetic spectra, while the fitting itself and the error estimations are done via the downhill simplex algorithm. An overview about the model atoms used is given in Table \ref{ionization+levels}. Including all significant elements (H, He, C, N, O, Ne, Mg, Al, Si, P, S, Fe, Ni) with their relevant model atoms in NLTE conditions is crucial to correctly model the structure of the stellar atmosphere and to improve the line fits in general. 

\begin{table}%[h]
\caption{Ionization stages for which detailed model atoms were used in the model atmosphere calculations. The number of levels (L) and super-levels (SL) are listed. For each element the ground state of the next higher ionization stage was included, but not listed here.}
\label{ionization+levels}
\begin{center}
%\begin{tabular}{c c}

\begin{tabular}{lrr|lrr}
\hline\hline
Element&L&SL&element&L&SL\\
\hline
H&16&1 &             Mg\,{\sc ii}&21&4\\
He\,{\sc i}&24&0&          Al\,{\sc iii}&19&4\\[3pt]
He\,{\sc ii}&20& 0 &    Si\,{\sc iii}&31&15\\ [3pt] 
C\,\sc{ ii}&34&5 &     Si\,{\sc iv}&19&4\\
C\,{\sc iii}&34&12&      P\,{\sc iv}&14&0\\
C\,{\sc iv}&35&2 &   P\,{\sc v}&12&4\\[3pt]
N\,{\sc ii}&32&10 &       S\,{\sc iii}&29&12\\
N\,{\sc iii}&39&9 &       S\,{\sc iv}&29&12\\
N\,{\sc iv}&34&14 &       S\,{\sc v}&20&5\\
N\,{\sc v}&21&4 &    S\sc{ vi}&13&3\\[3pt]
O\,{\sc ii}&36&12 &       Fe\,{\sc iii}&0&50\\
O\,{\sc iii}&28&13 &      Fe\,{\sc iv}&0&43\\
O\,{\sc iv}&31&8 &   Fe\,{\sc v}&0&42\\[3pt]
Ne\,{\sc ii}&23&9 &        Ni\,\sc{ iv}&0&38\\
Ne\,{\sc iii}&12&2 &   Ni\,{\sc v}&0&48\\

\hline
\end{tabular}
\end{center}
\end{table}

Before starting the determination of atmospheric parameters and the abundance analysis, it is important to validate the fitting method and check if the calculated model atmospheres are appropriate. While determining abundances is quite straightforward, the atmospheric structure is often much harder to model. 

\subsection{Model atmospheres and the effects of additional absorbers}\label{model_grids}
The determination of atmospheric parameters requires to  properly 
account for line blanketing effects. As more metals are added, the total opacity increases and the effects of line blanketing back warming, and surface cooling are becoming more and more important, changing the temperature and density stratification of the atmosphere. Because accounting for many metal ions is demanding, very few attempts have been undertaken  to model hot subdwarf atmospheres \citep[e.g. by][]{lanz1997}.

To investigate this issue, we compared the atmospheric structure delivered by different model grids. A model atmosphere with a temperature of 47\,000\,K, a surface gravity of log(g)=5.7 \footnote{Surface gravities are given in cgs units throughout this paper} and a helium enrichment of 100 times hydrogen by numbers (similar to the one of \uv, for which C is the most abundant metal at ten times the solar abundance) was calculated. The abundances of all metals are listed in Table \ref{sample-abus34}. The atmospheric structure was computed with an increasing number of opacity sources. The following steps were taken:
\begin{enumerate}
\item H, He
\item H, He, C
\item H, He, C, N, O
\item H, He, C, N, O, Ne, Mg, Al, Si, P, S
\item H, He, C, N, O, Ne, Mg, Al, Si, P, S, Fe, Ni
\end{enumerate}

\begin{figure}[ht]
\begin{center}
\includegraphics[width=\columnwidth]{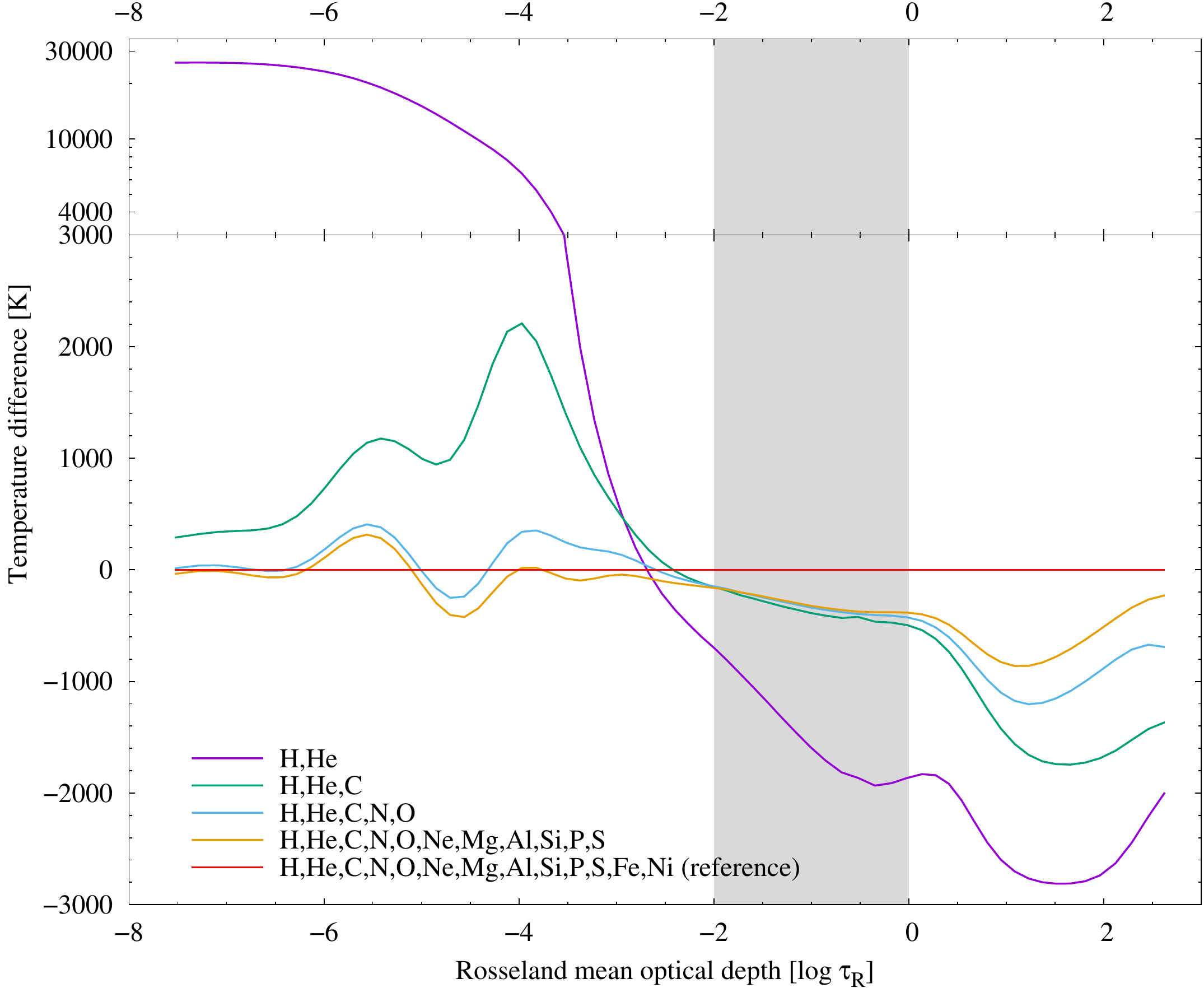}
\caption{
Temperature stratification of {\sc Tlusty} model atmospheres having different model atoms with respect to a fully line-blanketed reference model (No. 5). The shaded region marks the line forming region in the atmosphere. The y-axis shows the temperature difference with respect to the reference model, while the x-axis gives the Rosseland mean optical depth $\tau_{{\rm R}}$. The figure demonstrates the effects of metal opacities in NLTE atmospheres. The higher the opacity (the more opacity sources included) the cooler the upper atmosphere and the hotter the deeper layers (surface cooling and back warming). 
\label{tau_comp}}
\end{center}
\end{figure}

We use the full model (model 5) as our reference and plot the temperature distributions of the models 1 (H/He only) to 5 (full model) after subtracting that of the reference model in Fig. \ref{tau_comp}. The stratification of the H/He model (1) deviates drastically from that of the full model (5). 
As expected, adding absorbers cools down the outer atmospheric layers and heats up deeper layers significantly. However, in the line forming regions the temperature stratification of models 2 to 4 are almost identical and cooler than that of the full model (model 5) by 200K to 500K, only. We conclude that H/He/C composed model atmospheres with appropriately high abundances (He/H=100 and ten times solar carbon) produce a realistic temperature stratification in the line forming region. The same holds for N-rich He sdO stars if we replace carbon by nitrogen in the model calculations for grid No. 2.

{We considered an analysis of mock data more conclusive than the discussion of the temperature stratification. To this end we used the synthetic spectrum calculated from the full model (model 5) and fitted it with grids of synthetic spectra calculated from the less sophisticated models. The fit was performed using the H/He and the H/He/N grid. The parameters were chosen to be the final ones for \CD\ (T$_{\rm eff}$=42230\,K, $\log{g}$=5.6 and $\log{y}=\log{(n{\rm (He)}/n{\rm (H)}}=2.6$, see Sect. \ref{tghe}). Accordingly, the nitrogen abundance was about ten times solar and the abundances of other metals are as given in Table \ref{sample-abus1}.}

Despite the large differences in the temperature stratification of the H/He models, the best-fit effective temperature is hotter by 400K than that of the full model, only (see Table \ref{fig:mock_fit}). However, the surface gravity is significantly larger and the helium abundance lower by 0.1 dex and 0.8 dex, respectively.   

\begin{table}
\setstretch{1.3}
\caption{Spectral fit of a mock spectrum (parameters given in the 1. row) by a grid of H/He (2. row) and H/He/N (3. row) model spectra.\label{fig:mock_fit}}
\begin{center}
\begin{tabular}{l l l l}
\hline
\hline
Model & T$_{\rm eff}$ [K] & $\log$\,g & $\log$\,y \\ 
\hline
Mock data & 42230 & 5.60 & 2.60\\
H/He fit & 42600 & 5.70 & 1.79\\
H/He/N fit & 42290 & 5.58 & 2.40\\
\hline
\end{tabular}
\end{center}
\end{table}

The H/He/N grid reproduces the spectrum calculated from the fully blanketed model very well. The atmospheric parameters of the best-fit H/He/N model differ by as little as 10\,K, 0.02\,dex and 0.2\,dex for T$_{\text{eff}}$, $\log{g}$ and $\log{y}$, respectively. This reassures us that the use of model grids including the most abundant metal (C or N) at the appropriate high abundance is a good approximation to the computationally costly full models.

\section{Determination of effective temperature, surface gravity and abundances}
\label{atmos_params}

The effective temperature, surface gravity, and helium-to-hydrogen ratio are determined from the hydrogen and helium lines listed in Table \ref{list_h_he}. Another important atmospheric parameter is the abundance of the dominant metal (carbon and/or nitrogen, see Sect. \ref{atmospheres}).
Other metals also have an impact on the atmospheric temperature and density stratification, though at a lower level (see Sect. \ref{atmospheres}). Hence the following analysis strategy was designed.

We started with a pure H/He-grid in NLTE ranging from 35\,000\,K to 56\,000\,K in temperature and from 5.0 to 6.2 in $\log{g}$ to get a first estimate of the atmospheric parameters. The helium abundances covered the range from n(He)=n(H) to n(He)=2000 $\times$ n(H). The basic atmospheric parameters were determined with this grid.
Because of the strong carbon and/or nitrogen line spectrum, many helium lines are blended with C and/or N lines (see Figs. \ref{cd-helium} and \ref{uv-helium}), those blends had to be accounted for.
To that end a small sub-grid based on the preliminary atmospheric parameters was utilized to determine the carbon or nitrogen abundance of each star, depending on which element showed stronger lines by visual inspections of the spectra. The next step was to upgrade the H/He grid with carbon or nitrogen, using the previously determined value for the abundance. The H/He grid was upgraded to a H/He/C or a H/He/N grid (see Sect. \ref{model_grids}). The additional element was included in NLTE conditions. With this new grid, the atmospheric parameters ($T_{\text{eff}}$, log(g), n(He)/n(H)) were revised. This procedure improved 
the line fits considerably. 
In the next step we added additional metals by calculating new model atmospheres with solar abundance \citep{asplund2009} for each available additional ''visual/UVA'' element (O, Ne, Mg, Al, Si, P, S) as a starting point for the further analysis and the atmospheric parameters from the previous iteration step were utilized. For each element a wide range of abundances was chosen in order to avoid extrapolation in the process of fitting individual lines.
Afterwards, the abundances of all metals were calculated and fitted for one element after the other to the respective selected visual, UVA, and FUV lines.
The procedure involved a simultaneous fit to the lines of the particular ion and uncertainties were determined using the bootstrapping method \citep[see][]{napiwotzki1999}. After having determined the final abundance of one element, the next element was investigated. This means that all elements included in the model have an influence on the stratification of the atmosphere. Elements with a high expected abundance (i.e. from visual inspection of the spectra) were fitted first. These were the elements of C, N, O and Ne or Si, while elements with hardly any visible lines like S or P were fitted last.\\
To check for consistency between the visual, UVA and the FUV data and to improve the results, the fits were recalculated once more. This time iron and nickel were included from the beginning in the line formation and both the visual/UVA and FUV data were fitted for each ion. The final abundance is the error-weighted average of the ion-based abundances from the visual/UVA and the FUV. In those cases, where elements were only detectable in the visual/UVA, but not in the FUV, or vice versa, the abundance results are derived from only one wavelength range.

\section{Effective temperatures, surface gravity, and Helium-to-Hydrogen ratio}
\label{tghe}

\begin{figure*}
\begin{center}
\includegraphics[width=0.82\textwidth]{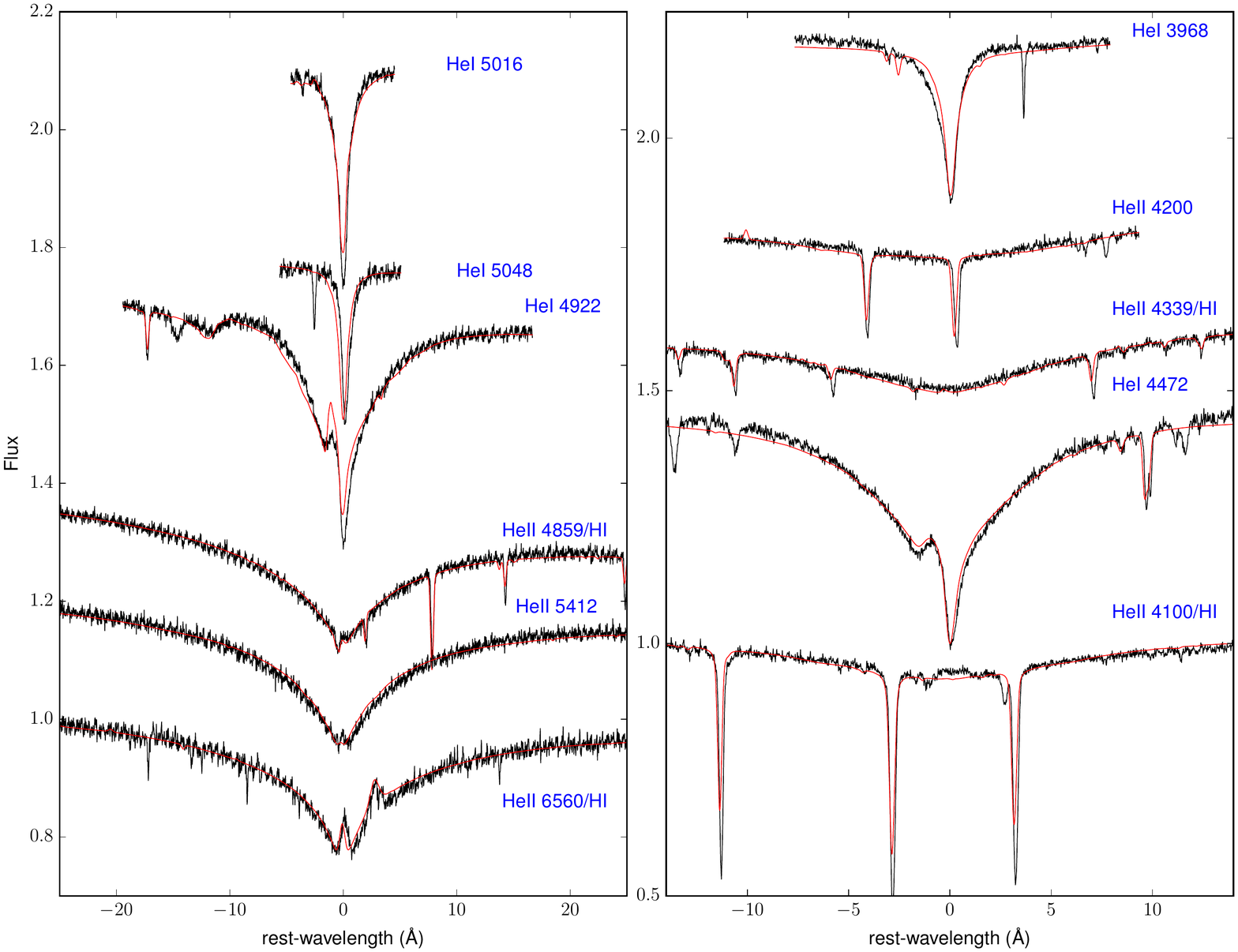}
\includegraphics[width=0.82\textwidth]{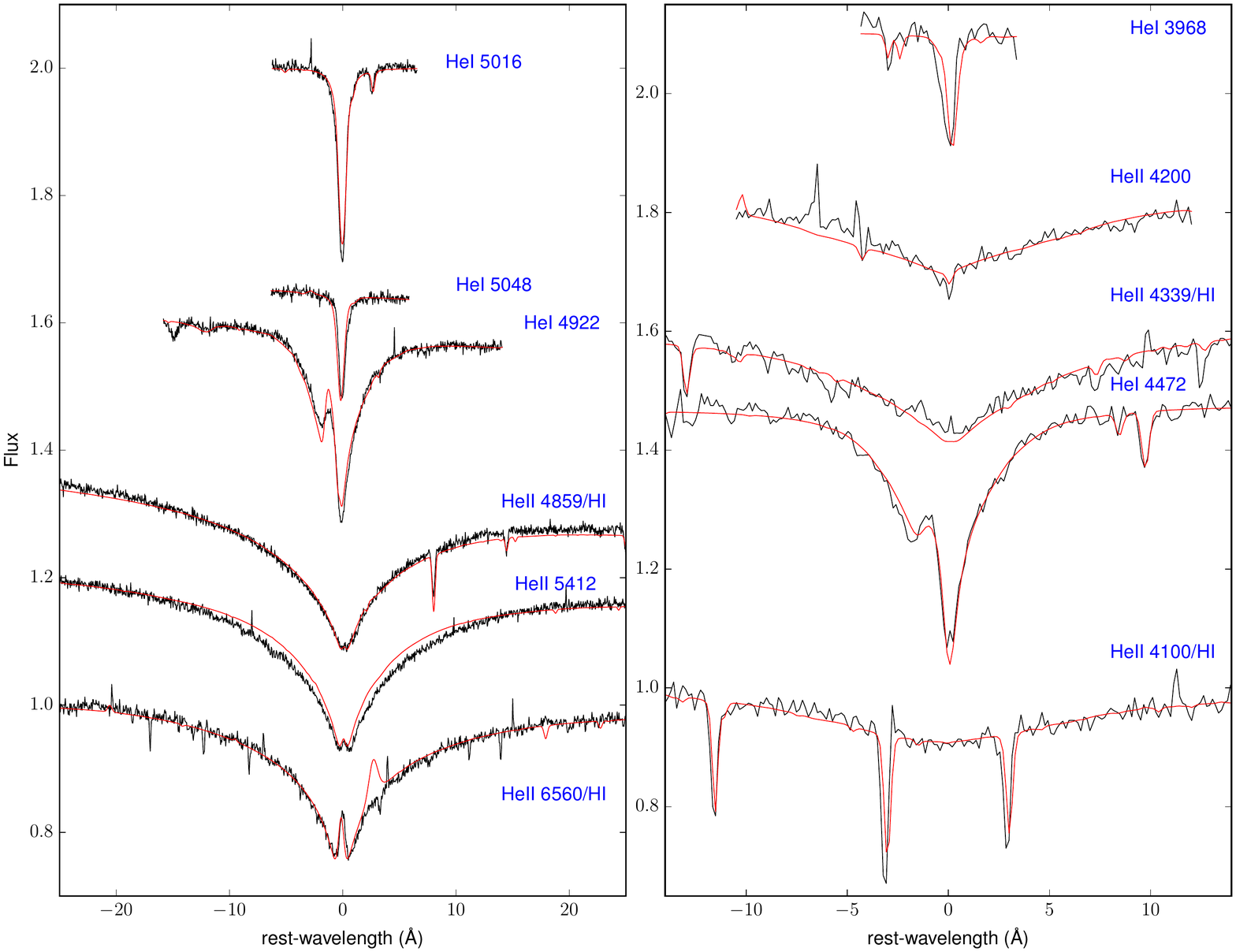}
\caption{{He\,{\sc i} and He\,{\sc ii} line profiles from the final models compared to the normalized observed spectra of \CD\ (UVES, top panels) and \lss\ (bottom panels, UVES left, CASPEC right). 
The spectra have been corrected for radial velocity. 
%and the fluxes of certain single lines were adjusted for a better visualization.
Because} it was not possible to match the He\,{\sc ii}/H\,{\sc i} 6560 line in \lss\ it was not included in the fit and is only shown for comparison. The very sharp absorption lines in the 6560 line wings are telluric lines. 
\label{cd-helium}}
\end{center}
\end{figure*}

%As described in section \ref{atmospheres} 
The first step was to determine the atmospheric parameters of each star from the visual helium/hydrogen lines (see Table \ref{list_h_he}). The first parameter estimate was done with a pure H/He grid. 
In the second step a H/He+C or H/He+N grid was used which were tailored to the specific object. 
  Figure \ref{cd-helium} shows the fit to the hydrogen and helium lines for \CD\ and \lss. The fits for \uv\ and \lsIV\ are shown in Fig.\ref{uv-helium} and Fig.\ref{lsiv-helium}. 
If possible, the same set of spectral lines was used in all stars. 
However, this was not always possible because of wavelength gaps, 
or artifacts in the spectra.
Because of the low hydrogen abundances the Balmer lines introduce small asymmetries to every second line in the He\,{\sc ii} Pickering series. At the appropriate He-to-H ratios all models predict H\,$\alpha$ emission.
For the nitrogen-rich \CD\ this component can be reproduced very well (see Fig. \ref{cd-helium}).   
Other Balmer lines have to be reproduced simultaneously and consistently as well. In this context it is worthwhile to note that the core of the He\,{\sc ii}/H\,$\beta$ line is blended by a nitrogen triplet, which is non-negligible contribution to the line profile and has to be accounted for at high nitrogen abundances.
However, the C-rich targets \lss\ and \uv\ do not show H\,$\alpha$ emission\footnote{The H$\alpha$ line is not observed for \lsIV.} components (see Figs. \ref{cd-helium} and \ref{uv-helium}). Hence, we are unable to reproduce this feature. Reducing the hydrogen content would remedy the issue, but the other lines of the Balmer series could not be reproduced. Hence, we rely on the higher Balmer lines.

\citet{Stroer2007} analyzed the high-resolution spectra of the SPY-sdO sample and found that  1-$\sigma$ statistical errors are typically
lower than 100\,K, 0.04, and 0.04 for T$_{\rm eff}$, $\log$\,g ,and
$\log$\,y, respectively. However, the error budget is dominated by systematic uncertainties. In the SPY project each target star was observed at least twice.
\citet{Stroer2007} estimated average systematic uncertainties by evaluating the distribution of differences in the fit parameters from individual spectra and find 1$\sigma$-values (sample standard deviation) of these distributions of $\Delta$T$_{\rm eff}$/T$_{\rm eff}$=0.026, $\Delta\log$\,g = 0.097 and $\Delta\log$\,y/$\log$\,y=0.097 for the sample. For our stars test calculations demonstrate that effective temperatures are better constrained by several ionization equilibria and the excellent quality of our spectra. However, the gravity uncertainties derived by \citet{Stroer2007} appear to be realistic, while that of the He/H ratio appears to be underestimated for the extremely helium-rich stars. The finally adopted uncertainties are given in Table \ref{sample-atmos} which summarizes the results for the effective temperature, surface gravity, helium-to-hydrogen number ratio ($\log(y)$), as well as the number ratio of the most abundant metal (($\log(X/H)$), that is N in the case of \CD, and C for the others) with respect to hydrogen. The rotation velocity as derived from metal lines (see Sect. \ref{vsini}) is also given. Because \lsIV\ is carbon and nitrogen-rich, we fitted the star with the N-rich model grid as well with the C-rich grid. The resulting atmospheric parameters are very similar (see Table \ref{sample-atmos}). 

%\subsection{Comparison with previous analyses}
%\label{previous}
%The atmospheric parameters derived in this work are in good agreement with the ones derived in previous analyses. Especially \lss, \uv\ and \lsIV\ show a very good compliance with the results of %\citet{Dreizler1993}. 
%The over- and under-abundances of certain metals is also reproduced by this work.

\uv\ has been analyzed by \citet{Hirsch2009} from the same FEROS visual spectra used here with synthetic spectra calculated with the TMAP code and hydrogen, helium and C composition, respectively. \citet{Hirsch2009} also analyzed \CD\ from a FEROS spectrum.
Hence, it is worthwhile to compare his results to ours (see Table \ref{sample-atmos}). 
%For \CD\ all parameter agree to within their mutual error ranges. %The same is true f
For \uv\ our analysis yields an effective temperature higher than that derived by \citet{Hirsch2009} by about 800\,K. For \CD\ both the effective temperature and the surface gravity are lower than derived by \citet{Hirsch2009}.

\begin{figure}%[h!]
\begin{center}
\includegraphics[width=\columnwidth]{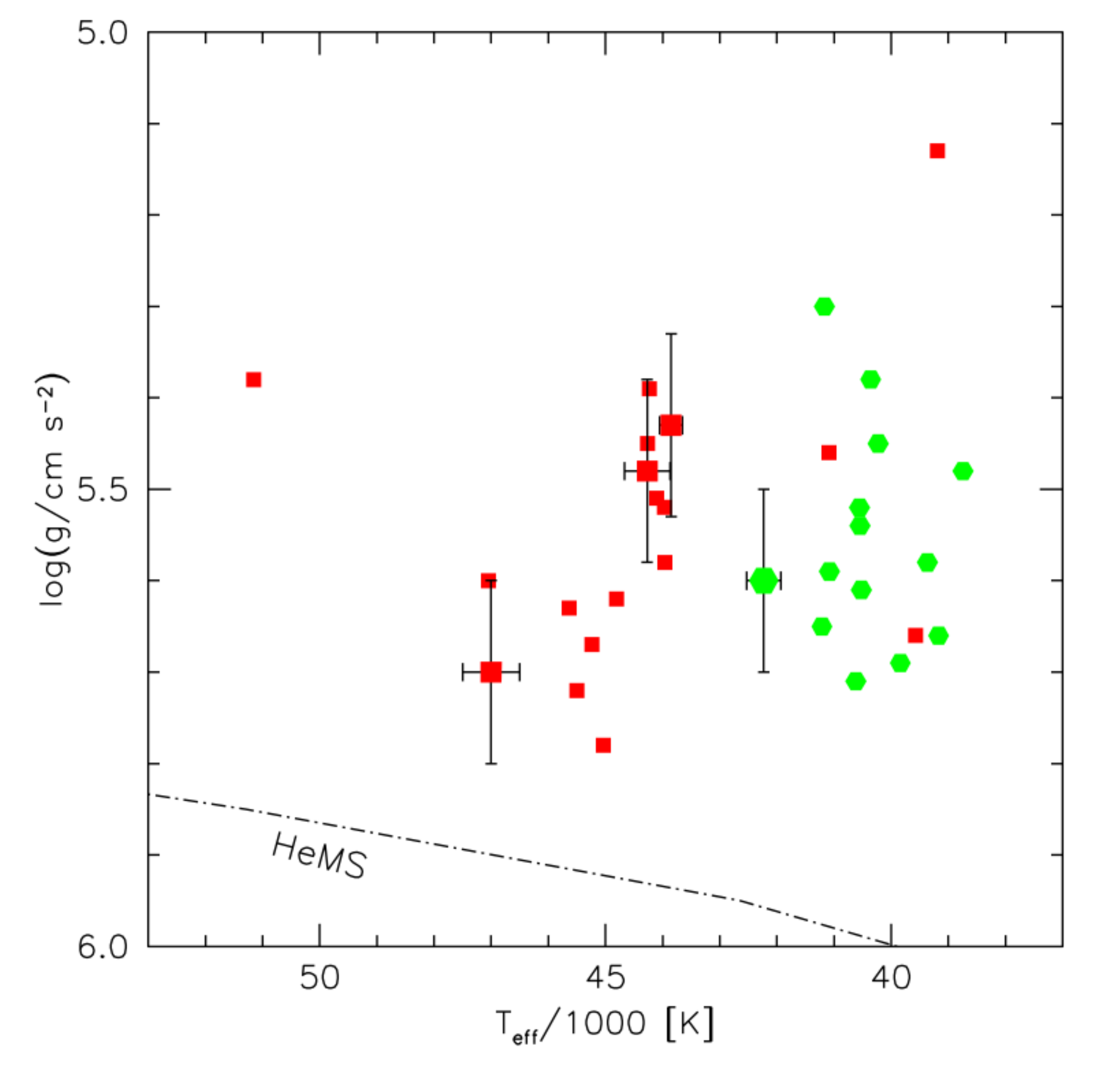}
\caption{Comparison of the position of the program stars in a T$_{\rm eff}$, $\log$\,g diagram to that the sample of \citet{Hirsch2009}. Stars with nitrogen dominated atmospheres are shown as green hexagons, while red square denote stars with carbon dominated atmospheres. The helium main sequence \citep{1971AcA....21....1P} %Paczynksi 1971
is shown for reference. } 
\label{tefflogg} 
\end{center}
\end{figure}

Figure \ref{tefflogg} compares the position of our program stars to the He-sdOs analyzed by \citet{Hirsch2009} in the T$_{\rm eff}$, $\log$\,g diagram. All stars lie slightly above the helium main sequence. The N-rich stars including \CD\ are cooler than the C-rich ones including \lss, \uv, and \lsIV, while both subclasses of He-sdOs have similar gravities.

\begin{table*}%[h!]
\setstretch{1.3}
\caption{Effective temperatures, surface gravities, H/He number ratio, abundance of the dominant metal X and projected rotational velocity v$_{\rm rot}\sin{i}$ of the sample stars, compared with the results of \citet{Hirsch2009}.\label{sample-atmos}}
\centering
\begin{tabular}{clcccccl}
\hline\hline
Name &$T_{\text{eff}}$ & $\log{g}$  & $\log{y}$&$\log(X/H)$ & v$_{\rm rot}\sin{i}$ & Spectrum & ref. \\ %$\log(X/H)$%\\ 
     &  [K]             &   [cgs]         &          &  &   [km\,s$^{-1}$]             &   \\
%\vspace{3px}
% table heading
\hline
\CD & $42230\pm300$  &  $5.60\pm0.1$  &$2.61\pm 0.20$   & $-0.31\pm 0.11$ ({N})    & $<$5 &  UVES  & 1\\
%    & $43170\pm550$  &$5.9\pm0.18$     &$2.70 \pm 0.20$  & $-0.18\pm 0.11$ (N)   & $<$5 &  FEROS &  1\\
    & $43000\pm250$  &$5.84\pm0.05$    &$2.75\pm 0.30$    & $-0.29\pm 0.17$ (N) & $<$10&  FEROS  &2\\[3pt]
%&$44000\pm1000$&$5.40_{-0.15}^{+0.30}$&--&\citet{bauer1995}&CASPEC&HHe\\
%\hline
\lss& $44270\pm400$  &$5.48\pm0.10$    &$2.17\pm0.25$    &   -0.01 $\pm 0.19$  (C)            &    $<5$   &  UVES  &  1\\[3pt]
%&$44500\pm1000$&$5.55\pm0.15$&2.0 (fixed)&\citet{Dreizler1993} (TMAP)&CASPEC&HHe\\
%\hline
\uv& $47000\pm500$   &$5.70\pm0.10$    &$2.0\pm0.3$      &       -0.11  $\pm 0.13$ (C)         &  $32\pm 3$    &  UVES  & 1\\
%&$46650\pm250$&$5.80\pm0.05$&2.0 (fixed)&\citet{Hirsch2009} (TMAP)& FEROS&HHe\\
   & $46170\pm250$   &$5.64\pm0.05$    & 1.91 $\pm 0.3$             &       -0.12 $\pm 0.2$(C)         & $30\pm 4$     &  FEROS  & 2 \\[3pt]
%&$46500\pm1000$&$5.55\pm0.15$&2.0 (fixed)&\citet{Dreizler1993} (TMAP)&CASPEC&HHe\\
%\hline
\lsIV& $43850\pm200$  &$5.43\pm0.1$   &$2.73\pm0.25$    &     +0.15  $\pm 0.13$    (C)      &  $<$5     &  UVES &1\\
     & $44000\pm250$  &$5.42\pm0.1$   &$2.76\pm0.20$    &      -0.13 $\pm$ 0.27    (N)       &  $<$5  &UVES& 1\\
%&$44500\pm1000$&$5.55\pm0.15$&2.0 (fixed)&\citet{Dreizler1993} (TMAP)&CASPEC&HHe\\
\hline
\multicolumn{3}{l}{$^1$=this work, $^2$=\citet{Hirsch2009}}\\
\end{tabular}

\end{table*}

\section{Angular diameters, interstellar reddening, and stellar masses}\label{sect:mass}
\label{angular}

\begin{figure*}
\begin{center}
\sidecaption
\includegraphics[width=12cm]{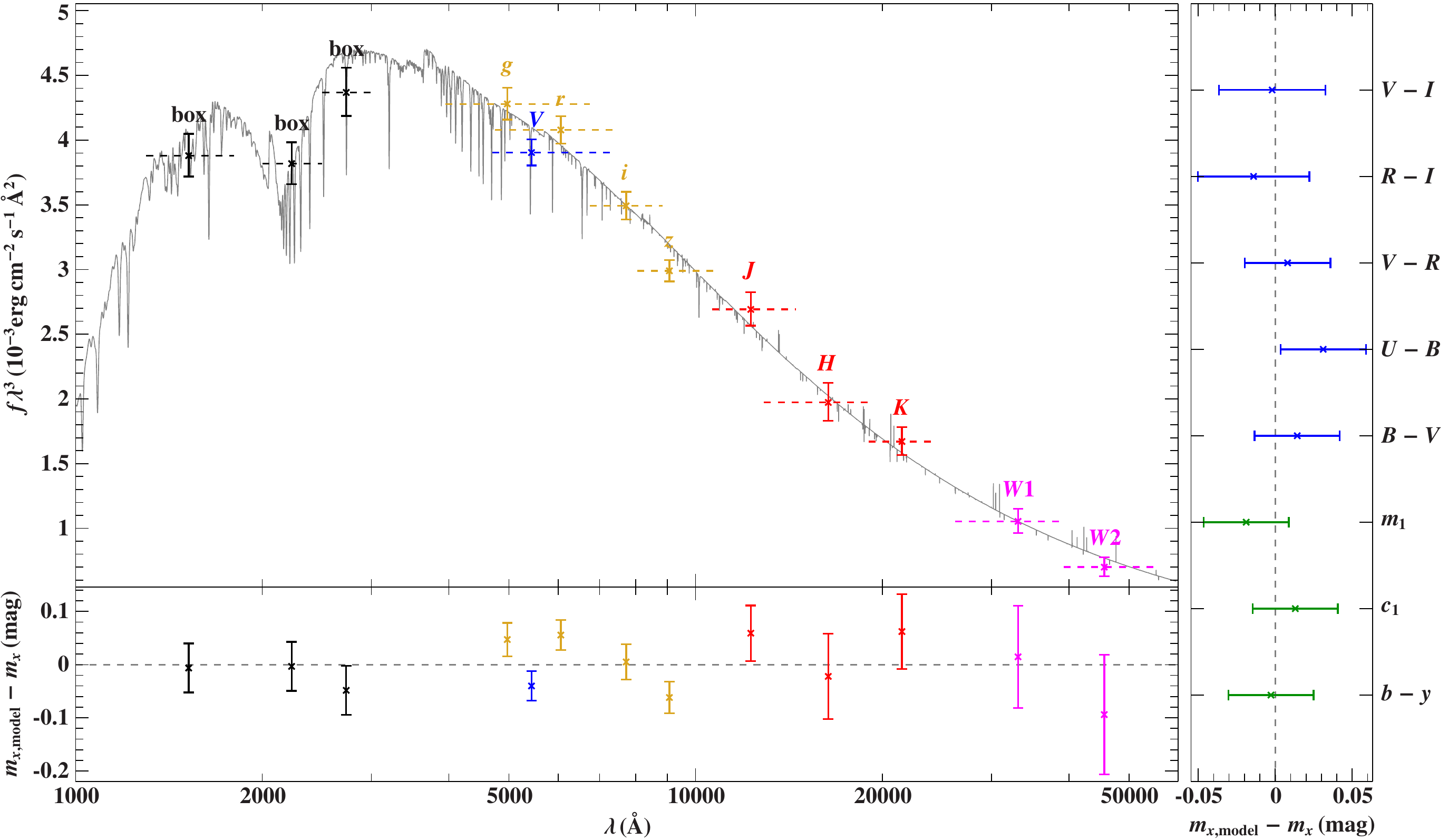}
\caption{Comparison of synthetic and observed photometry for \lss: The \textit{top panel} shows the spectral energy distribution. The three black data
points labeled ''box'' are fluxes converted from artificial magnitudes computed by means of box filters of the indicated width from a low-dispersion
IUE spectrum. Filter-averaged fluxes are shown as colored data points which were converted from observed magnitudes (the dashed horizontal lines indicate the respective filter widths), while the gray solid line represents a synthetic spectrum computed from a model atmosphere using the atmospheric parameters given in Table~\ref{sample-atmos}. 
The residual panels at the
bottom and right hand side  show the differences between synthetic and observed magnitudes/colors. The following color codes are used to identify the photometric systems: Johnson-Cousins (blue), Str\"omgren (green), SkyMapper(yellow), 2MASS (red), WISE (magenta).} 
\label{sed_fit_lss} 
\end{center}
\end{figure*}

\begin{table*}
\caption{Angular diameter $\Theta$, interstellar reddening A$_{\text{V}}$, interstellar reddening to infinity A$_{\text{V}_S}$\citep{schlafly}. 
%photometric distances and {distances from the Galactic plane} z for an adopted mass of 0.47 M$_{\odot}$ from the photometric fits of the SEDs. 
The Gaia parallaxes $\varpi$ and the stellar masses are given in the last two columns. } 
\label{sed_fit_results}
\setstretch{1.3}
\begin{center}
%\begin{tabular}{lllllll}
%\hline
%Star& $\Theta$ &  A$_{\text{V}}$& A$_{\text{V}_S}$ & d(photo) [pc] & z[pc] & plx [mas] & d(plx)[pc]\\
%\hline
%\CD&$3.5150\cdot10^{-11}$& 0.083& 1.678 &$227_{-35}^{+43}$ & -22$_{-3}^{+4}$ & 4.8619$\pm$0.0642 & 206$\pm 3$ \\
%\lss&$1.311\cdot10^{-11}$& 0.342& 1.683 &$718\pm 166$ & 68$\pm 16$& 1.4618$\pm$0.0461 & 684$\pm 22$ \\
%\lss&$4.6488\cdot10^{-11}$&0.354&1.683&$200_{-41}^{+52}$ &-18$_{-4}^{+5}$ & \\
%\lsIV&$1.7919\cdot10^{-11}$&0.162&0.196&$550_{-83}^{+98}$ &-179$_{-27}^{+32}$ & 2.041 $\pm$0.0737 & 490$\pm 16$ \\
%\uv&$1.7278\cdot10^{-11}$&0.063&0.063&$419_{-86}^{+108}$ & +179$_{-40}^{+51}$ &  2.1574$\pm$0.0737 & 464$\pm 15$\\
%\hline
\begin{tabular}{llllll}
\hline\hline
Star& $\Theta$ &  A$_{\text{V}}$& A$_{\text{V}_S}$ & Parallax $\varpi$ [mas] & Mass [M$_\odot$]\\
\hline
\CD&$3.561\cdot10^{-11}$& 0.056& 1.678 & 4.8619$\pm$0.0642 & 0.38 $\pm$ 0.09\\
\lss&$1.298\cdot10^{-11}$& 0.339& 1.683 & 1.4618$\pm$0.0461 & 0.43$\pm$ 0.10\\
\lsIV&$1.809\cdot10^{-11}$&0.164& 0.196& 2.041  $\pm$0.0628 & 0.38 $\pm$ 0.09 \\
\uv&$1.742\cdot10^{-11}$& 0.066& 0.063&  2.1574$\pm$0.0737 & 0.59 $\pm $ 0.14\\
\hline
\end{tabular}
\end{center}
\end{table*}

Once the atmospheric parameters have been determined, photometric measurements allow additional parameters to be determined, in particular the stellar angular diameter and the interstellar extinction. If the distance is known, the stellar mass can be derived. To this end we constructed spectral energy distributions from photometric measurements ranging from the ultraviolet (IUE) to the infrared \citep[J,H,K; 2MASS ,UKIDS, ][]{2006AJ....131.1163S,2007MNRAS.379.1599L} and \citep[W1,W2;  WISE,][]{2013yCat.2328....0C}. Magnitudes and colors in the Johnson \citep{1997A&AS..124..349M,2007AJ....133.2502L}, Str\"omgren \citep{2015A&A...580A..23P}, Tycho \citep{2000A&A...355L..27H}, and SDSS \citep{2016yCat.2336....0H,
2018PASA...35...10W} photometric systems were fitted along with the SEDs \citep[for details see][]{2018OAst...27...35H}. %Heber et al. 2018, Open Astronomy
The photometric filter systems are inhomogeneous with respect to their bandwidth. Also the precision of the measured magnitudes may differ for the different filters. Ultraviolet spectra from IUE cover the wavelength range from 1150\AA\ to 3150\AA\ at a spectral resolution of 6\,\AA, whereas visual and infrared magnitudes are derived both from narrow band (e.g. Str\"omgren) and wide-band (e.g. Johnson, 2MASS, WISE) photometry.
Because we have to combine UV spectrophotometry from IUE with broad and intermediate band visual and infrared
photometry a suitable set of filters were defined to derive UV-magnitudes from IUE spectra that comprises of three box filters, which cover the spectral ranges $1300$--$1800$\,\AA, $2000$--$2500$\,\AA, and $2500$--$3000$\,\AA.
Interstellar extinction is accounted for by multiplying the synthetic flux with a reddening factor $10^{-0.4 A(\lambda)}$ using the extinction curve of  \citet{1999PASP..111...63F}. 
We match the synthetic SED calculated from the final model to the observed one by adjusting the angular diameter and the interstellar reddening parameter E(B-V) assuming an extinction parameter $R_V = A(V) / E(B-V)$ = 3.15.
In Figures \ref{sed_fit_lss} (\lss) and \ref{sed_fit_other} (\CD, \lsIV, and \uv) we display the residuals (O-C) of the SED and color fits and plot the SEDs as flux density times the wavelength to the power of three (F$_\lambda \lambda^3$) as a function of wavelength to reduce the steep slope of the SED on such a broad wavelength range.

The resulting angular diameter and interstellar reddening parameters are
listed in Table \ref{sed_fit_results}.  
Because of the quite low Galactic {distances z} of the program stars {from the Galactic plane}, we expect interstellar reddening to be substantial. Therefore, we also give the reddening to infinity from \citet{2011ApJ...737..103S}. The reddening for \uv\ is consistent while for the three others it is found to be significantly less than derived from the maps of \citet{2011ApJ...737..103S} indicating that most of the reddening takes place beyond the stars' distances. {Reddening parameters for \uv\ and \lsIV\ can be estimated from the 3D dust maps of \citet{2018MNRAS.478..651G} and are consistent with those derived from the SED.}
Figures \ref{sed_fit_lss} (\lss) and \ref{sed_fit_other} (\CD, \lsIV, and \uv) demonstrate that the synthetic SEDs reproduce the observed ones all the way from  the ultraviolet to the infrared. No infrared excess becomes apparent for any of our program stars. For \lsIV\ this is at odds with the results by \citet{ulla1998} who found an infrared excess for \lsIV\ and concluded that this star has a companion of spectral type F4 to F6. However, their infrared photometry are brighter by up to two magnitudes than the 2MASS photometry we used (see Table \ref{magnitudes}). We have no reason to believe that the star is variable in light and conclude that the claimed IR excess is spurious.

\begin{table}

\caption{Comparison of photometric magnitudes J, H and K from \citet{ulla1998} and \citet{cutri2003}.}\label{magnitudes}
\setstretch{1.3}
\begin{center}
\begin{tabular}{ccc}
\hline\hline
&\citet{ulla1998}&\citet{cutri2003}\\
\hline
J&$11.15\pm0.03$&$12.632\pm0.03$\\
H&$10.98\pm0.05$&$12.813\pm0.02$\\
K&$10.90\pm0.05$&$12.916\pm0.04$\\
\hline
\end{tabular}
\end{center}
\end{table}

The recent, second data release of Gaia \citep{2018A&A...616A...1G} provided high-precision parallaxes $\varpi$ for the program stars. 
  Hence we can derive the stellar mass from the parallax, the angular diameter and the surface gravity via 
  \begin{equation}\label{eq:one}
  M = g\,\Theta^2/(4G\,\varpi^2). 
  \end{equation}
  
  The uncertainty of the mass {is}  
  \begin{equation}\label{eq:two}
  {\Delta M}/{M} = d\ln\,M =
  \sqrt{(A\Delta\log\,g)^2 + 4(\frac{\Delta\Theta}{\Theta})^2 + 4(\frac{{\Delta}\varpi}{\varpi})^2)}, 
  \end{equation}
   
   ($A=\ln{10}$),
  which for the program stars is dominated by the uncertainty of the gravity.
The masses, listed in Table\,\ref{sed_fit_results}, scatter around the canonical mass for the core helium flash with \CD\ {and \lsIV\ }at the low end and \uv\ at the high one.  

%\clearpage

\section{Metal abundances}
\label{sect:results}

\begin{table*}%[h!]
\setstretch{1.3}
 \caption{Surface abundances of \CD\ (left hand panel) and  for \lss\ (right hand panel) as derived from visual/UVA and FUV data. The values are given as $\log{(n({\rm X})/n({\rm H}))}$. A \glqq -\grqq\ indicates that no usable lines of the element were found. Abundance are listed as derived from each ion accessible. 
A bold element symbol indicates the abundance from all ionization stages combined. The finally adopted abundances are the error-weighted averages of the results from the visual/UVA and FUSE spectra, if available. Abundances with uncertainties exceeding 0.5 dex are given in parentheses (see text). 
}\label{sample-abus1}
\begin{minipage}{\columnwidth}
\begin{tabular}{cccc}
\hline\hline
\multicolumn{4}{c}{\CD}\\
Element & Visual/UVA & FUSE & Adopted\\
\hline\\
%\vspace{3pt}
He\,{\sc i} \& He\,{\sc ii} &$2.61\pm0.20$&$-$&$2.61\pm0.20$\\[3pt]
%\hline
C\,{\sc iii}&$-2.30\pm0.11$&$-1.72\pm0.15$&\\
C\,{\sc iv}&$-2.09\pm0.04$&$-1.84\pm0.21$&\\
\textbf{C}&$-2.14\pm0.13$&$-1.76\pm0.18$&$-2.01\pm0.16$\\[3pt]
N\,{\sc ii}&$-0.25\pm0.06$&$-$&\\
N\,{\sc iii}&$-0.24\pm0.02$&$-0.37\pm0.12$&\\
N\,{\sc iv}&$-0.27\pm0.02$&$-0.41\pm0.12$&\\
\textbf{N}&$-0.25\pm0.10$&$-0.39\pm0.12$&$-0.31\pm0.11$\\[3pt]
O\,{\sc ii}&$-2.09\pm0.04$&$-$&\\
O\,{\sc iii}&$-2.29\pm0.05$&$-2.12\pm0.08$&\\
O\,{\sc iv}&$-$&$-2.58\pm0.28$&\\
\textbf{O}&$-2.19\pm0.11$&$-2.24\pm0.14$&$-2.21\pm0.13$\\[3pt]
Ne\,{\sc ii}&$-0.98\pm0.10$&$-0.62\pm0.14$&$-0.83\pm0.12$\\[3pt]
Mg\,{\sc ii}&$-1.56\pm0.10$&$-$&$-1.56\pm0.10$\\[3pt]
Al\,{\sc iii}&$-2.61\pm0.10$&$-2.93\pm0.32$&$-2.68\pm0.21$\\[3pt]
Si\,{\sc iii}&$-1.57\pm0.10$&$-1.75\pm0.14$&\\
Si\,{\sc iv}&$-1.78\pm0.15$&$-1.75\pm0.18$&\\
\textbf{Si}&$-1.65\pm0.16$&$-1.75\pm0.16$&$-1.70\pm0.16$\\[3pt]
P\,{\sc iv}&$-$&$-3.92\pm0.11$&\\
P\,{\sc v}&$-$&$-3.83\pm0.13$&\\
\textbf{P}&$-$&$-3.88\pm0.15$&$-3.88\pm0.15$\\[3pt]
S\,{\sc iii}&$-1.79\pm0.03$&$-$&\\
S\,{\sc iv}&$-2.22\pm0.07$&$-2.09\pm0.10$&\\
S\,{\sc v}& -- &$-1.87\pm0.41$&\\
S\,{\sc v}& --& --&\\
\textbf{S}&$-1.93\pm0.11$&$-2.05\pm0.27$&$-1.96\pm0.19$\\[3pt]
\textbf{Fe}&$-$&$-2.01\pm0.27$&$-2.01\pm0.27$\\[3pt]
\textbf{Ni}&$-$&$-2.49\pm0.23$&$-2.49\pm0.23$\\
\hline
\end{tabular}
\end{minipage}
\begin{minipage}[t]{\columnwidth}
\begin{tabular}{cccc}
\hline\hline
\multicolumn{4}{c}{\lss}\\
Element &Visual/UVA &FUSE & Adopted\\
\hline\\
%\vspace{3pt}
He\,{\sc i} \& He\,{\sc ii} &$2.17\pm0.25$&$-$&$2.17\pm0.25$\\[3pt]
C\,{\sc iii}&$-0.01\pm0.06$&$0.11\pm0.11$&\\
C\,{\sc iv}&$-0.14\pm0.08$&$0.07\pm0.34$&\\
\textbf{C}&$-0.06\pm0.12$&$0.09\pm0.25$&$-0.01\pm0.19$\\[3pt]
N\,{\sc ii}&$(-1.29\pm0.71)$&$-$&\\
N\,{\sc iii}&$-1.19\pm0.06$&$-1.07\pm0.36$&\\
N\,{\sc iv}&$-1.01\pm0.21$&$-1.05\pm0.11$&\\
\textbf{N}&$-1.15\pm0.17$&$-1.06\pm0.26$&$-1.11\pm0.22$\\[3pt]
O\,{\sc ii}&$(-1.60\pm0.80)$&$-$&\\
O\,{\sc iii}&$-1.90\pm0.22$&$-1.75\pm0.21$&\\
O\,{\sc iv}&$-$&$(-1.40\pm0.59)$&\\
\textbf{O}&$-1.90\pm0.24$&$-1.75\pm0.23$&$-1.82\pm0.24$\\[3pt]
Ne\,{\sc ii}&$-0.70\pm0.11$&$-0.76\pm0.31$&$-0.71\pm0.21$\\[3pt]
Mg\,{\sc ii}&$-1.96\pm0.29$&$-$&$-1.96\pm0.29$\\[3pt]
Al\,{\sc iii}&$-1.93\pm0.31$&$-2.25\pm0.30$&$-2.09\pm0.30$\\[3pt]
Si\,{\sc iii}&$-2.23\pm0.21$&$-2.19\pm0.25$&\\
Si\,{\sc iv}&$-2.23\pm0.14$&$-2.08\pm0.03$&\\
\textbf{Si}&$-2.23\pm0.20$&$-2.09\pm0.17$&$-2.16\pm0.19$\\[3pt]
P\,{\sc iv}&$-$&$-4.02\pm0.19$&\\
P\,{\sc v}&$-$&$-3.80\pm0.11$&\\
\textbf{P}&$-$&$\it{-3.80\pm0.15}$&$-3.80\pm0.15$\\[3pt]
S\,{\sc iii}&$-$&$-$&\\
S\,{\sc iv}&$-2.46\pm0.06$&$-2.54\pm0.13$&\\
S\,{\sc v}&$-$&$-2.32\pm0.39$&\\
S\,{\sc vi}&$-$&$(-2.41\pm0.70)$&\\
\textbf{S}&$-2.46\pm0.11$&$-2.49\pm0.28$&$-2.47\pm0.19$\\[3pt]
\textbf{Fe}&$-$&$-2.42\pm0.18$&$-2.42\pm0.18$\\[3pt]
\textbf{Ni}&$-$&$-2.62\pm0.29$&$-2.62\pm0.29$\\
\hline
\end{tabular}
\end{minipage}
\end{table*}

Having determined the atmospheric parameters we proceed to derive metal abundances.
We used the NLTE synthetic spectra calculated with {\sc Tlusty}/{\sc Synspec} to derive the elemental abundance of C, N, O, Ne, Mg, Si, P, S, Fe, and Ni from selected lines and 
line blends were added subsequently in the spectral synthesis (see Sect. \ref{line_id}). Finally, the synthetic spectra were compared to the observed spectra to identify outliers. 
Many ions display spectral lines in the visual/UVA as well as in the  FUSE spectral range. This allows a comparison of abundances derived from different spectral ranges to be made, which is a crucial cross-check for the reliability of FUV results which suffer from line crowding and uncertainties of continuum adjustment. For iron and nickel we have to rely on the analyses of FUV spectra. In Fig. \ref{LSS-C}  we compare the final synthetic spectrum to the observed spectra of \lss\ by selecting lines of different strengths, ionization stages and from different wavelength regimes to demonstrate the quality of fits achieved.

Several chemical elements (C,N,O, Si, P, S, Fe, and Ni) display spectral lines arising from two or three stages of ionization. Because ionization equilibria are very sensitive to effective temperature, the inspection of the chemical abundances derived from different ions is an important check on the effective temperatures derived from the He\,{\sc i} and He\,{\sc ii} lines. As demonstrated in Tables \ref{sample-abus1} and \ref{sample-abus34}   
the ionization equilibria for almost all elements are intact for the four stars, irrespective of whether they were obtained from the visual or the FUSE spectra. Elemental abundances are given as the error averaged mean values derived from different stages of ionization if available.
The comparison of abundances derived from the visual/UVA to that from the FUV is also presented in Table \ref{sample-abus1} and \ref{sample-abus34}. There seems to be a slight tendency for abundances from the FUV to be larger than those from the visual/UVA by about 0.2 dex in the case of \uv. However, for the other three stars the difference between FUSE and visual abundance scatter around zero by about 0.2 dex in most cases. Hence, we conclude that abundances derived from FUSE and and visual/UVA spectra are consistent to within error margins. The finally adopted values are the error weighted averages from both spectral ranges, the errors are the average from the single errors. The finally adopted abundances were derived from error weighted averages of abundances from the different wavelength regimes if available.     
Test calculation have shown that the uncertainties of the atmospheric parameters have a minor effect on the elemental abundances, except for O, Ne and Ni lines in the FUV. A very conservative estimate of the corresponding error is +-0.1dex for the results from the visual/UVA. The same number is appropriate for the FUV  range for all elements but O, Ne, and Ni, for which we adopt 0.15 dex. The adopted abundance error is calculated by adding these numbers in quadrature before averaging the abundances from the two wavelength ranges. Abundances with uncertainties exceeding 0.5 dex are not included in the averaging procedure.

\begin{table*}%[h!]
\setstretch{1.3}
\caption{Same as Table \ref{sample-abus1}, but for \uv\ (left hand panel) and \lsIV\ (right hand panel).}
%\caption{Left hand panel: Surface abundances of \uv\ as derived from the visual FEROS spectrum (second column) and the FUSE FUV spectrum (third column). The values are given as $\log{(n({\rm X})/n({\rm H}))}$. A bold element symbol indicates the abundance from all ionization stages combined.
% A \glqq --\grqq\ indicates that no usable lines of this element were found. The finally adopted abundances are the error-weighted averages of the results form the FUSE and visual spectra, if available.
%Right hand panel: Same, but for \lsIV.}
\label{sample-abus34}
\begin{minipage}{\columnwidth}
\begin{tabular}{cccc}
\hline\hline
\multicolumn{4}{c}{\uv}\\
Element & Visual/UVA & FUSE & Adopted\\
%\vspace{3px}
% table heading
\hline
He\,{\sc i} \& He\,{\sc ii}&$2.0\pm0.3$&$-$&$2.0\pm0.3$\\[3pt]
%\hline
C\,{\sc iii}&$-0.44\pm0.07$&$0.09\pm0.06$&\\
C\,{\sc iv}&$0.11\pm0.05$&$-0.13\pm0.27$&\\
\textbf{C}&$-0.17\pm0.06$&$0.05\pm0.19$&$-0.11\pm0.13$\\[3pt]
%\hline
N\,{\sc ii}&$-$&$-$&\\
N\,{\sc iii}&$-1.69\pm0.04$&$-1.49\pm0.08$&\\
N\,{\sc iv}&$-1.99\pm0.37$&$-1.53\pm0.11$&\\
\textbf{N}&$-1.70\pm0.23$&$-1.51\pm0.14$&$-1.58\pm0.19$\\[3pt]
%\hline
O\,{\sc ii}&$-2.07\pm0.37$&\\
O\,{\sc iii}&$-2.34\pm0.16$&$-1.94\pm0.13$&\\
O\,{\sc iv}&$-$&$(-2.18\pm0.57)$&\\
\textbf{O}&$-2.25\pm0.28$&$-1.94\pm0.16$&$-2.05\pm0.22$\\[3pt]
%\hline
Ne\,{\sc ii}&$-1.03\pm0.20$&$-0.96\pm0.08$&\\
Ne\,{\sc iii}&$-$&$-0.94\pm0.05$&\\
\textbf{Ne}&$-1.03\pm0.22$&$-0.95\pm0.12$&$-0.97\pm0.16$\\[3pt]
%\hline
Mg\,{\sc ii}&$-2.41\pm0.16$&$-$&$-2.41\pm0.16$\\[3pt]
%\hline
Al\,{\sc iii}&$-3.05\pm0.23$&$(-2.91\pm0.62)$&$-3.05\pm0.23$\\[3pt]
%\hline
%Si\,{\sc iii}&$(-2.03\pm0.91)$&$-2.43\pm0.30$&\\
Si\,{\sc iii}&$-$&$-2.43\pm0.30$&\\
Si\,{\sc iv}&$-2.58\pm0.08$&$(-3.15\pm0.54)$&\\
\textbf{Si}&$-2.58\pm0.13$&$-2.43\pm0.32$&$-2.53\pm0.22$\\[3pt]
%\hline
P\,{\sc iv}&$-$&$-4.33\pm0.27$&\\
P\,{\sc v}&$-$&$-4.01\pm0.12$&\\
\textbf{P}&$-$&$-4.12\pm0.22$&$-4.12\pm0.22$\\[3pt]
%\hline
S\,{\sc iii}&$-3.38\pm0.14$&$-$&\\
S\,{\sc iv}&$-$&$-3.18\pm0.21$&\\
S\,{\sc v}&$-$&$-2.77\pm0.19$&\\
S\,{\sc vi}&$-$&$(-2.69\pm1.03)$&\\
\textbf{S}&$-3.38\pm0.17$&$-2.94\pm0.22$&$-3.18\pm0.20$\\[3pt]
%\hline
%Fe\,{\sc iii}&$-$&$-2.23\pm0.10$&\\
%Fe\,{\sc iv}&$-$&$(-2.70\pm0.68)$&\\
%Fe\,{\sc v}&$-$&$-2.54\pm0.21$&\\
%\textbf{Fe}&$-$&$-2.33\pm0.18$&$-2.33\pm0.18$\\
\textbf{Fe}&$-$&$-2.68\pm0.25$&$-2.68\pm0.25$\\[3pt]
%\hline
%Ni\,{\sc iv}&$-$&$-3.01\pm0.41$&\\
%Ni\,{\sc v}&$-$&$-3.19\pm0.26$&\\
%\textbf{Ni}&$-$&$-3.13\pm0.37$&$-3.13\pm0.37$\\
\textbf{Ni}&$-$&$-3.11\pm0.26$&$-3.11\pm0.26$\\
\hline
\end{tabular}
\end{minipage}
\begin{minipage}{\columnwidth}
\begin{tabular}{cccc}
\hline\hline
\multicolumn{4}{c}{\lsIV}\\
Element & Visual/UVA & FUSE & Adopted\\
%\vspace{3px}
% table heading
\hline
He\,{\sc i} \& He\,{\sc ii}&$2.73\pm0.25$&$-$&$2.73\pm0.25$\\[3pt]
%\hline
C\,{\sc iii}&$0.17\pm0.03$&$0.12\pm0.10$&\\
C\,{\sc iv}&$-$&$0.09\pm0.15$&\\
\textbf{C}&$0.17\pm0.10$&$0.11\pm0.16$&$0.15\pm0.13$\\[3pt]
%\hline
N\,{\sc ii}&$-0.01\pm0.04$&$-$&\\
N\,{\sc iii}&$-0.13\pm0.21$&$-0.18\pm0.21$&\\
N\,{\sc iv}&$-0.21\pm0.07$&$-0.33\pm0.47$&\\
\textbf{N}&$-0.09\pm0.19$&$-0.22\pm0.36$&$-0.13\pm0.27$\\[3pt]
%\hline
O\,{\sc ii}&$-0.70\pm0.03$&$-$&\\
O\,{\sc iii}&$-0.75\pm0.06$&$-0.81\pm0.05$&\\
O\,{\sc iv}&$-$&$-0.89\pm0.26$&\\
\textbf{O}&$-0.72\pm0.11$&$-0.82\pm0.18$&$-0.76\pm0.15$\\[3pt]
%\hline
Ne\,{\sc ii}&$0.07\pm0.02$&$0.25\pm0.31$&\\
Ne\,{\sc iii}&$-$&$0.15\pm0.27$&\\
\textbf{Ne}&$0.07\pm0.10$&$0.17\pm0.31$&$0.09\pm0.21$\\[3pt]
%\hline
Mg\,{\sc ii}&$-1.21\pm0.10$&$-$&$-1.21\pm0.10$\\[3pt]
%\hline
Al\,{\sc iii}&$-2.13\pm0.10$&$(-2.21\pm0.60)$&$-2.13\pm0.10$\\[3pt]
%\hline
Si\,{\sc iii}&$-1.21\pm0.06$&$-1.07\pm0.14$&\\
Si\,{\sc iv}&$-1.17\pm0.03$&$-0.96\pm0.24$&\\
\textbf{Si}&$-1.18\pm0.11$&$-1.03\pm0.21$&$-1.12\pm0.16$\\[3pt]
%\hline
P\,{\sc iv}&$-2.87\pm0.10$&$-3.40\pm0.15$&\\
P\,{\sc v}&$-$&$-3.02\pm0.35$&\\
\textbf{P}&$-2.87\pm0.15$&$-3.29\pm0.27$&$-3.02\pm0.21$\\[3pt]
%\hline
S\,{\sc iii}&$(-3.53\pm0.87)$&$-$&\\
S\,{\sc iv}&$-$&$-2.18\pm0.19$&\\
S\,{\sc v}&$-$&$-1.91\pm0.19$&\\
S\,{\sc vi}&$-$&$(-2.37\pm0.85)$&\\
\textbf{S}&$(-3.53\pm0.87)$&$-2.05\pm0.21$&$-2.05\pm0.21$\\[3pt]
\textbf{Fe}&$-$&$-1.89\pm0.16$&$-1.89\pm0.16$\\[3pt]
%\hline
%Ni\,{\sc iv}&$-$&$-2.32\pm0.15$&\\
%Ni\,{\sc v}&$-$&$-2.69\pm0.42$&\\
%\textbf{Ni}&$-$&$-2.41\pm0.30$&$-2.41\pm0.30$\\
\textbf{Ni}&$-$&$-2.20\pm0.31$&$-2.20\pm0.31$\\[3pt]
\hline
\end{tabular}
\end{minipage}
\end{table*}

\subsection{Abundance by element}

The available spectra and NLTE model atoms for TLUSTY/SYNSPEC enabled us to derive elemental abundance for C, N, O, Ne, Mg, Al, Si, P, S, Fe, and Ni. Tables \ref{sample-abus1} and \ref{sample-abus34} lists the resulting number abundances ratios with respect to hydrogen. 

%The sample can be divided in two subgroups.
\subsubsection{Carbon, nitrogen, and oxygen}\label{sect:cno}

\begin{sidewaysfigure*}
\includegraphics[width=1\textwidth]{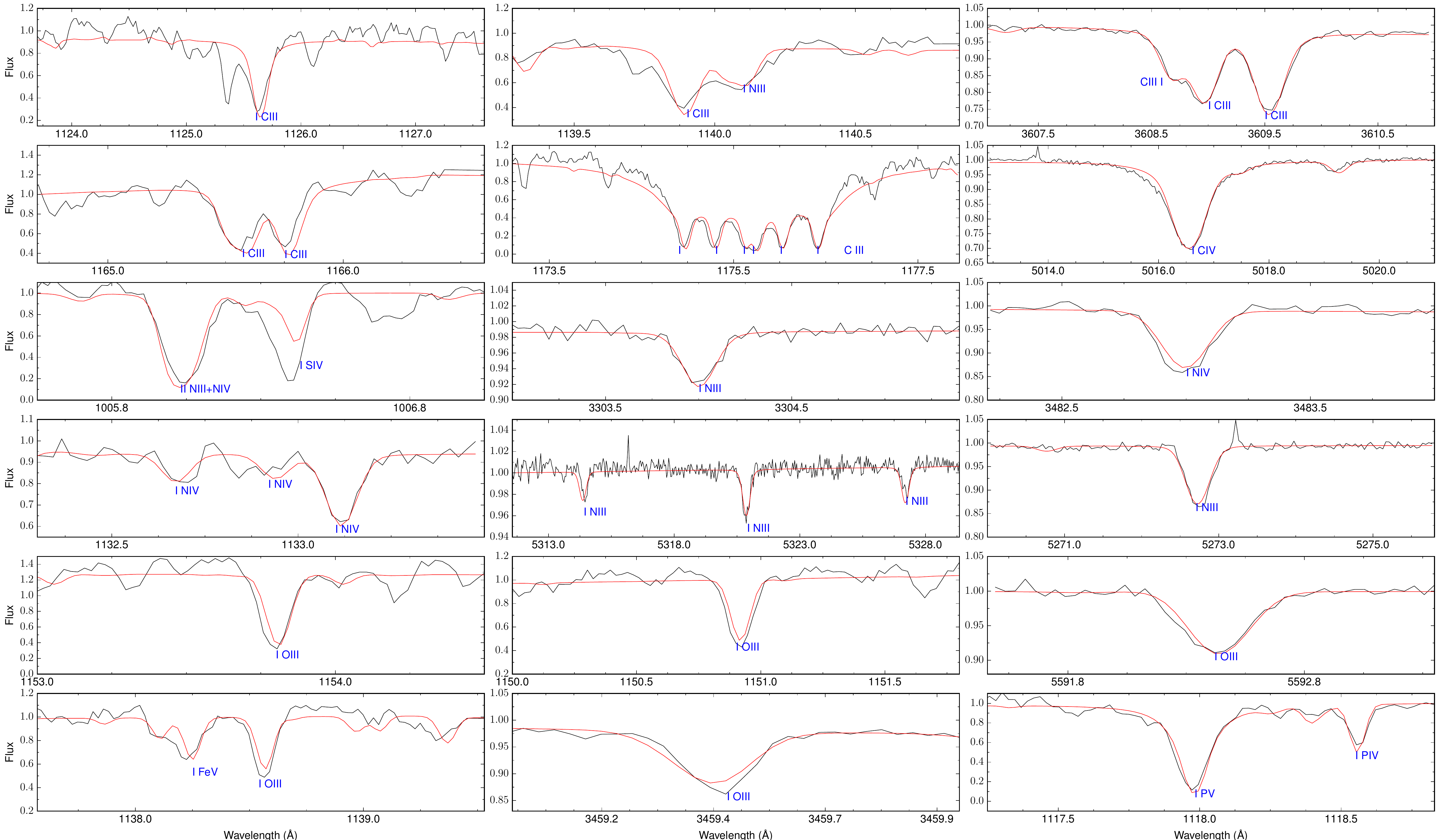}
\caption{Comparison of the final synthetic spectrum to the normalized visual, UVA, and the FUSE FUV spectrum of \lss. Upper panels: 
C\,{\sc iii} and C\,{\sc iv} lines. The continuum adjustment of the strong 
C{\sc iii}, 1176 \AA, septuplet is problematic, which may point to a line broadening issue because the model line wings are too strong but the cores are well matched. Middle panels: N\,{\sc iii} and N\,{\sc iv} lines. Lower panels: Spectral lines of O{\sc iii} {and {the P\,{\sc iv}\&P\,{\sc iv} lines} in the P$@$FUV panel} are well reproduced{, while we consider the match for the Fe{\sc v} in the lower left O$@$FUV panel line reasonable}. %S{\sc v}, Fe{\sc v}, and Ni{\sc iv}, CROPPED from the Diss. Fig. 6.37  TO BE REPLACED by the same for LSS or another star from the paper.
} 
\label{LSS-C}
\end{sidewaysfigure*}

There are plenty of lines both in the FUV and the visual for carbon and nitrogen in the C\&N-type stars \uv\, \lss, and \lsIV. However, only a few carbon lines are found in the \CD\ spectra. Fits to selected carbon lines of different ionization stages are displayed in Fig. \ref{LSS-C} for \lss. 
{C\,{\sc ii} 4267\AA\ may be an important spectroscopic diagnostic when observed in such hot stars. The line is faintly present in the spectra of the C-rich stars, but not in the C-poor \CD\
(see Fig.~\ref{fig:c2} in the appendix).} 

For \CD\ the many visual and FUV nitrogen lines are even stronger than for the former stars. Both elements are present in several stages of ionization. In particular, nitrogen lines from N\,{\sc ii}, N\,{\sc iii}, and N\,{\sc iv} are present.  Figure \ref{LSS-C} illustrates the match of synthetic spectra to observations for a few spectral lines for \lss. The final model spectra reproduce the N\,{\sc iii} and N\,{\sc iv} as well as the C\,{\sc iii} and C\,{\sc iv} lines very well, indicating that their ionization equilibria are matched and the effective temperature is consistent.

Oxygen lines are weak in all program stars.  The O\,{\sc ii} ion is observable only in \CD. O\,{\sc iii} is available both in the FUV, in the UVA and visual spectrum (see Fig. \ref{LSS-C}). {Accordingly, the oxygen abundances are low for all stars. It is worthwhile to note that the oxygen abundance of \lsIV\ is significantly higher than those derived for the other stars.}

\begin{figure*}%[h!]
\begin{center}
\includegraphics[width=0.8\textwidth]{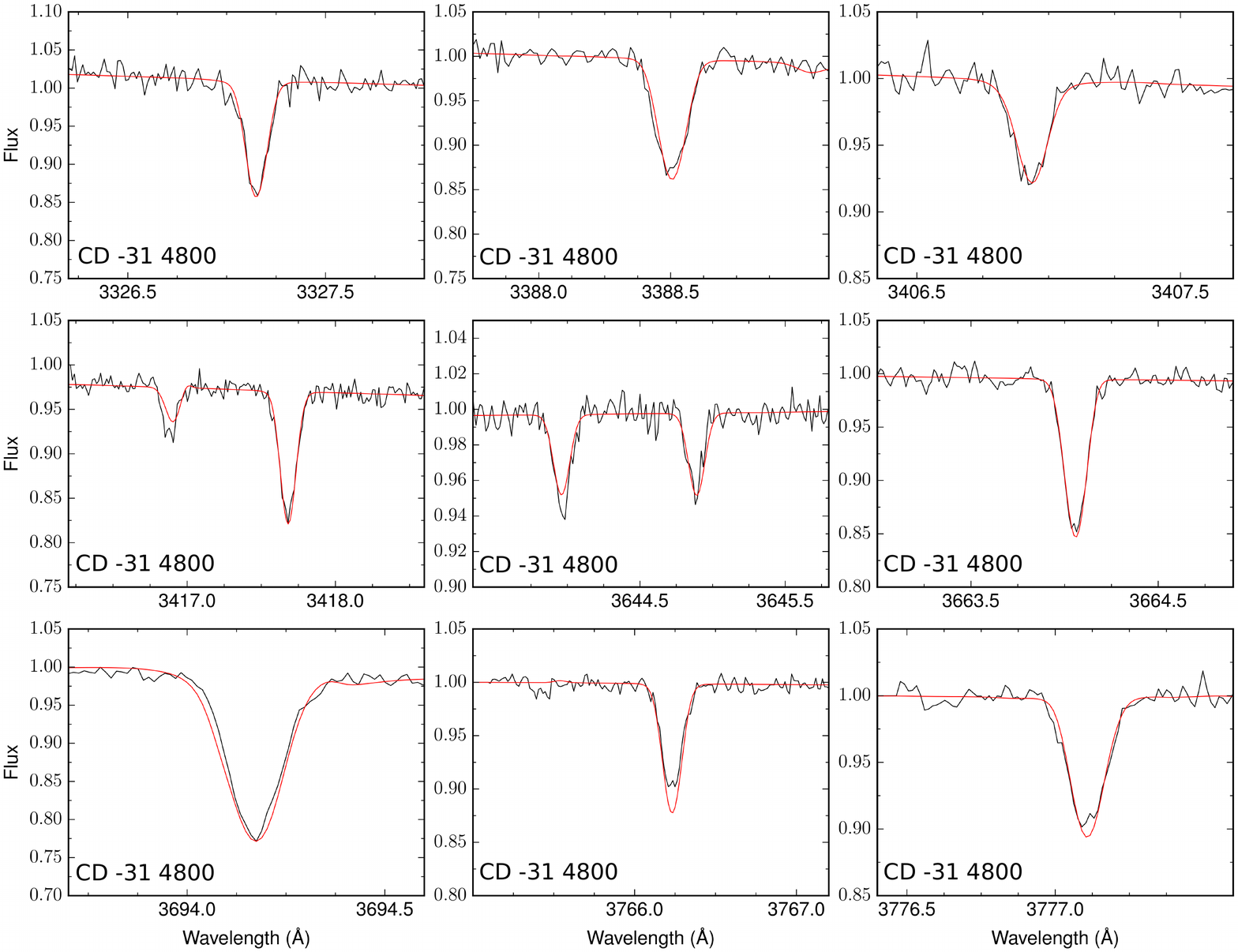}
\includegraphics[width=0.8\textwidth]{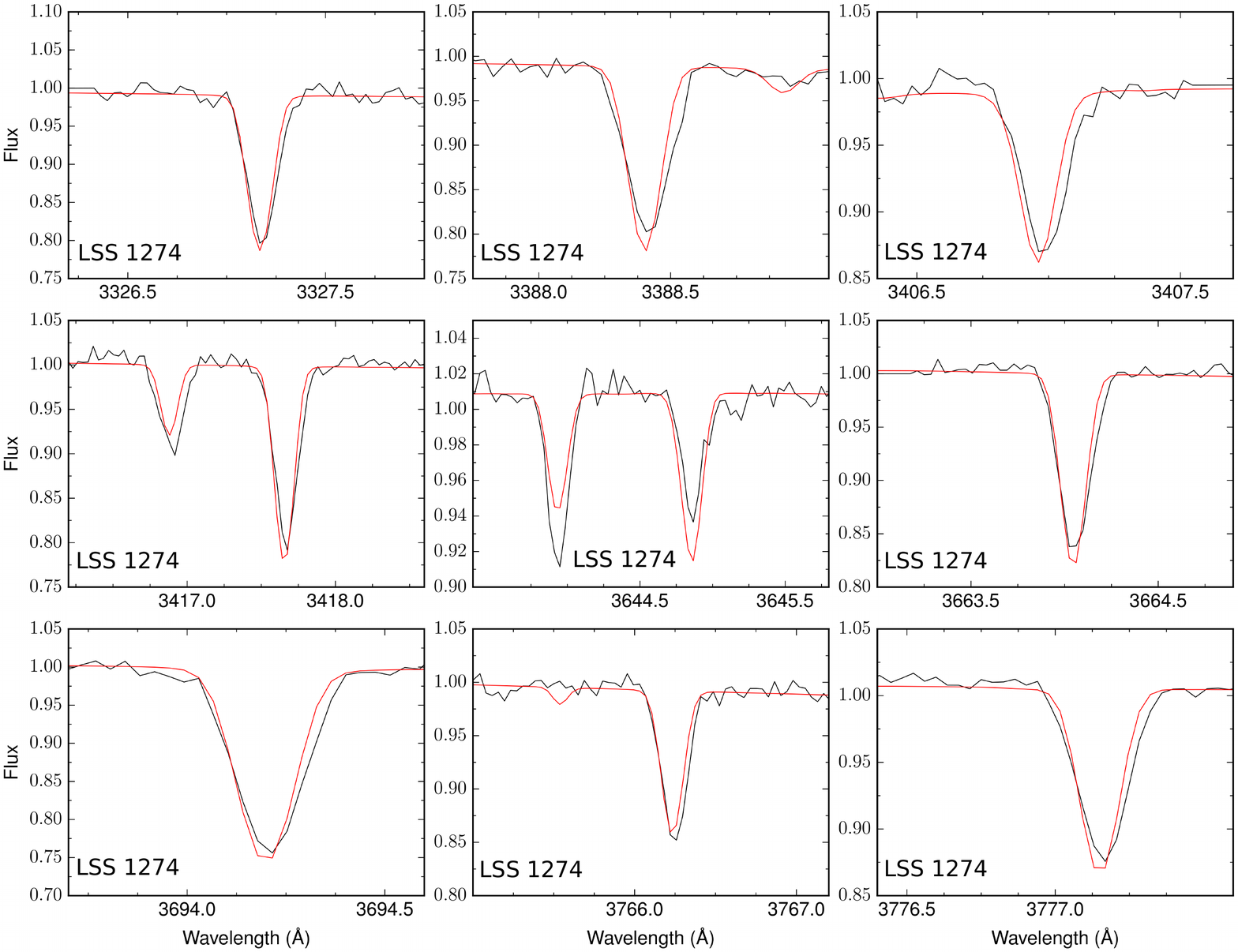}
\caption{Ne\,{\sc ii} lines: Comparison of the final synthetic spectrum to the normalized UVA spectrum. Upper half: \CD, lower half: \lss. 
We note that the strength of all lines are weaker in \CD\ than in \lss\ because of the higher neon abundance. of the latter. 
\label{cd3148_neon}} 
\end{center}
\end{figure*}

\subsubsection{Neon\label{neon}}

Many lines of neon (Ne\,{\sc ii}) lie in the UVA regime, which is difficult to observe from the ground. We take advantage of the capability of the UVES spectrograph to observe this spectral range.
The Ne\,{\sc ii} line strength of the N-type He-sdO \CD\ are considerably weaker than that of the C-types, indicating quite different neon abundances. This is illustrated in Fig. \ref{cd3148_neon}, where the  Ne\,{\sc ii} line fits for \CD\ are compared to those of \lss. While the neon abundance of \CD\ turns out to be solar, it is about five times solar in \lss\ and even almost ten times in \lsIV.

\subsubsection{Magnesium, aluminum, silicon, phosphorus, and sulfur }

In the visual spectrum the only magnesium line is Mg\,II 4481\AA. Since no Mg lines in the FUV could be used, the magnesium abundance is based on a single line, which, however, is well reproduced by the final synthetic spectra in all program stars. For aluminum, the lines at 3601.63\AA\ and 5696.6\AA\ of Al\,{\sc iii} were visible in all program stars, but no other ionization stage is observable. The FUV supplements two additional Al\,{\sc iii} lines (see Table \ref{2xx}). 

A variety of lines of silicon can be found in the visual, the Si\,{\sc iii} lines being faint or absent in the hotter stars. 
Additional Si lines in the FUV were used for all program stars. As for carbon and nitrogen, we find Si\,{\sc iii} and Si\,{\sc iv} lines to be matched very well by the spectral line fit, indicating that the effective temperature is correct.
All visual phosphorus lines were relatively weak and only usable in some of the sample stars. The strong phosphorus lines located in the FUSE range are the best choice to derive the phosphorus abundance. % (see Fig. \ref{LSS-p-uv} for \lss). 
Both P\,{\sc iv} and P\,{\sc v} are matched well.
The same holds for sulfur, for which the FUV lines of S\,{\sc iv} and S\,{\sc v} are the best choice.

\subsubsection{Iron and nickel}

The iron group elements dominate the ultraviolet spectral range of the program stars. We concentrate on the iron and nickel lines for three and four times ionized species. In Table \ref{2xx} the selected iron and nickel lines are listed. Some lines are given as single lines as well as part of a wider fit window with multiple lines. Some lines listed as multiplets in Table \ref{2xx}, cannot be resolved in single lines. Iron and nickel lines were fitted in the same way as spectral lines from other chemical elements.

\begin{table*}
\setstretch{1.3}
\caption{Logarithmic surface abundances by mass of the analyzed elements. The values have been derived from Table \ref{sample-abus1}. Hydrogen was the reference element in the calculations of the model atmospheres. The last column gives the solar mass abundances taken from \citet{asplund2009} as a reference.\label{beta-ratio} }
\centering
\begin{tabular}{cccccc}
\hline\hline
Element &\CD&\lss&\uv&\lsIV&solar\\
%\vspace{3px}
% table heading
\hline
\input{beta_all.tex}

\hline
\end{tabular}
\end{table*}

\subsection{Elemental composition of the program stars}

 In Table \ref{beta-ratio} the elemental composition of the atmospheres of the program stars are summarized as mass fractions and compared to the solar composition \citep{asplund2009}.
As expected for He-sdOs, helium is dominant (97 to 99\%  by mass), hydrogen is strongly depleted in all stars, in fact only 0.06\% to 0.2\% of the atmospheric mass is hydrogen and carbon, nitrogen, or neon may be more abundant than hydrogen. 
The carbon mass fractions range from 1\% to 2\% in the C-rich stars, while nitrogen is the second most abundant element (0.4\%) in \CD. Neon provides 1\% of the atmospheric mass for \lsIV. The abundance pattern of \CD\ is shown in Fig. \ref{mass_abu} and compared to those of the other three program stars in Fig. \ref{mass_abu}. 

 In \CD, a clear CNO bi-cycle abundance pattern is observed. Nitrogen is enriched, while carbon and oxygen are depleted. Neon is slightly above solar, while the other intermediate mass elements are below solar by a factor of about 2. Iron is depleted compared to the Sun while Ni is slightly above solar.
 
\lss, \lsIV, and \uv\ are enriched in carbon with respect to the sun, while oxygen is depleted in all three stars. \lsIV\ is as strongly enriched in nitrogen as is \CD, while \lss\ is  somewhat less enriched than \CD. The nitrogen abundance of \uv\ is close to solar. 
%The carbon to nitrogen ratio, however, differs widely from star to star ([C/N]=1.7 for \lsIV, 11 for \lss, and 25 for \uv).
%The level of C- and N-enrichment is different for the three stars. While \uv\ has a almost solar nitrogen level and a strong carbon enrichment, \lsIV\ features more nitrogen than carbon. 
The carbon-to-nitrogen ratio by mass is $X_{\rm C}/X_{\rm N}$=1.7 for \lsIV, 11 for \lss, and 25 for \uv. 
Neon is enriched with respect to the Sun in \lss, \uv, and \lsIV\  by factors between three and ten, while the other intermediate mass elements all scatter around the solar abundance level. In general, the abundances in \lsIV\ seem to be a little bit higher when compared to the other two stars. While iron  is subsolar, the level of nickel is solar in \lsIV\ and in \uv\ and supersolar in \lss.

\begin{figure}%[h!]
\begin{center}
%\resizebox{\hsize}{!}{
\includegraphics[width=\columnwidth]{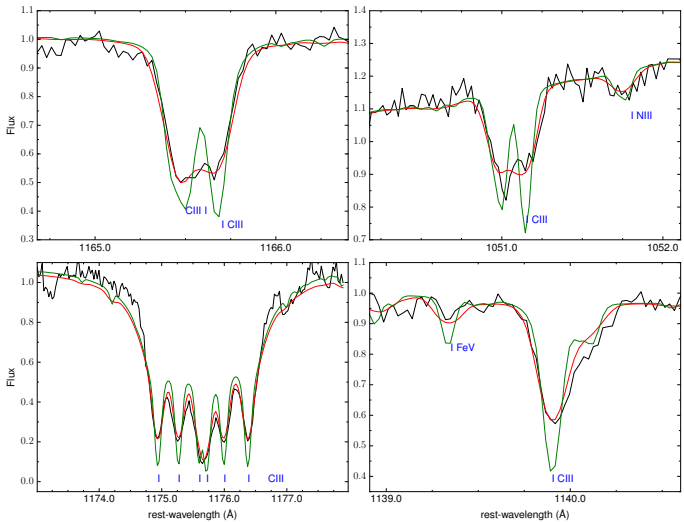}
%}
\caption{Fits to C\,{\sc iii} lines in the FUSE FUV spectrum of \uv. The red profile is calculated with the derived projected rotational velocity of 30 km\,s$^{-1}$.  
The green line indicates the line profiles without rotation for comparison. 
As in the case of \lss\ the continuum adjustment for the C{\sc iii} septuplet at 1176\,\AA\ is problematic (see Fig.~\ref{LSS-C}).
\label{uv-carbon-UV} }
\end{center}
\end{figure}

\subsection{Projected rotation velocity}\label{vsini}
\CD, \lss\ and \lsIV\ show no rotational broadening. For these stars, the projected rotational velocity is lower than $v_{\text{rot}}\sin{i}$ = 5 km\,s$^{-1}$, the detection limit.  Only for \uv\  rotational line broadening is detected. The projected rotation velocity was measured from the carbon lines in the visual spectrum of \uv\ to be $v_{\text{rot}}\sin{i}= 35\pm4$ km\,s$^{-1}$ (see Fig. \ref{uv-helium}). The rotational broadening is also obvious in the FUSE FUV spectrum (see Fig. \ref{uv-carbon-UV}) and was derived to be 30 $\pm4$ km\,s$^{-1}$. Hence the projected rotational velocity from the FUV is consistent with the visual one within error margins.

%\clearpage

\section{Evolutionary status: Mergers vs. flashers}\label{discussion}

Confronting the properties of the program stars with evolutionary models should allow their origin to be constrained. We introduce the two rivalling scenarios and summarize the main observational results, before discussing the scenarios in depth.

\begin{figure*}%[h!]
\begin{center}
%\resizebox{\hsize}{!}{
%\includegraphics[width=\columnwidth]{fig8_cd31.pdf}
%\includegraphics[width=\columnwidth]{fig8_new.pdf}
\includegraphics[width=\columnwidth]{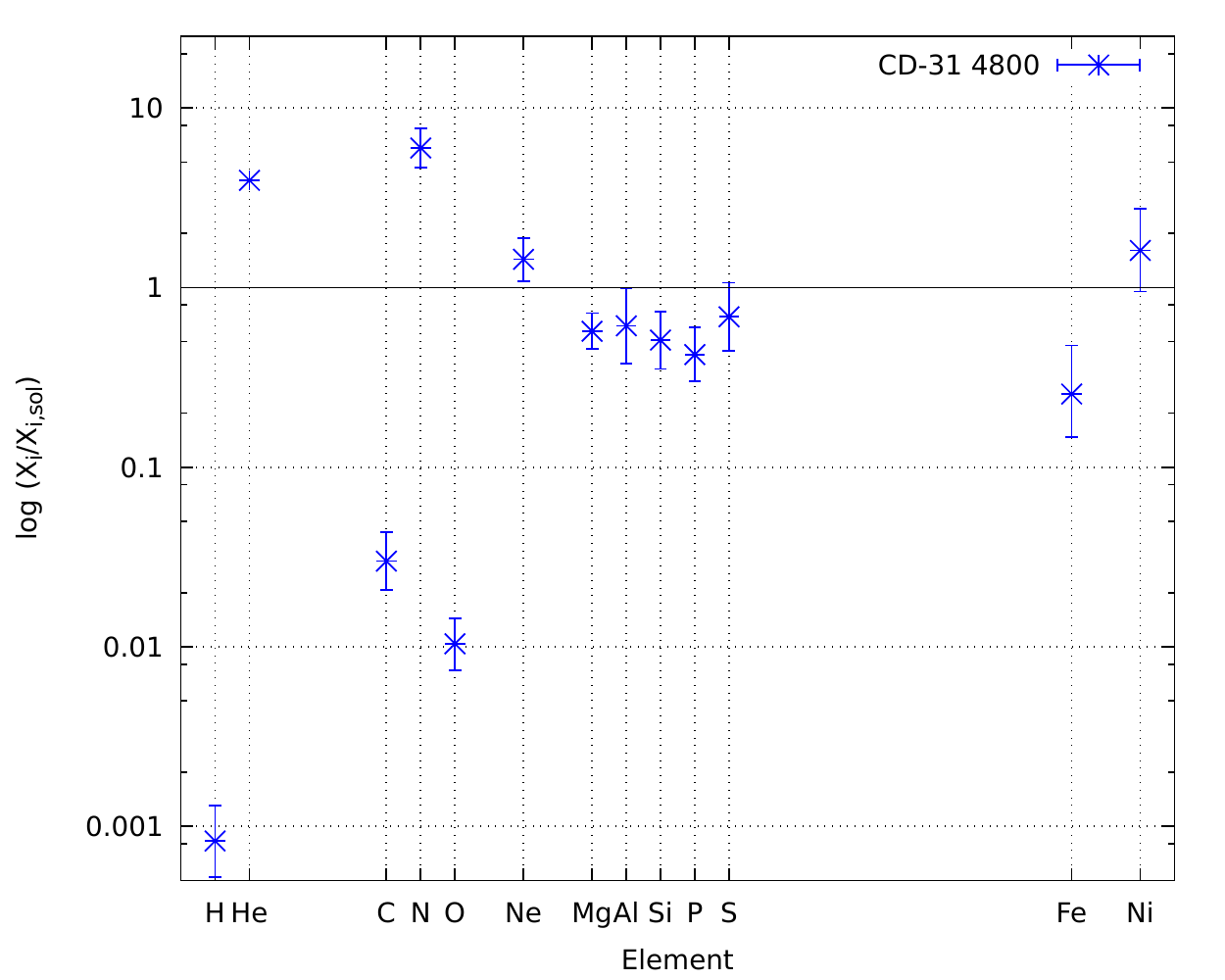}
\includegraphics[width=\columnwidth]{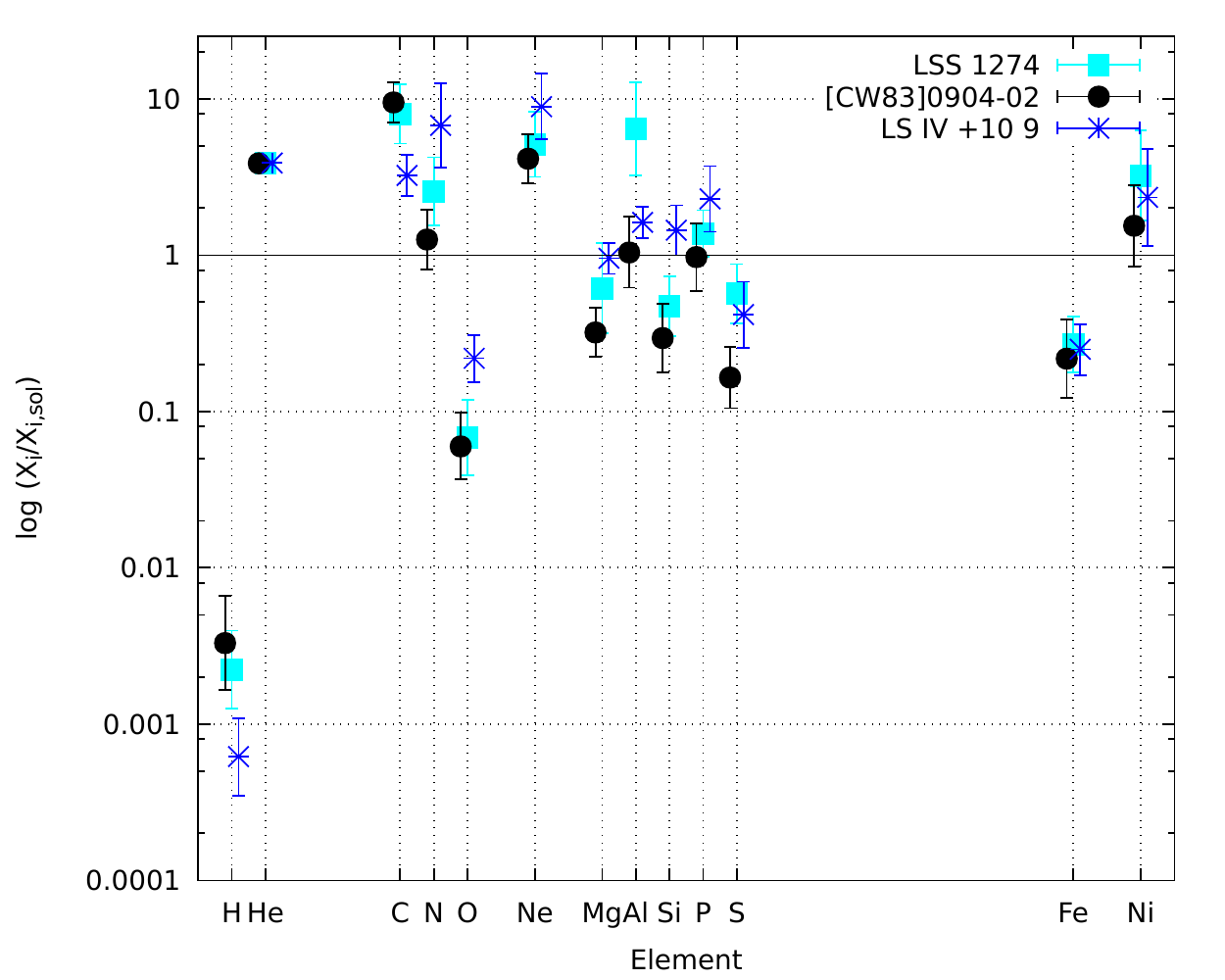}
\caption{Abundance pattern of \CD\ (left hand panel) and \lss, \uv, and \lsIV\ (right hand panel) with respect to solar composition.\label{mass_abu} }
\end{center}
\end{figure*}

Two rivaling scenarios have been proposed to explain the origin of He-sdO stars, the helium WD merger \citep{webbink1984} and the late hot flasher scenario \citep{1997fbs..conf....3S,Miller2008}. % Sweigart, 1997, Miller Bertolami et. 2008
In the merger scenario a binary system consisting of two He WDs loses orbital energy and angular momentum by the emission of gravitational waves and the orbit shrinks until the lighter component overfills its Roche lobe and disrupts, forming an accretion disk and possibly a hot corona around the more massive component. Once this material has been accreted by the massive He WD, helium is ignited in the core and the remnant emerges as a hot subluminous star. The surface composition may be altered by thermonuclear reactions and mixing during the accretion process. Depending on the initial masses of the stars and the rate of accretion, thermonuclear processes may occur and change the chemical composition of the companion \citep{zhang2012}.

The late hot flasher scenario does not require binary evolution and can also happen in single stars.
If a low-mass star loses a sufficient amount of material on the red giant branch (RGB) before the helium core flash occurs, a delayed core helium flash kicks in when the star descends along the hot WD cooling track \citep{Castellani1993}. Hot subdwarfs formed via this scenario can be found at the hot end of the extreme horizontal branch, close to the helium main sequence. The depletion of hydrogen can be explained by the presence of a sub-surface convection zone emerging during the helium flash that pushes hydrogen into deeper layers where it is diluted and can be burnt. Depending on the epoch when the core flash occurs, that means on or before reaching the WD cooling sequence, theoretical models predict different abundance patters to emerge in the stellar atmospheres \citep{2001ApJ...562..368B,2003ApJ...582L..43C,2004ApJ...602..342L}. 

\begin{figure}%[h!]
\begin{center}
\resizebox{\hsize}{!}{\includegraphics{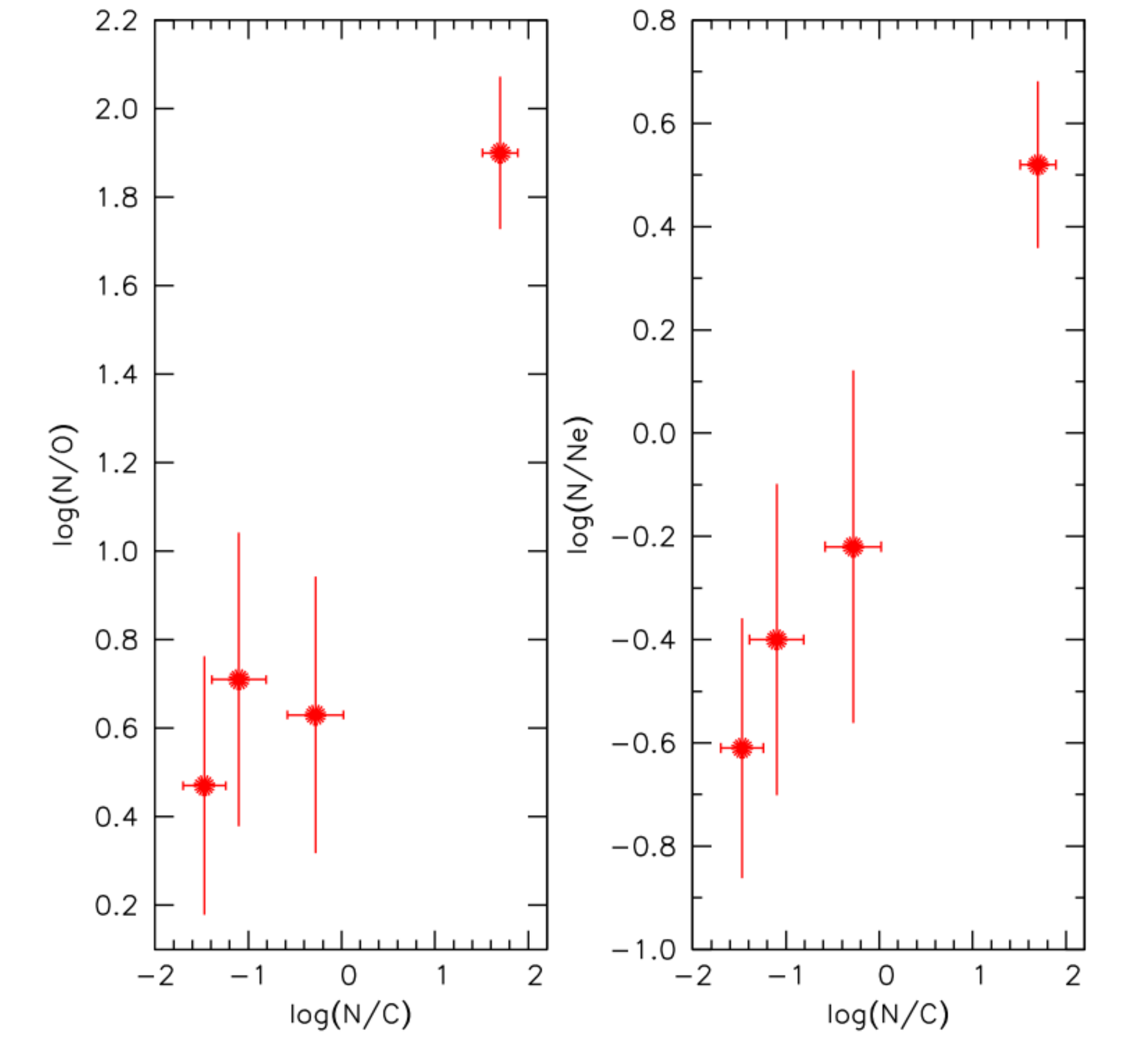}}
\caption{Abundance ratios by number: N/O versus N/C (left hand panel) and N/Ne vs. N/C (right hand panel).
}\label{abu_ratio}
\end{center}
\end{figure}

The main observational results can be used to narrow down the parameter space for the evolutionary models: 
\begin{enumerate}
\item The abundance patterns:
\begin{itemize}
\item Hydrogen is a trace element and detected at a mass fraction of 0.06\% to 0.2\%.  
\item The overall metallicity of the stars is solar or slightly below solar (factor 2). Metal poor models can, therefore, be discarded. 
\item Signatures of nucleosynthesis are obvious for C, N, O, and Ne.
\item \CD\ displays C and O depletion and, N enrichment with respect to the Sun characteristic for processing by the CNO bicycle (see Fig. \ref{mass_abu}, left hand panel).
\item all C-rich stars have enhanced C, N, and Ne, while O is depleted with respect to the Sun (see Fig. \ref{mass_abu}, right hand panel).
\end{itemize}
\item Abundance ratios for N/O and N/Ne show a trend to increase with increasing N/C (Fig. \ref{abu_ratio}) for  the four stars.
\item The stellar masses are consistent with the canonical mass of 0.47 M$_\odot$, which is in line with all variants of hot flasher models.
\begin{itemize}
\item \uv\ is possibly more massive ($\approx$0.6 M$_\odot$), while the others are somewhat lighter ($\approx$0.4 M$_\odot$).
\item \CD, \lss, and \lsIV\ may have formed from mergers of pairs of low mass He WD (0.15 to 0.35 M$_\odot$), that is their progenitors could involve at least one He WD of extremely low mass \citep[ELM, M$<0.2$ M$_\odot$, see][]{2017ASPC..509...85H}, while \uv\ could result from normal He WDs.
% From Zhang & Jeffery's models only the 0.2+0.2 Model would be relevant.
\item \uv\ could be a merger from more 'normal mass' He WDs. 
%0.3+0.3 would do and any other combination of masses that adds up to 0.6
\end{itemize}
\item Rotation: Projected rotation velocities are below the detection limit (5\,km\,s$^{-1}$) except for \uv, which rotates at v$_{\rm rot}\sin$ i=30\,km\,s$^{-1}$.
\end{enumerate}

\subsection{Formation through He-WD mergers}

 Stellar mergers of two He WDs have been suggested to explain the origin of He-sdO and -sdB stars.  
\citet{zhang2012} modelled the post-merger evolution of a helium plus helium double WD for four equal mass pairs (0.25M$_\odot$+0.25M$_\odot$; 0.3M$_\odot$+0.3M$_\odot$; 0.35M$_\odot$+0.35M$_\odot$; 0.4M$_\odot$+0.4M$_\odot$) in 1D geometry. The result of the merging process depends on the mass transfer rate. In the slow merger model, a debris disk forms from which the primary WD accretes. Temperatures are too low to allow nuclear burning to take place. The material accumulating at the surface of the primary is that of the disrupted He WD, that is rich in helium and nitrogen, but poor in C and eventually oxygen. If the mass transfer rate is high the less massive WD
quickly transfers its entire mass to the companion's surface directly. The heating forces the material to expand and a hot corona quickly builds up. The high temperatures enable helium burning  and a strong convection zone forms to possibly dredge up fresh carbon to the surface. The efficiency of the dredge up increases with merger mass. The third model investigated by \citet{zhang2012} is a composite one assuming both processes to occur. The surface abundance pattern from a composite merger should be a mix of the patterns of both slow and fast
models. High-mass mergers models predict carbon-rich material to be dredged-up, whereas low-mass ones do not. Accordingly, the composite merger models of \citet{zhang2012} predict the atmospheres to become C-rich if the mass of the merger remnant exceeds 0.65 M$_\odot$, but to be N-rich for lower masses.

%In a composite merger, the secondary star is disrupted by tidal forces and %forms an accretion disk around the primary from which material is slowly %accreted by the primary. A significant part of the companion's remnant may form %a hot corona around the primary, in which temperatures can get high enough for %thermonuclear reactions to occur. 

\begin{figure}%[h!]
\begin{center}
\resizebox{\hsize}{!}{\includegraphics[angle=270]{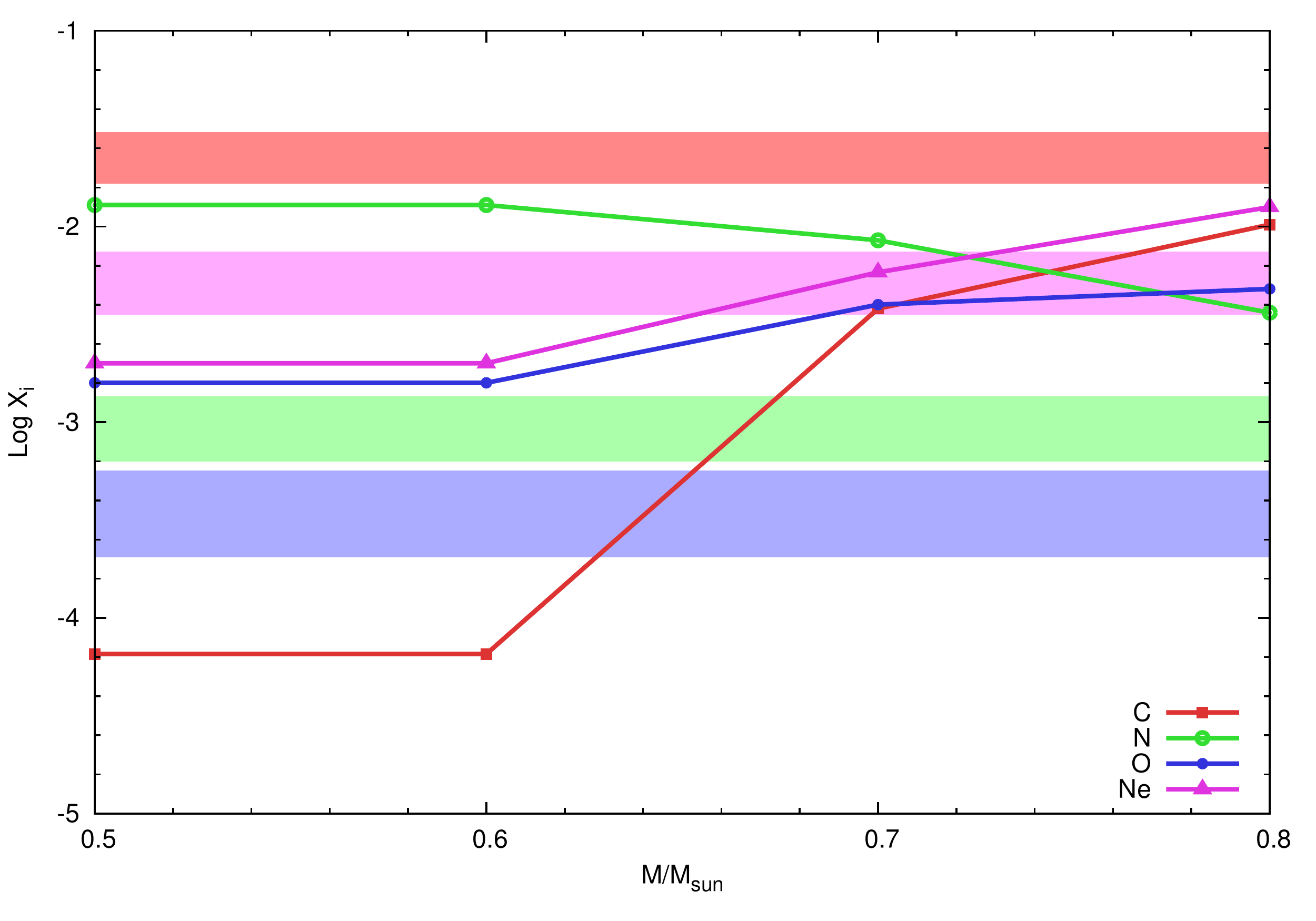}}
%\resizebox{\hsize}{!}{\includegraphics[width=\textwidth]{{}}
\caption{Chemical abundances by mass for carbon ($^{12}$C+$^{13}$C), nitrogen ($^{14}$N), oxygen ($^{16}$O+$^{18}$O), and neon ($^{20}$Ne+$^{22}$Ne) predicted by composite merger models \citep{zhang2012} as a function of the merger mass (solid colored lines) for a metallicity of $z=0.02$, The horizontal bands represent the abundances and their uncertainties determined for \uv. 
\label{composite_CN}}
\end{center}
\end{figure}

The N-rich, C- and O-poor composition  of \CD\ (Fig. \ref{mass_abu}) resembles that of a helium WD, that is the core composition of a RGB star, as demonstrated in Table~\ref{he_wd}. The abundances predicted for the typical composition of the He-cores  of RGB stars of different masses (M$_{\rm He-core}$=0.2 and 0.45 M$_\odot$) and from different initial masses (1.0 and 1.5 M$_\odot$) of the progenitors and the appropriate metallicity \citep[z=0.01,][]{2016A&A...588A..25M}. The C, N, and Ne abundances of \CD\ are close to those predicted.
However, for oxygen the model abundance is
much higher than in \CD, which indicates that the NO-
cycle might had been operating at higher temperatures in \CD\ than in the models.
Nevertheless, we conclude that the abundance of \CD\ are consistent with the slow (cold) merger model of a double He-WD binary. However, the He-sdO star resulting from such a merger would be expected to show some rotation \citep{peter2012}.

% which is not the case for \CD. 
Indeed, \citet{2018MNRAS.476.5303S} %Schwab
predicts surface rotation  of a 30 and 100 km\,s$^{-1}$ for the merger remnant after merging He-WD binaries of 0.2 + 0.3 M$_\odot$  and 0.3 + 0.4 M$_\odot$, respectively. Because the projected rotation velocity of \CD\ is small, the inclination of the rotation axis would need to be small if \CD\ resulted from a He-WD merger.

The high C abundance in the other three stars can not be explained by a cold merger. However, hot and composite merger models may be appropriate. 
Figure \ref{composite_CN} %and Fig.  \ref{composite_ONe} 
shows predictions for the  abundances of C, N, O, and Ne in a composite merger model as a function of the final merger mass and for solar metallicity \citep[z=0.02, ][]{zhang2012} and compares them to the abundances determined for \uv. The carbon abundance is much too low for stellar masses of less than 0.7 M$_\odot$, but comes close to the observed ones for more massive models. The same holds for neon, while the predicted N abundance is higher than the observed one for all masses, although it comes closer to the observed one for high masses. The predicted O abundance is much higher than the observed one as is the case for \CD\ and the cold merger model prediction. For the composite merger the discrepancy for oxygen even increases with increasing mass.
Hence, we conclude that the abundance pattern predicted by the composite merger models of \citet{zhang2012} for equal mass He-WD mergers are inconsistent with the observed abundance patterns of the $\approx$0.6 M$_\odot$ star \uv\
(Fig.~\ref{composite_CN}). Because \lss, and \lsIV\ are of lower mass, their abundances are differ strongly from the predictions. Additional calculations are required for pairs of unequal mass He-WDs. Three-dimensional modelling would also be necessary for more robust predictions. 

The only program star, for which we obtained a significant projected rotation velocity is \uv\ (31\,km\,s$^{-1}$), close to the prediction of the model for a 0.2 + 0.3 M$_\odot$ He-WD merger of \citet{2018MNRAS.476.5303S}. Therefore, the measured rotational velocity would point to the composite merger scenario for \uv, as this is able to explain the observed rotation, although its surface abundance is not fully consistent with the model prediction. 

\begin{table*}
\caption{Comparison of the abundances (logarithmic mass fractions) derived for \CD\ to %the composition 
%{of the He-WD model from \cite{zhang2012}} and 
the typical composition of the He-cores  of RGB stars (of different mass M$_{\rm He-core}$), coming from progenitors of different initial masses (M$_{\rm ZAMS}$) and a metallicity of z=0.01\citep[as computed by][]{2016A&A...588A..25M}.}\label{he_wd}
\setstretch{1.3}
\begin{center}
\begin{tabular}{lcccc}
\hline\hline
& C& N & O & Ne\\
\hline
CD$-$31 4800 &$-4.14$ $\pm0.16$ & $-2.38$ $\pm0.11$  & $-4.22$ $\pm0.13$& $-2.74$ $\pm0.12$ \\\hline
%He-WD & $-4.22$ & $-1.90$ & $-2.79$ & $-2.73$ & $-3.43$\\\hline
 \multicolumn{5}{c}{RGB Cores} %$, Z_{\rm ZAMS}=0.01$}
 \\
$M_{\rm He-core} (M_{\rm ZAMS})$& C& N & O & Ne \\\hline
 0.20$M_\odot$ (1$M_\odot$)  & -4.30 & -2.22 & -3.13 & -3.01\\   
 0.45$M_\odot$ (1$M_\odot$)  & -4.20 & -2.20 & -3.39 & -3.01\\   
 0.20$M_\odot$ (1.5$M_\odot$)  & -4.17 & -2.19 &-3.53 & -3.01\\   
 0.45$M_\odot$ (1.5$M_\odot$)  &  -4.13 & -2.19 & -3.69 & -3.01\\\hline
% \multicolumn{5}{c}{RGB Cores, $Z_{\rm ZAMS}=0.02$}\\
%$M_{\rm He-core} (M_{\rm ZAMS})$& C& N & O & Ne \\\hline
% 0.20$M_\odot$ (1$M_\odot$)  & -4.20 & -1.97 & -2.50 & -2.71 \\   
% 0.45$M_\odot$ (1$M_\odot$)  & -3.99 & -1.92 & -2.81  & -2.71 \\
% 0.20$M_\odot$ (1.5$M_\odot$)  & -3.97  & -1.90 & -3.00 & -2.71 \\
% 0.45$M_\odot$ (1.5$M_\odot$)  & -3.88 & -1.89 & -3.22 & -2.71 \\   
%\hline

\end{tabular}
\end{center}
\end{table*}

\subsection{Formation through late hot flashes}

The Helium core flash terminates the H-shell burning of red giant stars. In the canonical pictures the flash occurs at the 
tip  of  the  RGB.  However, the H-rich envelope may be lost before the onset of  the  He-core  flash, for instance by binarity or enhanced winds due to stellar rotation 
\citep{1997fbs..conf....3S,2015Natur.523..318T}
%(Sweigart 1997; Tailo et al. 2015)
or the ingestion of a substellar companion \citep{1998AJ....116.1308S,2014A&A...570A..70S} or after a common envelope event \citep{2018MNRAS.475.4728B}. %Byrne et al. 2018
% or interaction with a stellar companion \citepads{2018MNRAS.475.4728B}. % Soker 1998, (Schaffenroth et al. 2014), Byrne et al. 2018
 The remnant would then evolve toward high temperature and finally enter the
WD cooling sequence. If a He-core flash occurs during this evolution, it is termed a hot flasher. Two types of hot flashers have to be considered, an early one, that occurs before the WD cooling sequence has been reached, and a late hot-flasher when the 
star is close or already on the WD cooling sequence. During the He-core flash a convective zone develops. In an early hot-flasher model, as in the canonical model, this zone does not reach the H-rich envelope and the surface abundances are not altered. Therefore, we can exclude early hot-flasher to explain He-sdOs. However, in a late hot-flasher, which takes place when the star embarks on the WD cooling curve,
the  He-flash  driven  convective  zone gets into contact with the H-rich envelope. 
Depending on whether H-burning takes place or not, hot-flasher models 
 are divided in deep-mixing and shallow-mixing events.
After the He-flash the surface becomes cooler because the stellar envelope expands triggering an outer convective zone to emerge which moves inwards. In the shallow mixing (SM) case the He-core flash driven convective zone eventually splits giving rise to
a convective zone that persists in the outer region of the core until it merges 
with the outer convective zone of the envelope, which leads to a dredge up of material from  the core to the surface. However, no  H-burning occurs,  but  the H content at the surface decreases. In the deep mixing (DM) case, the convective region reaches the H-rich layer shortly after the He-flash maximum and H is mixed leading to violent H-burning. As a consequence hydrogen almost vanishes from the surface and processed material is dredged up to the surface \citep[see ][ for details]{battich2018}.

\begin{figure}[h]
\begin{center}
\resizebox{\hsize}{!}{\includegraphics{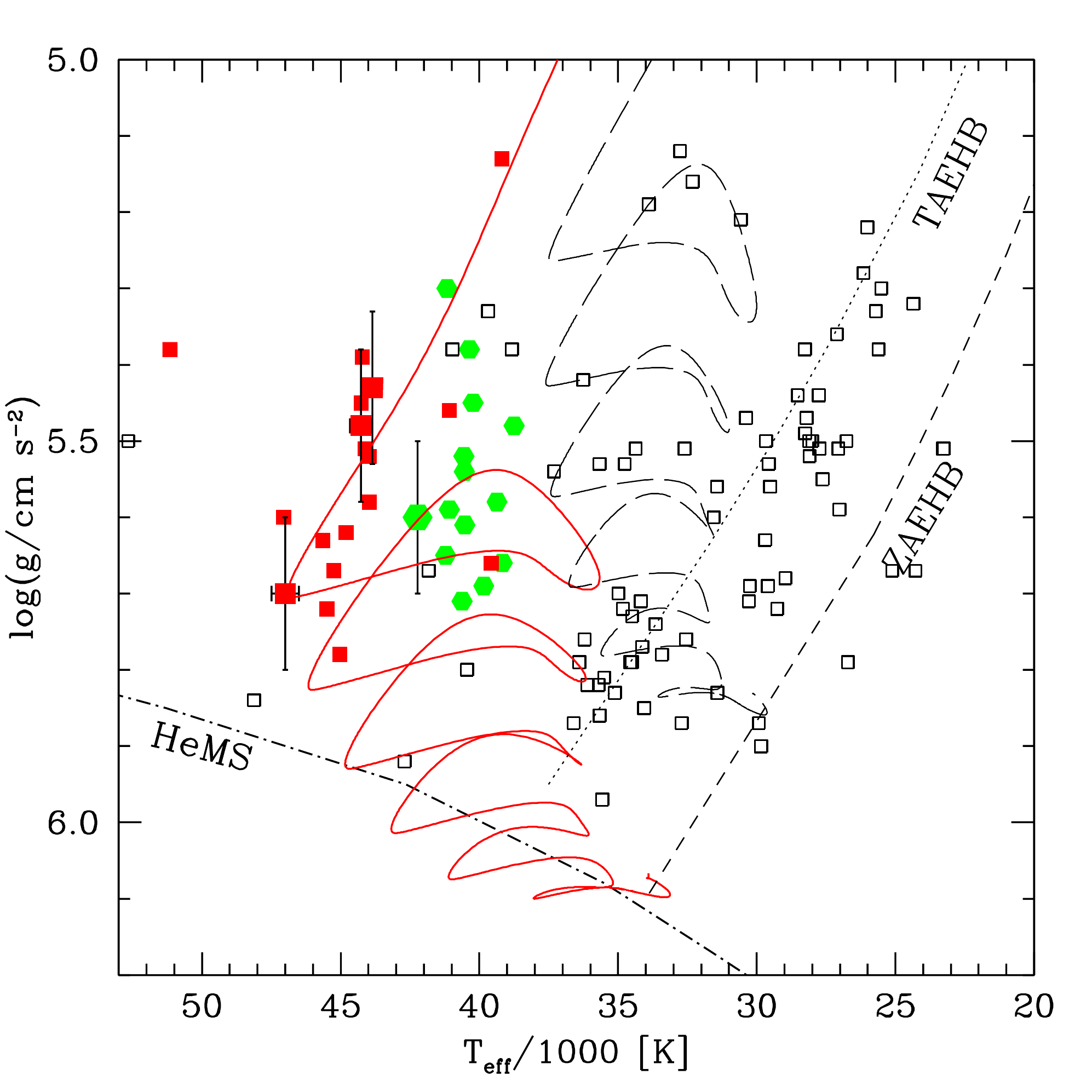}}
\caption{Comparison of the positions of the program stars (plotted with error bars) in a Kiel diagram 
(T$_{\rm eff}$, $\log$ g) with different evolutionary tracks for hot flashers \citep{battich2018} and the hot subdwarfs from the SPY project \citep{2005A&A...430..223L,Hirsch2009}. The hydrogen-rich hot subdwarfs are marked by open squares, while C-rich He-sdOs are are shown as red filled squares and N-rich as green filled hexagons.  
The dashed (black) track is for a shallow mixing case and a stellar mass of 0.46794 M$_\odot$, while the full drawn (red) one represent predictions from the deep-mixing hot flasher scenario for a mass of 0.46599 M$_\odot$. Also shown are the zero- (ZAEHB) and terminal age EHB \citep[TAEHB, adopted from ][]{1993ApJ...419..596D}
% Dorman et al.  
and the helium main sequence \citep{1971AcA....21....1P}.}\label{kiel}
\end{center}
\end{figure}

Figure  \ref{kiel} compares the positions of the program stars in the T$_{\rm eff}$--$\log$ g diagram to evolutionary tracks of the hot flasher scenario with shallow and deep mixing. The shallow-mixing tracks are too cool to meet the observed positions and the remaining hydrogen content is too high. Shallow mixing can, therefore, be excluded. The tracks for deep mixing, however, evolve through the region occupied by the program stars and, hence, are viable.

\begin{figure}[h]
\begin{center}
\resizebox{\hsize}{!}{\includegraphics{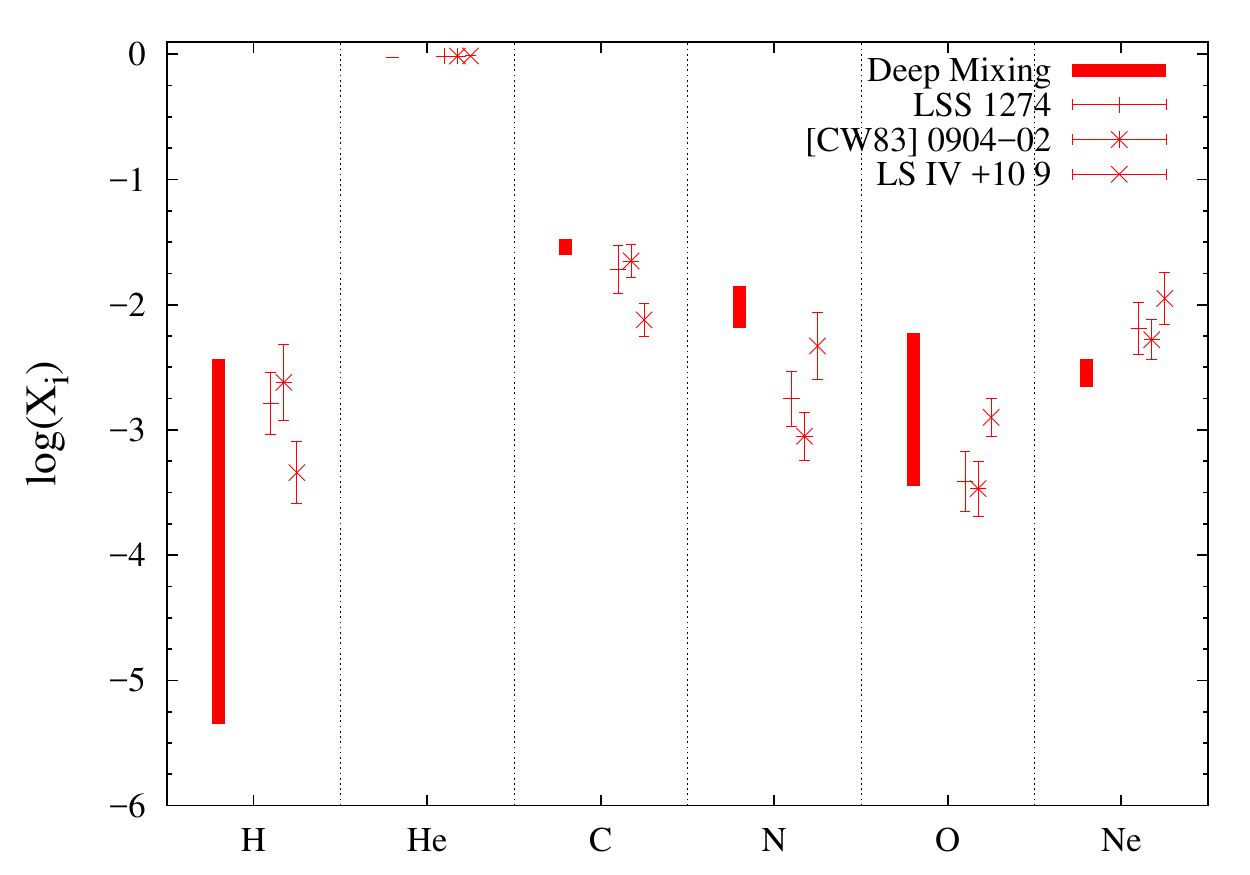}}
%\resizebox{\hsize}{!}{\includegraphics{abundances-markus3.eps}}
\caption{H, He, C, N, O and Ne abundances of our stars as compared with the predictions from the deep-mixing late hot-flasher scenario \citep[red bars, ][]{battich2018}.
% and the typical abundances in the He-cores of RGB stars (blue bars).
}
{\label{comp:abu}}
\end{center}
\end{figure}

First, we compare in Fig.  \ref{comp:abu} the common pattern of hydrogen and light metal abundances of \lss, \uv\ and \lsIV\ with those predicted by the stellar evolution simulations of \citet{battich2018}. In Fig. \ref{comp:abu} we have chosen tracks with an initial metallicity and helium abundance of z=0.02 and y=0.285, respectively. 
The observed hydrogen abundance is at the high end of the prediction while the oxygen abundance is at the low one. The C and N abundances are slightly lower, while the Ne abundance is somewhat higher than predicted. Given the uncertainties involved in 1D evolutionary models we consider the observed abundance pattern to be consistent with the DM late flasher predictions. Qualitatively, the origin of the abundance pattern emerging from the flasher models of \citet{battich2018} can be described as follows. 

When the He flash starts 
 $^{14}$N from the He-core is burnt with helium through the reactions
\begin{equation}\label{eq:three}
\isotope[14][7]{N}\, (\isotope[4][2]{He},\gamma) \,\isotope[18][9]{F} \,(e^+ \nu_e)\, \isotope	[18][8]{O}\,(\isotope[4][2]{He},\gamma)\, \isotope[22][10]{Ne},
\end{equation}
leading to  a $^{18}$O- and  $^{22}$Ne-enrichment of the He-flash convective zone. Later, the C- and Ne-rich He-flash driven convective zone reaches the other, upper layers of the star, leading to a strong enrichment in both C and Ne at the photosphere. In addition, when this happens the remaining H-envelope is mixed down into the hot He-burning interior, where it is burned with carbon. As a consequence the deep-mixing scenario is also expected to produce a strong N-enrichment of the surface through the reactions 
\begin{equation}\label{eq:four}
\isotope[12][6]{C}\,(\isotope[1][1]{H},e^+ \nu_e)\,\isotope[13][6]{C}\,(\isotope[1][1]{H},\gamma)\,\isotope[14][7]{N}.
\end{equation}

\begin{figure}[h]
\begin{center}
\resizebox{\hsize}{!}{\includegraphics{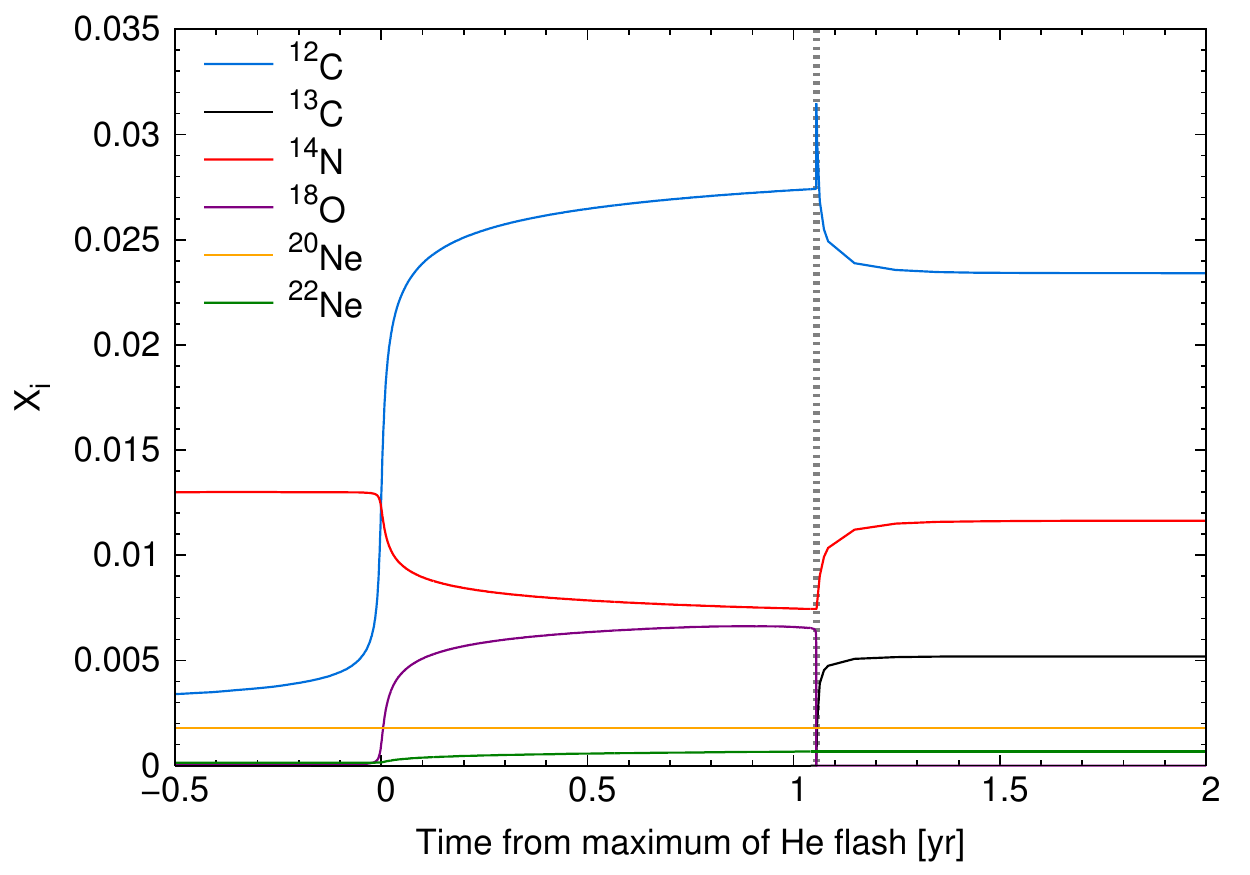}}
\caption{Development of isotopic abundances during He- and H-flashes. The abundances correspond to the inner part of the star where He- and H-flashes take place. The dashed vertical (gray) line marks the onset of H flash. The abundances of $^{16}$O, $^{17}$O and $^{15}$N isotopes (not showed) are always below $5\times 10^{-4}$.
\label{flash}}
\end{center}
\end{figure}

Fig.  \ref{flash} shows the development of the abundances of the important elements between the He- and H-flash in the simulation of a deep mixing event.
The helium flash takes place at $t=0$ and $^{12}$C is significantly increased, at the expense of helium. $^{18}$O and $^{22}$Ne are enriched on the cost of $^{14}$N. Shortly after about one year the hydrogen flash occurs and oxygen and $^{12}$C is enriched for a short time before it is transformed to $^{14}$N by the reaction stated above. Immediately after the H-flash the He-flash convective zone splits, with the outermost convective zone being driven by H-burning. Due to the lower temperatures of H-burning the  $^{22}$Ne abundances do not change in the outermost convective region after the split. On the contrary, $^{22}$Ne abundances in the innermost convective zone increase continuously during the He-flash, and consequently, the later the convective zone splits, the higher the final $^{22}$Ne surface abundance of the models. 

The correlation of abundance ratios in Fig. \ref{abu_ratio} might be
interpreted in the context of hot-flasher models.  $^{22}$Ne is
created at expenses of $^{14}$N at the same time that $^{12}$C is
created at expenses of $^{4}$He. This means that while the $^{14}$N
abundance decreases, the $^{12}$C and $^{22}$Ne abundances increase,
and naturally, that explains that N/Ne correlates with N/C in time.
This correlation is present in the models in spite of the N created
during the H-flash \citep{battich2018}.  However, in the models too
much N is produced during the H-flash, while not enough $^{22}$Ne is
produced (see Fig. \ref{comp:abu}). This might indicate that the inner
and outer convective zones during the Hot-Flasher event might stay
connected longer than predicted by 1D stellar evolution models.  The
relatively low N/O values in our C-rich He-sdO stars can only be reproduced by
hot-flasher models with a high $^{18}$O abundance at the moment of the
H-flash.  As before, the nitrogen-to-oxygen ratio of the models might also be improved if
convective zones stay connected after the H-flash.

The correlation between N/C -- N/Ne and N/C -- N/O may be telling that,
for \lss, \uv\ and \lsIV, the material that appears in the surface was
processed in a late deep-mixing He-flash event. Three-dimensional modelling would also be helpful for more robust predictions.

\section{Summary and conclusions}\label{sect:summary}

%We carried out detailed quantitative spectral analyses from high-resolution spectroscopy of four prototypical He-sdO stars. Visual spectra were used to determine the atmospheric parameters.  In order to identify the chemical signatures of processes that led to formation of the He-sdO stars, UVA and far-UV spectra from the FUSE mission were combined with the visual spectra to determine the abundances of the most important chemical elements.
 
%\CD\ belongs to the N strong-lined subclass of He-sdOs whereas, \lss, \uv\ and \lsIV\ are C strong-lined. Because close binary evolution plays a vital role for the evolution of sdB stars, we searched for any evidence of binarity. Neither any radial velocity variations nor any photometric signature of a companion could be found.
%{\sc Tlusty200/Synspec49} were used to compute line blanketed NLTE model atmospheres and synthetic spectra. Atmospheric parameters and the abundances patterns were derived. 
%Line crowding in the FUV complicates the spectral analysis, therefore, where lines are available in both the visual/UVA and FUV spectra, we derived abundances from both spectral ranges separately. 
%The consistent visual/UVA and FUV abundances were averaged.
 
%Because line crowding in the FUV may complicate the spectral analysis, abundance were derived from visual/UVA and FUV spectra separately if lines were available in both spectral ranges. Because the results were consistent, abundances were averaged.

{We studied four prototypical He-sdO stars, the N-rich \CD\ and the C\&N-rich \lss, \lsIV, and \uv\ from high quality spectra using state-of-the-art NLTE model atmosphere and spectral synthesis tools. The following results may be highlighted:

\begin{enumerate}
 
%All stars have been studied previously with less sophisticated model atmospheres. 
%Therefore, we embarked on an investigation of systematic effects caused by limitations of the model atmospheres used. 
%We started by calculating a model grid of H/He composition, as used in previous studies, and added additional elements. 

\item The analysis made use of data set of unprecedented quality, spectral resolution and coverage by combining ground-based visual and UVA data with FUV observations from the FUSE satellite. 

\item Final models included C, N, O, Ne, Mg, Al, Si, S, Fe, and Ni represented by the most detailed model atoms available. Because of the enrichment of either nitrogen or carbon, it turns out, that models including either C or N at the appropriate high abundance reproduce the temperature stratification of full models to within about 1\% %$\approx$500K 
in the line forming regions. Therefore, derived atmospheric parameters differ by less than the observational uncertainties.   
%Using mock data we find that atmospheric parameters derived from such models differ from that derived from the full models by less than the uncertainties encountered in the analysis of observed spectra. 
Hence grids of synthetic spectra of H/He/N or H/He/C composition are sufficient to derive the atmospheric parameters of He-sdO stars. In order to avoid inconsistencies the chemical abundances are derived from full models at fixed effective temperatures and gravities by adding element by element in a sequence of decreasing solar abundance.

\item Chemical abundances:

\begin{itemize}
\item[] Hydrogen is a trace element in all four program stars. Correlations among C, N, O, and Ne abundances are found, that means the O/N and Ne/N ratios correlate positively with the N/C.  
For \CD\ the He and N enrichment is accompanied by C and O depletion, a typical signature of hydrogen burning in the CNO bi-cycle. While the Ne abundance is solar. Neon is enriched in the C\&N-rich stars but shows a large scatter, extending up to
 a factor of ten with respect to the sun in \lsIV. 

\item[] Elements heavier than Ne are subsolar by about a factor of two, except for Si which is close to solar. The abundance patterns of the C-rich He-sdOs are more complex. A somewhat subsolar metallicity is accompanied by C and N-enrichment and O-deficiency, less pronounced than in \CD. 
%Neon is mildly to strongly enriched, up to 
The nickel-to-iron ratio is significantly supersolar. 

\end{itemize}

\item Trigonometric parallaxes measured by the Gaia satellite allows us for the first time to derive stellar masses, which turns out to be consistent with predictions from canonical evolutionary models. 
From the analysis of spectral energy distributions we rule out the presence of late-type stellar companions.

\item The analyses aimed at constraining helium WD merger scenarios on the one hand and hot flasher scenarios on the other:

\begin{itemize}
\item[] {\bf Merger Scenarios:} The slow, cold merger slowly accreting from a disk, predicts no nuclear burning effecting the surface composition of the remnant. A fast merger predicts a hot corona to form and helium burning to occur. The composite merger scenario considers both processes to occur simultaneously. 
Except for the low oxygen abundance observed, the abundance pattern of \CD\ is consistent with predictions of models of cold, slow helium WD mergers, that is the atmospheric composition is that of the interior of a lighter helium WD, that has been accreted by the more massive one to form the He-sdO. Carbon-rich He-sdOs can be formed in a composite merger, if their masses are significantly larger than the canonical one. Derived masses for \lss\ and \lsIV\ are lower than the canonical value with \uv\ being the only relatively massive object of the sample. The remnant of a He-WD merger is expected to retain significant angular momentum and rotate at 30 to 100 km\,s$^{-1}$. However, \uv\ is the only program star that rotates at the expected rate and may therefore be a viable merger remnant candidate. 

\item[] {\bf Hot flasher scenarios:} Early flasher as well as late flasher models with shallow mixing can be excluded because both predict much larger fractions of hydrogen to remain on the stellar surface. Therefore the late hot flasher scenario with deep mixing is the only viable one. The occurrence of convective zones predict 
the dredge-up of carbon during the helium flash as well as hydrogen burning during the mixing episodes. New evolutionary calculations predict abundance patterns similar to those determined for the C-rich program stars. The C-deficient \CD, however, would be difficult to reconcile in the context of the flasher scenario.
\end{itemize}
\end{enumerate}
}

In order to distinguish between the rivalling formation scenarios, it is of utmost importance to determine the stellar masses. The merger scenario predicts larger masses in particular for C-rich objects, exceeding about 0.65\,M$_\odot$. %We made use of trigonometric parallaxes from Gaia's second  data release to derive masses from the atmospheric parameters and the stars' spectral energy distributions. The results are close to the canonical mass for the He core flash. 
{Hence, the masses derived from parallax and gravity may be tale telling.}
However, due to the uncertainties of the surfaces gravities the error margin is too large to be conclusive, although it is worthwhile to note that \uv\ is the most massive program star and, therefore, a promising candidate for a composite He-WD merger. The derived C, N, O, and Ne abundances, in particular their correlations, provide important constraints for the merger models. A more accurate determination of the surface gravity will be needed to narrow down the uncertainties of the stellar masses, which are predominantly of systematic nature. Therefore, we started a project to obtain very high S/N, medium resolution spectra with very wide wavelength coverage for differential spectral analyses in order to reduce the systematic uncertainties of the surface gravities. In addition such differential quantitative spectral analyses have to be carried out for a larger sample to establish the mass distribution of He-sdO stars. 

\begin{acknowledgements}
MS acknowledges funding by the Deutsche Zentrum f\"ur Luft- und Raumfahrt (grant No. 50 OR 1406). PN acknowledges funding by the Deutsche Forschungsgemeinschaft (grant No. HE 1356/45-2) and by the Grant Agency of the Czech Republic (grant No. GACR 18-20083S).
This research has made use of ISIS functions provided by ECAP/Remeis obervatory.
We thank Sabine Moehler for her help in reducing and correcting the UVES spectrum of \CD.
Some of the data presented in this paper were obtained from the Mikulski Archive for Space Telescopes (MAST). STScI is operated by the Association of Universities for Research in Astronomy, Inc., under NASA contract NAS5-26555. Support for MAST for non-HST data is provided by the NASA Office of Space Science via grant NNX09AF08G and by other grants and contracts.
Based  on observations  at the Paranal  Observatory  of the  European  Southern  Observatory  for  program  number  69.C-0171(A), 072.D-0290(A), 077.C-0547(A) and 67.D-0047(A). Based  on  observations  at  the  La  Silla  Observatory  of  the  European  Southern Observatory  for programs  number 074.B-0455.
This research has used the services of {\sc Astroserver.org}.
TB and M3B are partially supported by ANPCyT through grant PICT-2014-2708, by the MinCyT-DAAD bilateral cooperation program through grant DA/16/07, and by a Return Fellowship from the Alexander von Humboldt Foundation.
M.L. acknowledges support from the Alexander von Humboldt Foundation.

\end{acknowledgements}
%%%%%%%%%%%%%%%%%%%%%%%%%%%%%%%%%%%%%%
\bibliographystyle{aa} 
\bibliography{references.bib} 
%%%%%%%%%%%%%%%%%%%%%%%%%%%%%%%%%%%%%%

\begin{appendix}
\section{Additional material}

In this appendix we give details about the spectroscopic data used (Sect.\ref{specdata}), list the lines used for the determination of the atmospheric parameters in Sect. \ref{line_list_hhe} and illustrate the quality of fit for \uv\ in Sect. \ref{hhe_uv}. We present a reanalysis of CASPEC spectra in  \ref{sect:caspec} and show the SED fits for \uv\ and \lss\ in Sect. \ref{sec:appendix_sed}. Finally, we list the metal lines used for the abundance analyses in Sect. \ref{line_list_metals}.

\subsection{Spectroscopic data}\label{specdata}
Table \ref{sample-overview} lists the observational data for the four program stars with the available information.
\begin{table*}[h!]
\caption{List of observational data for the four He-sdO stars. We list the name of the star, the instrument used, the Program ID, the available wavelength range, the exposure time and the average resolving power for the observations we included in our analysis. For \lsIV, only the blue part of the UVES spectrum was available in the ESO Data Archive.
}
\label{sample-overview}
\centering
\setstretch{1.3}
\begin{tabular}{l l l c r r}%looks funny
\hline\hline
Star& Instrument&Program ID&Wavelength range [\AA]&Exposure time [s]&R\\
\hline
\CD&UVES&69.C-0171(A)&3281 -- 4562&2700&68642\\
\CD&UVES&69.C-0171(A)&4583 -- 6686&2700&107200\\
\CD&FEROS&074.B-0455(A)&3527...9216&600&48000\\
\CD&FUSE&p2050603000&905 -- 1185&11747&22000\\[3pt]       
\uv&FEROS&074.B-0455&3527 -- 9216&1000&48000\\
\uv&FEROS&074.B-0455&3527 -- 9216&1000&48000\\       
\uv&FUSE&p2051302000&905 -- 1185&3157&22000\\
\uv&FUSE&p2051303000&905 -- 1185&2385&22000\\[3pt]
\lss&UVES&077.C-0547(A)&4726 -- 6835&2100&66320\\
\lss&UVES&077.C-0547(A)&4726 -- 6835&2100&66320\\
\lss&UVES&077.C-0547(A)& 3024 -- 3884&2100&49620\\
\lss&UVES&077.C-0547(A)& 3024 -- 3884&2100&49620\\
\lss&CASPEC& & 3925 -- 49{53} &  & 18000 \\
\lss&FUSE&p2051701000&905 -- 1185&13927&22000\\
\lss&FUSE&p2051702000&905 -- 1185&7935&22000\\[3pt]
\lsIV&UVES&67.D-0047(A)&3281 -- 4562&1500&53750\\
\lsIV&CASPEC& & 3930 -- 4888 &  &  18000 \\
\lsIV&FUSE&p2050901000&905 -- 1185&6639&22000\\
\lsIV&FUSE&p2050903000&905 -- 1185&45577&22000\\
\hline
\end{tabular}
\end{table*}

\begin{table}[h!]
\caption{Visual hydrogen and helium lines used for the determination of atmospheric parameters.}\label{list_h_he}
\begin{center}
\setstretch{1.3}
\begin{tabular}{cc}
\hline\hline
&wavelength [\AA]\\
\hline
He\,{\sc i}&3888.65\\
He\,{\sc ii}\,/\,H\,{\sc i}&3968.44\,/\,3970.08\\
He\,{\sc ii}\,/\,He\,{\sc i}&4025.61\,/\,4026.19\\
He\,{\sc ii}\,/\,H\,{\sc i}&4100.05\,/\,4101.71\\
He\,{\sc ii}&4199.84\\
He\,{\sc ii}\,/\,H\,{\sc i}&4338.68\,/\,4340.47\\
He\,{\sc i}&4387.93\\
He\,{\sc i}&4437.55\\
He\,{\sc i}&4471.48\\
He\,{\sc ii}&4541.59\\
He\,{\sc ii}&4685.70\\
He\,{\sc i}&4713.15\\
He\,{\sc ii}\,/\,H\,{\sc i}&4859.32\,/\,4861.35\\
He\,{\sc i}&4921.93\\
He\,{\sc i}&5015.68\\
He\,{\sc i}&5047.74\\
He\,{\sc ii}&5411.51\\
He\,{\sc i}&5875.75\\
He\,{\sc ii}\,/\,H\,{\sc i}&6560.09\,/\,6562.79 \\
He\,{\sc i}&6678.15\\
\hline
\end{tabular}
\end{center}
\end{table}

\subsection{Hydrogen and helium lines}\label{line_list_hhe}

Hydrogen and helium lines used to obtain the atmospheric parameters are listed in Table \ref{list_h_he}.

\subsection{Hydrogen and helium line profile fits for \uv\ and \lsIV }\label{hhe_uv}
\begin{figure*}[h!]
\begin{center}
\includegraphics[width=0.82\textwidth]{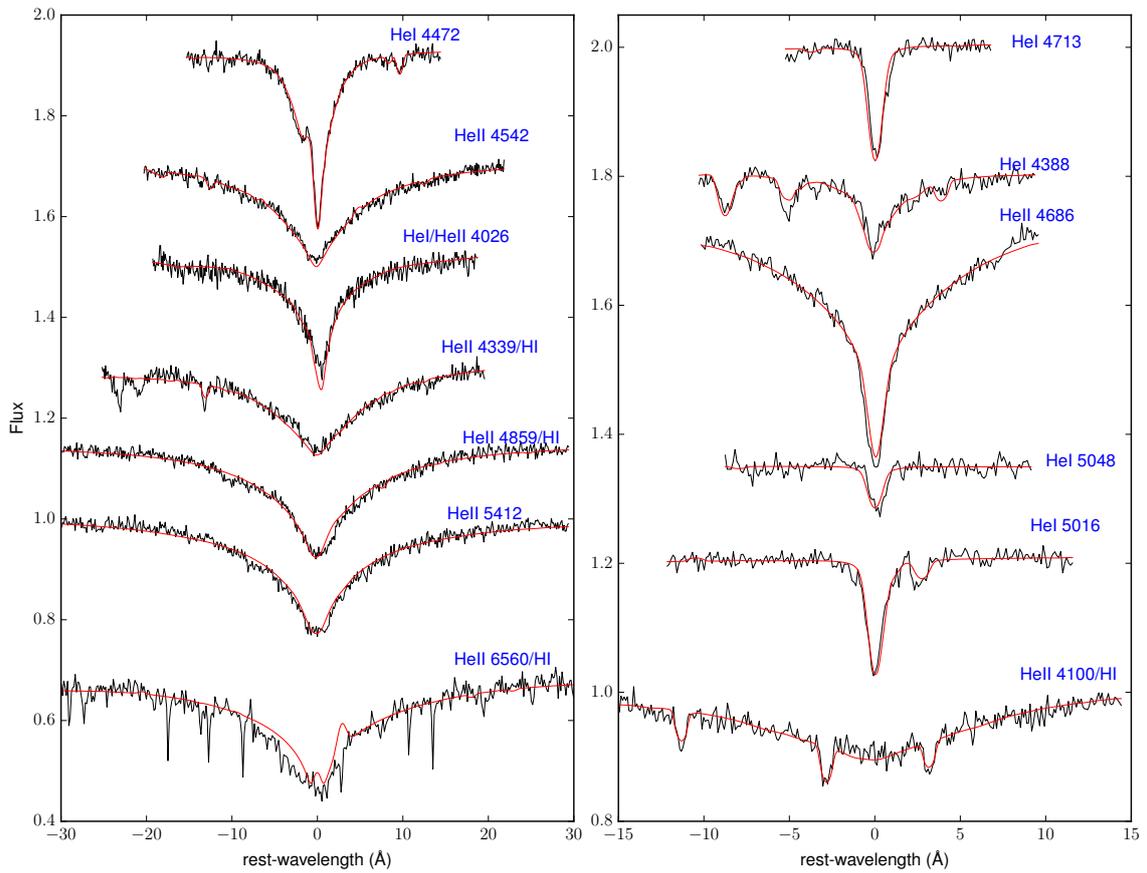}
\caption{{He\,{\sc i} and He\,{\sc ii} line profiles from the final model compared to the normalized observed spectrum of \uv.}
 The spectra have been corrected for radial velocity.
{We note that telluric lines in the He\,{\sc ii}/H\,{\sc i}, 6560\AA\ line are particularly strong and also affect the core of that line, in particular near the predicted H $\alpha$ emission component.
Because it was not possible to match the He\,{\sc ii}/H\,{\sc i} 6560 line it was not included in the fit and is only shown for comparison here.} 
The rotational velocity of \uv\ leads to a notable broadening and smearing of the line profiles.\label{uv-helium} }
\end{center}
\end{figure*}

Figure \ref{uv-helium} shows fits with a H/He/C grid to an UVES spectrum of \uv. For details see Sect. \ref{atmos_params}.

\begin{figure*}[h!]
\begin{center}
\includegraphics[width=0.6\textwidth]{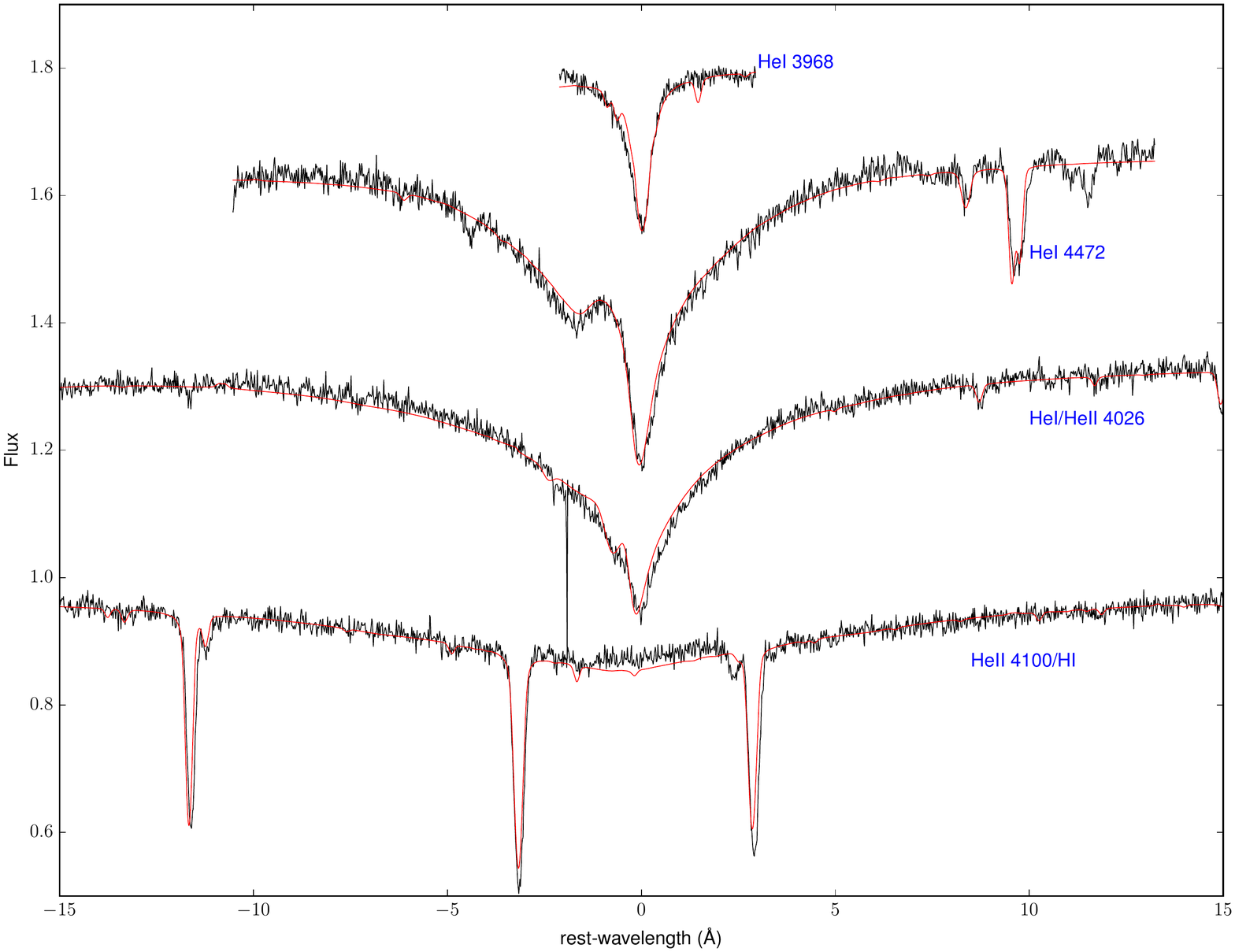}
\caption{{He\,{\sc i} and He\,{\sc ii} line profiles from the final models compared to the normalized observed spectrum of \lsIV.} The spectra have been corrected for radial velocity.
\label{lsiv-helium} }
\end{center}
\end{figure*}

\subsection{A reanalysis of CASPEC spectra}\label{sect:caspec}

Previous spectral analyses of our program stars \citep{bauer1995} and \citep{Dreizler1993} used atmospheric models and analysis strategies different from ours. In order to investigate the systematic differences arising, we reanalyzed the visual spectra used in the previous investigations, that is making use  
of the same intermediate resolution (R=18\,000) CASPEC spectra.

In his analysis of \lsIV, \lss, and \uv, \citet{Dreizler1993} fixed the helium abundance at 100 times that of hydrogen to determine the effective temperature and gravity, which we adopt here, too.

For \CD\ the helium abundance was fixed to $\log{y}=2.6$ as this matched the data best.
 
We compare our results for the atmospheric parameters obtained with the pure H/He grid and a H/He+X grid, where X is the most abundant element metal (either nitrogen for \CD\ or carbon for the other stars) in Table \ref{comp-results}. 
 
\begin{table*}
\setstretch{1.3}
\caption{Comparison of atmospheric parameters to literature values determined from CASPEC spectra using grids with different metallicity.\label{comp-results}}
\begin{center}
\begin{tabular}{clccc}
\hline\hline
\multirow{3}{*}{Star}&\multirow{3}{*}{Parameters}&H/He &H/He  ({\sc Tlusty})&H/He+X ({\sc Tlusty})\\
\cline{3-5}
&&\citet{bauer1995}&\multirow{2}{*}{this work}&\multirow{2}{*}{this work}\\
&&\citet{Dreizler1993}&&\\
\hline
&$T_{\rm eff}$ [K]&$44000\pm2000$&$42790\pm700$&$42570\pm500$\\
\CD&$\log{g}$ [cm\,s$^{-2}$]&$5.4\pm0.3$&$5.41\pm0.08$&$5.50\pm0.07$\\
&$\log{y}$ (fixed)& &2.6&2.6\\
%\hline
&$T_{\rm eff}$ [K]&$44500\pm1000$&$43200\pm600$&$43700\pm500$\\
\lss&$\log{g}$ [cm\,s$^{-2}$]&$5.55\pm0.15$&$5.22\pm0.05$&$5.25\pm0.10$\\
&$\log{y}$ (fixed)&2.0&2.0&2.0\\
%\hline
&$T_{\rm eff}$ [K]&$44500\pm1000$&$45880\pm500$&$45420\pm750$\\
\lsIV&$\log{g}$ [cm\,s$^{-2}$]&$5.55\pm0.15$&$5.48\pm0.05$&$5.58\pm0.08$\\
&$\log{y}$ (fixed)&2.0&2.0&2.0\\
%\hline
&$T_{\rm eff}$ [K]&$46500\pm1000$&$48090\pm800$&$48090\pm750$\\
\uv&$\log{g}$ [cm\,s$^{-2}$]&$5.55\pm0.15$&$5.40\pm0.07$&$5.49\pm0.07$\\
&$\log{y}$ (fixed)&2.0&2.0&2.0\\
\hline
\end{tabular}
\end{center}
\end{table*}

From Table \ref{comp-results} it is apparent that the results from our H/He grid are all in accordance with the results from previous investigations and the error ranges overlap for all parameters. 

The effects of the additional metals in the H/He+C and H/He+N grids 
can be explained by the effects of back-warming. The atmospheric parameters derived from the analyses of CASPEC spectra and those derived from of modern visual spectra agree also very well (cf. Table \ref{sample-atmos}). 

\subsection{The C\,{\sc ii}, 4267\AA\ line}\label{sect:c2}

\begin{figure}%[h!]
\begin{center}
\includegraphics[width=\columnwidth]{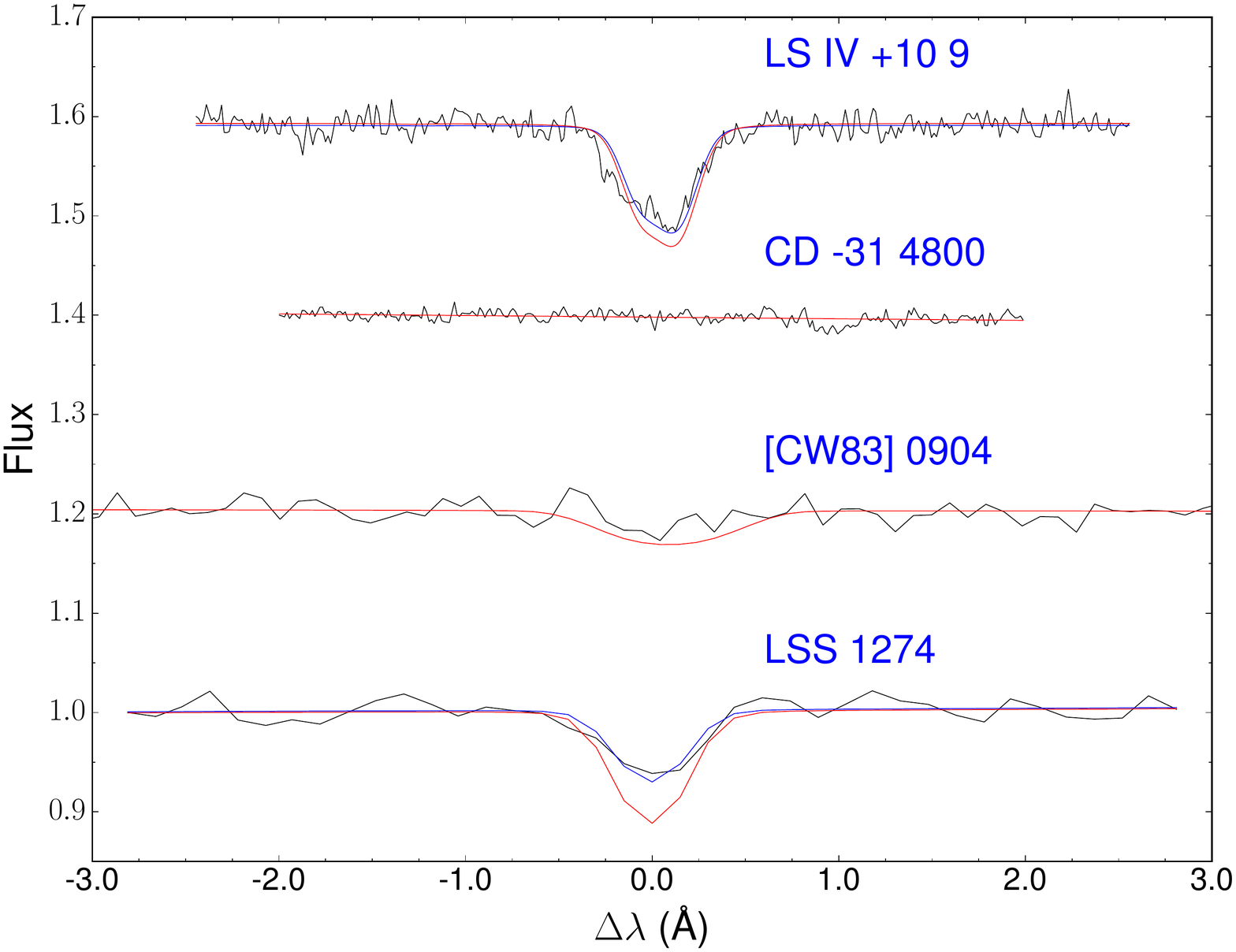}
\caption{{Comparison of the observed C\,{\sc ii}, 4267\AA\ line to the prediction from the final model. Synthetic line profiles calculated for the C abundances listed in Tables \ref{sample-abus1} and \ref{sample-abus34} are plotted in red. For
\lsIV\ and \lss\ the blue profiles represent the best fits to the line.}} 
\label{fig:c2}
\end{center}
\end{figure}

The spectral line of singly ionized carbon (C\,{\sc ii}) at 4267\AA\ is present in the visual spectra of the three C-rich stars (see Sects.~\ref{sect:cno}), but is very weak. {Nevertheless,} lines of ionized carbon were included in models containing H,He,C,N, and O, but had to be excluded in the full model, because the very low population of C\,{\sc ii} atomic levels led to numerical instabilities and prevented the models to converge. However, the carbon abundance determined in the latter model (from C\,{\sc  iii} and C\,{\sc iv} lines only) differs from the one of the former model (C\,{\sc ii}, C\,{\sc iii}, and C\,{\sc iv}) by 0.02 dex, only, which is much less than the abundance uncertainties and, therefore, insignificant.
{In order to evaluate C\,{\sc ii}, 4267\AA\ (see Sect.\ref{sect:cno}) line profiles from the final models and adopted C abundances (Tables \ref{sample-abus1} and  \ref{sample-abus34}) are compared to the observed spectra in Fig.~\ref{fig:c2}. For \CD\ the line is predicted to be too weak to be observable, consistent with observations. The line is hardly visible in the spectrum of \uv\ because of the star's rotation.
For \lsIV\ 
 the observed profile is well reproduced, while the line is predicted stronger than observed for \lss.
% A fit to match the hardly detectable line in \uv\ requires a C abundance lower by 0.37 +-0.23 dex than from C III/IV, }
A fit to match the individual line (see Fig.~\ref{fig:c2}) results in $\log(n_{\rm C}/n_{\rm H}$)=0.11$\pm$0.10 for \lsIV\ and log(n$_{\rm C}$/n$_{\rm H}$)=-0.35$\pm$0.09 for \lss,
which are close to the adopted C/H ratios of 0.15 dex and -0.01 dex, respectively, to within the scatter range of individual C\,{\sc iii} and C\,{\sc iv} lines.}

\subsection{SED and photometric fit for \CD, \lsIV,  and \uv}\label{sec:appendix_sed}
\begin{figure*}%[h!]
\begin{center}
\includegraphics[width=0.7\textwidth]{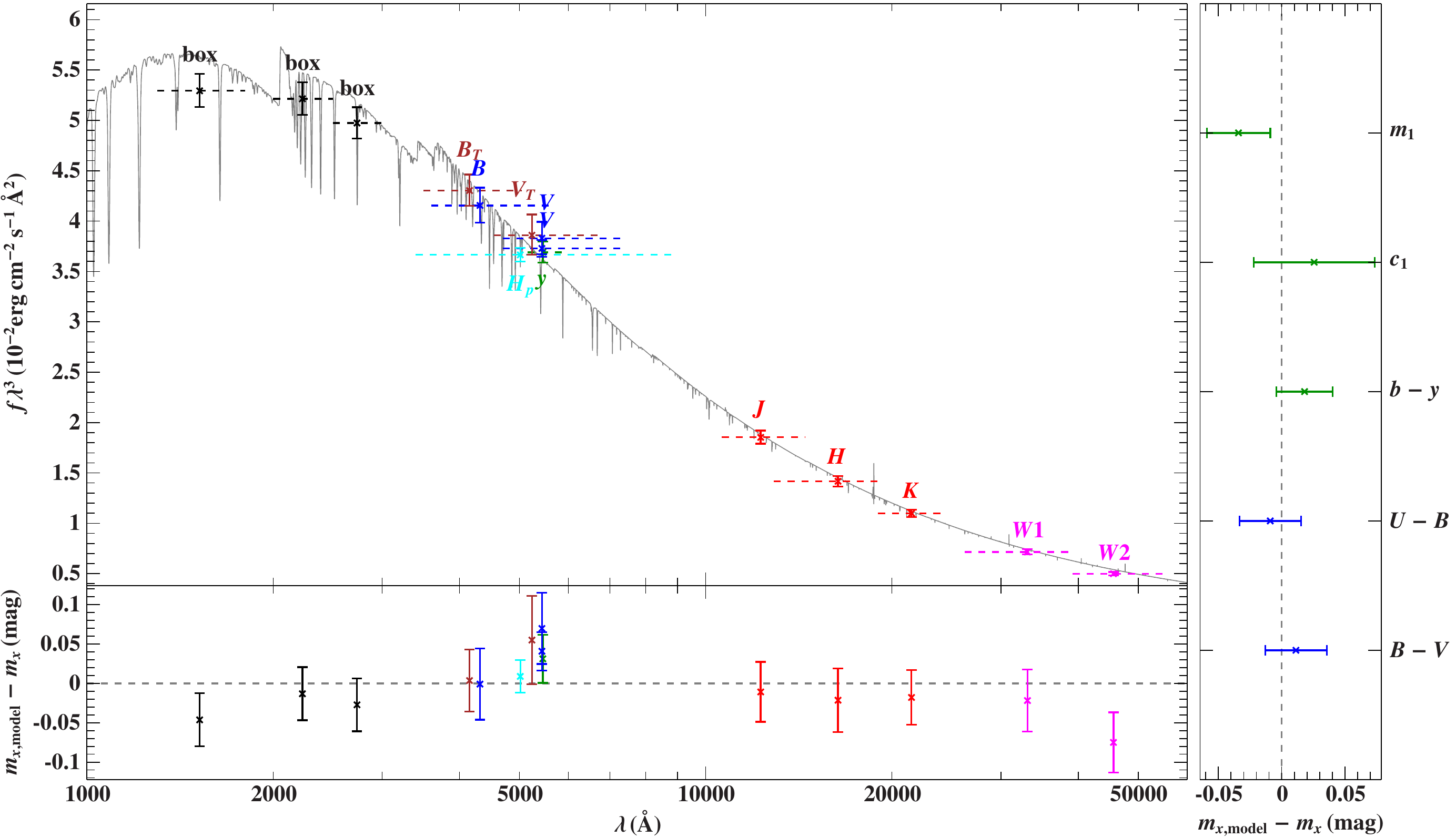}
\includegraphics[width=0.7\textwidth]{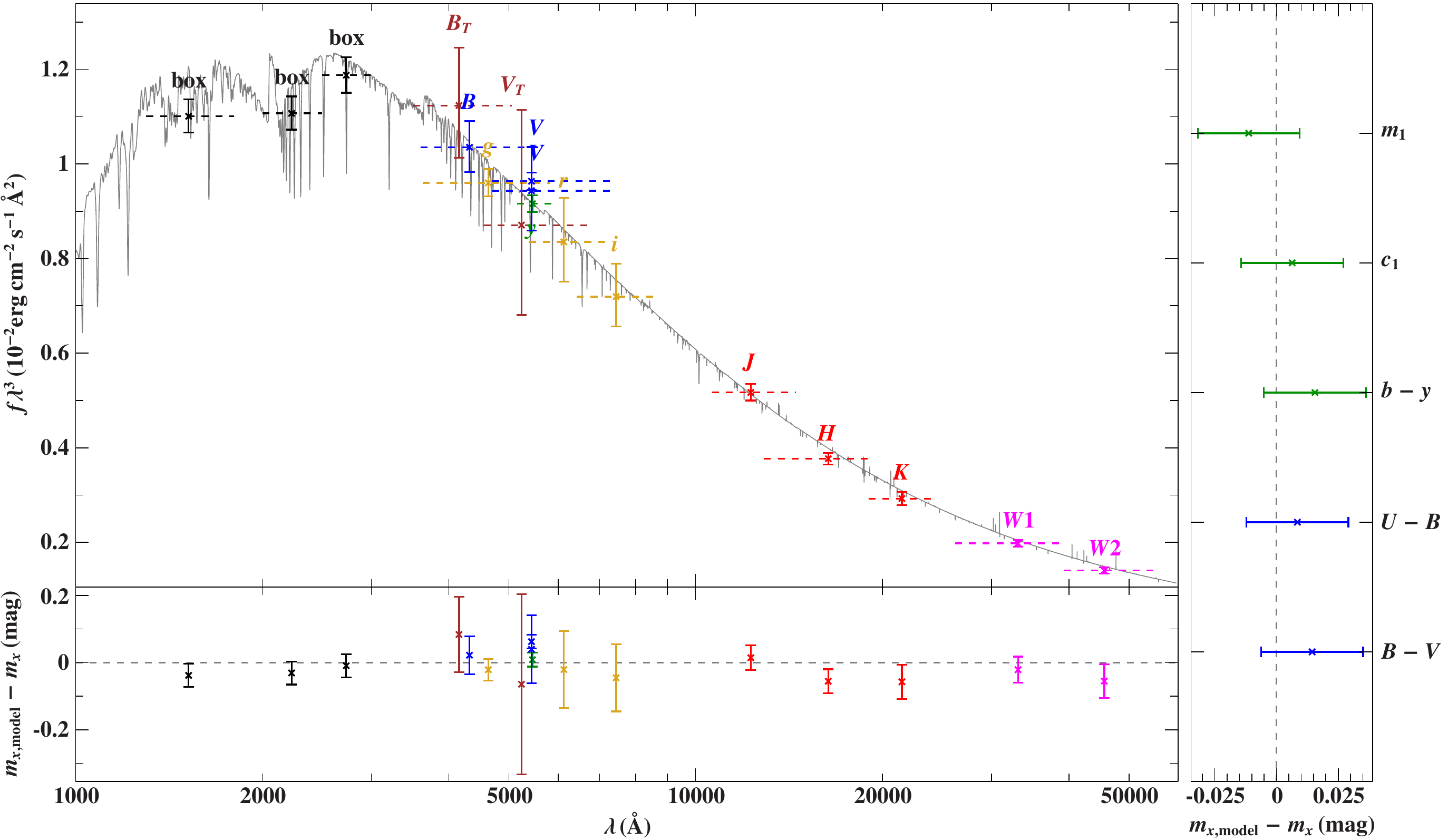}
\includegraphics[width=0.7\textwidth]{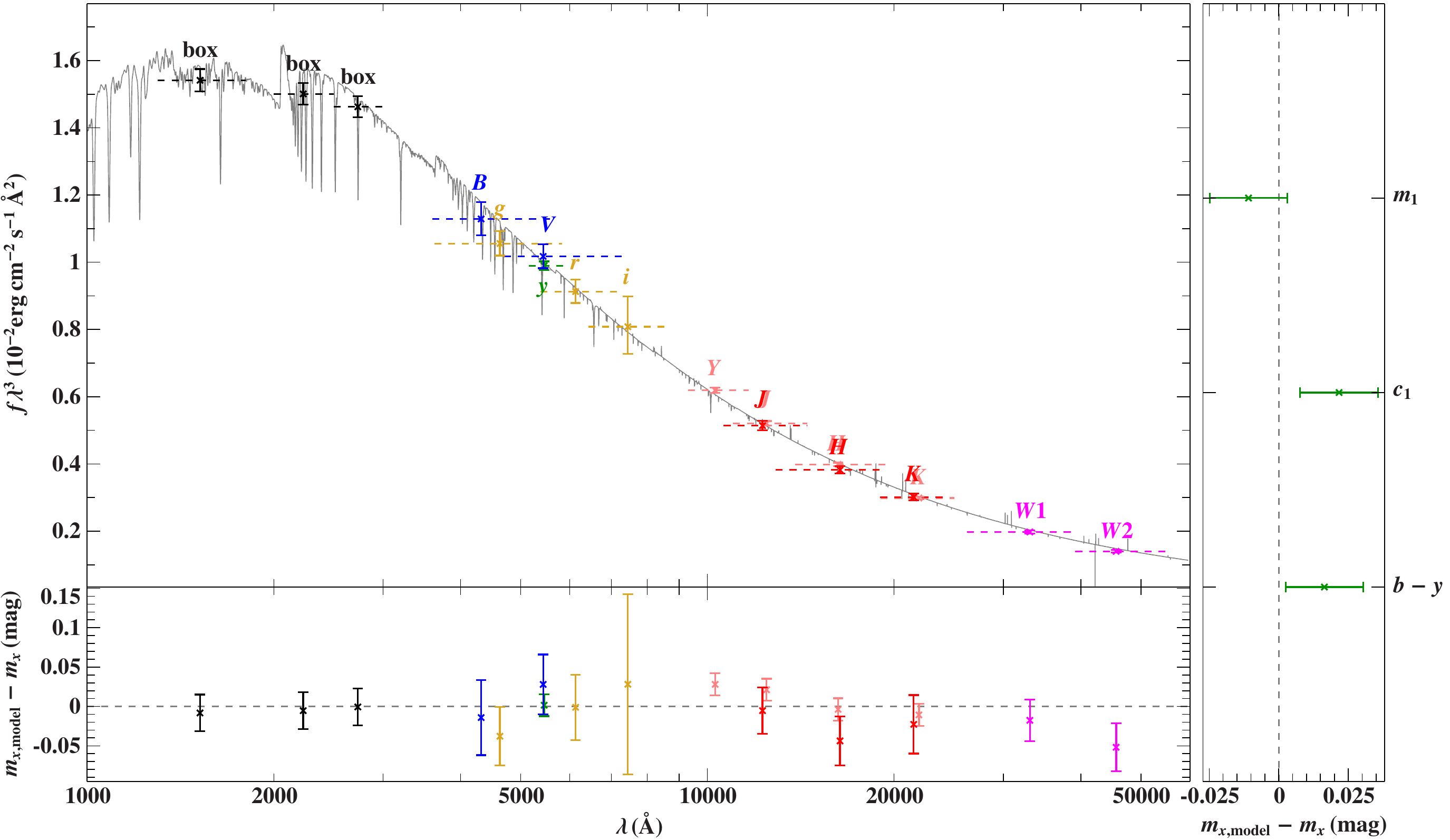}
\caption{Matching the spectral energy distributions and colors of \CD\ (upper panel), \lsIV\ (middle panel), and \uv\ (lower panel). The synthetic SEDs (grey) are overlayed to the fluxes derived from observed apparent magnitudes in different photometric systems. The residuals for the magnitudes and colors are given in the panels below (SED) and to the right (colors) of the main panel.} 
\label{sed_fit_other}
\end{center}
\end{figure*}

Figure \ref{sed_fit_other} shows the fits to the spectral energy distribution of \CD, \lsIV, and \uv\ (from top to bottom) . Overlayed on the SED (gray) are the measurements for different magnitudes in different systems. The residuals for the magnitudes and colors are given in the smaller panels below and to the right of the SED. For details see Sect. \ref{angular}.

\subsection{Line list for abundance analysis}\label{line_list_metals}

In Table \ref{1xx} and Table \ref{2xx} we list the spectral lines used for the metal abundance analysis for the visual/UVA spectral and the FUV spectral range respectively.

\begin{table*}[h!]
\caption{Visual and UVA spectral lines used for the abundance determination. {The profiles of some spectral lines were calculated together, either because they are closely blended lines or resolved multiple lines, for which no lines from other species intervene (e.g. Ne\,{\sc ii}, 3643.93/3644.86\AA, see Fig. \ref{cd3148_neon}). The wavelengths of such ''combined'' lines} are marked by asterisks.}\label{1xx}
\begin{minipage}{0.45\textwidth}
\begin{center}
\begin{tabular}{cc}
\hline\hline
\rule{0pt}{15pt}Element+Ion. stage&Wavelength [\AA]\\
\hline
%C\,{\sc ii}&3876.19\\
%C\,{\sc ii}&3920.69\\
%C\,{\sc ii}&*\\
%*&4267.00\\
%*&4267.26\\
%C\,{\sc ii}&5145.16\\
%C\,{\sc ii}&6578.05\\
%C\,{\sc ii}&6582.88\\
%\hline
C\,{\sc iii}& 3883.81\\
%C\,{\sc iii} + He\,{\sc i}&*\\
%*&3888.65\\
%&3892.51\\
C\,{\sc iii} & * \\
*& 3608.78\\
* &3609.05\\
* &3609.07\\
* & 3609.62\\
* & 3609.68\\
%C\,{\sc iii}&3609.05\\
%C\,{\sc iii}&3609.52\\
C\,{\sc iii}&4056.06\\
C\,{\sc iii}&*\\	
*&4067.94\\
*&4068.91\\
*&4070.26\\
%C\,{\sc iii}&4070.01\\ % not identifiable
%C\,{\sc iii}+He\,{\sc i}&*\\
%*&4120.85\\
C\,{\sc iii}&4121.85\\ %He I Blend 
C\,{\sc iii}&4186.90\\	
C\,{\sc iii}&4247.31\\
C\,{\sc iii}& 4325.56 \\ %revised wavelength from 4325.7
C\,{\sc iii} &4361.85\\	
%C\,{\sc iii}&4669\\
C\,{\sc iii} &4382.90\\
%C\,{\sc iii}+He\,{\sc i}& *\\
%*&4387.93\\ % this He I the C III line is 5 AA away
%*&4382.90\\
C\,{\sc iii}&4516.79\\ %4516.3 Korrigiert NIST
%C\,{\sc iii} + C\,{\sc iv}&4649\\
%*&4647.42\\
%*&4658.30\\
C\,{\sc iii}&*\\
*&4647.42\\
*&4650.25\\
*&4651.02\\
*&4651.47\\
*&4652.05\\
%C\,{\sc iii}&5260\\
C\,{\sc iii}&5695.92\\
%C\,{\sc iii} &5826.42\\	
%C\,{\sc iii} &5827.12\\
%C\,{\sc iii}&5826.42\\
C\,{\sc iii}& *\\
*&6149.28\\
*&6154.16\\
*&6155.12\\
*&6156.69\\
*&6160.01\\
%C\,{\sc iii}&6350.77\\
%\hline
% He\,{\sc i} 5016 + C\,{\sc iv}&5015.69\\
C\,{\sc iv} &4658.30\\
C\,{\sc iv}& *\\
*&5016.62\\ 
*&5018.40\\ 
%C\,{\sc iv}&5801.33\\ % RV use
%C\,{\sc iv}&5811.98\\ % RV use
C\,{\sc iv}& *\\
*&5801.33\\
*&5811.97\\
%\hline
N\,{\sc ii}&3838.37\\
N\,{\sc ii}&3842.19\\
N\,{\sc ii}&3995.00\\
N\,{\sc ii}&4176.16\\
N\,{\sc ii}&4227.74\\
N\,{\sc ii}&4447.03\\
N\,{\sc ii}&4694.27\\
N\,{\sc ii}&4779.72\\
N\,{\sc ii}&4788.13\\
N\,{\sc ii}&4803.29\\
N\,{\sc ii}&4987.38\\
N\,{\sc ii}&5002.80\\
N\,{\sc ii}&5007.32\\
N\,{\sc ii}&5010.62\\
N\,{\sc ii}&5025.66\\
N\,{\sc ii}&5045.10\\
N\,{\sc ii}&5495.67\\
\hline
\end{tabular}
\end{center}
\end{minipage}
\begin{minipage}{0.45\textwidth}
\begin{center}
\begin{tabular}{cc}
\hline \hline
\rule{0pt}{15pt}Element+Ion. stage&Wavelength [\AA]\\
\hline
N\,{\sc ii}&5666.63\\
N\,{\sc ii}&5676.02\\
N\,{\sc ii}&5679.56\\
N\,{\sc ii}&5686.21\\
N\,{\sc ii}&5710.77\\
N\,{\sc ii}&5931.78\\
%\hline
N\,{\sc iii}&3004.03 \\ % outside wavelength range
N\,{\sc iii}&3330.11\\
N\,{\sc iii}&3342.09\\
N\,{\sc iii}&*\\
*&3353.98\\
*&3354.32\\
N\,{\sc iii}&*\\
*&3360.98\\
*&3365.80\\
N\,{\sc iii}&*\\
*&3365.80\\
*&3367.30\\
*&3374.07\\
N\,{\sc iii}&*\\	
*&3754.62\\
*&3757.65\\
*&3762.60\\
N\,{\sc iii}&*\\
*&3771.05\\
*&3771.36\\
N\,{\sc iii}&3792.97\\
N\,{\sc iii}&3938.52\\	
N\,{\sc iii}&3942.88\\	
N\,{\sc iii}& *\\	
*&3998.63\\
*&4003.58\\
N\,{\sc iii}&4097.33\\
N\,{\sc iii}&4103.39\\
%\hline
%N\,{\sc iii}&4103.43\\ % same line but Ritz wavelength
N\,{\sc iii}&4195.70\\
%N\,{\sc iii}&4195.76\\ % same line
N\,{\sc iii}&4200.02\\	
N\,{\sc iii}&4215.77\\ %wavelength revised from 4215.77
N\,{\sc iii}&4318.78\\
N\,{\sc iii}&*\\
*&4318.78\\
*&4321.22\\
*&4325.43\\ % Blend C III 4325.43 in C-rich stars
*&4327.88\\
N\,{\sc iii}&*\\	
*&4332.91\\
*&4337.01\\
N\,{\sc iii}&*\\
* &4345.68 \\	
*& 4345.81\\ %moved here because it lies within this window
*& 4351.11\\
N\,{\sc iii}& *\\
*&4514.86\\
*&4518.91\\
N\,{\sc iii}&4523.56\\
N\,{\sc iii}&4530.86\\
N\,{\sc iii}&4534.58\\ %out of sequence move here
\hline

\end{tabular}
\end{center}
\end{minipage}
\end{table*}

\begin{table*}[h!]
Visual and UVA spectral lines used for the abundance determination. (continued).

\begin{minipage}{0.45\textwidth}

\begin{center}
\begin{tabular}{cc}
\hline\hline
\rule{0pt}{15pt}Element+Ion. stage&wavelength [\AA]\\
\hline
%N\,{\sc iii}&3998.63\\
%N\,{\sc iii}&4534.58\\ %out of sequence move down 	
%N\,{\sc iii}&4547.30\\ %out of sequence move down
%N\,{\sc iii}&4325.43\\ % listed twice 
%N\,{\sc iii}&4332.91\\ % listed twice 
%N\,{\sc iii}&4345.81\\ % lies within 4345.68 - 4345.81\\
%N\,{\sc iii}&4351.11\\% listed twice
%N\,{\sc iii}&4514.86\\ %listed twice	
N\,{\sc iii}&4547.30\\ %out of sequence move here
N\,{\sc iii}&*\\	
*&4589.18\\
*&4591.98\\
N\,{\sc iii}& 4634.14\\
N\,{\sc iii}+ He\,{\sc ii}\&H$\beta$& *\\ % He II/Hbeta 4859.32
*&4858.82\\
*&4861.22\\
*&4867.15\\
%N\,{\sc iii}&4867.15\\ listed twice
N\,{\sc iii}&4873.10\\
N\,{\sc iii}&4884.14\\
N\,{\sc iii}&*\\	
*&5260.86\\
*&5270.57\\
*&5272.68\\
*&5282.43\\ % possibly 5282.54 (revised by me) U.H.
N\,{\sc iii}& * \\
*&5297.75\\
*&5298.95\\
%N\,{\sc iii}&5314.35\\ %listed twice	
N\,{\sc iii}*\\	
*&5314.35\\
*&5320.32\\
*&5327.18\\
N\,{\sc iii}&*\\	
*&5817.79\\
*&5820.57\\
N\,{\sc iii}&5847.94\\	
N\,{\sc iii}&5954.70\\
%\hline
%N\,{\sc iv}&3478.71\\ %listed twice
N\,{\sc iv}&*\\
*&3478.71\\
*&3482.99\\
*&3484.96\\
%N\,{\sc iv}&3484.96\\ listed twice
N\,{\sc iv}&4057.76\\
N\,{\sc iv}&6380.77\\
%\hline
O\,{\sc ii}&3759.87\\	
O\,{\sc ii}&3857.16\\	
%O\,{\sc ii}&3961.59\\	%neither in NIST nor in Kurucz
O\,{\sc ii}&4069.89\\	
O\,{\sc ii}&4072.15\\	
O\,{\sc ii}&4075.86\\	
O\,{\sc ii}& *\\
*&4307.23\\
*&4312.10\\
*&4313.44\\
*&4317.14\\
*&4317.70\\
*&4319.63\\
*&4325.75\\
O\,{\sc ii}&4349.42\\	
O\,{\sc ii}&4336.87\\
O\,{\sc ii}&6152.56\\	
%O\,{\sc ii}&4419.90\\ %neither in NIST nor in Kurucz	
\hline
\end{tabular}
\end{center}
\end{minipage}
\begin{minipage}{0.45\textwidth}
\begin{center}
\begin{tabular}{cc}
\hline\hline
\rule{0pt}{15pt}Element+Ion. stage&wavelength [\AA]\\
\hline
O\,{\sc iii}&3340.74\\
O\,{\sc iii}&3459.48\\
O\,{\sc iii}&3715.08\\
O\,{\sc iii}&3757.21\\
O\,{\sc iii}&3759.87\\
O\,{\sc iii}&5592.37\\
O\,{\sc iii}&6507.55\\
%\hline
Ne\,{\sc ii}&3323.73\\
Ne\,{\sc ii}&3327.15\\
Ne\,{\sc ii}&3388.41\\
Ne\,{\sc ii}&3406.94\\
Ne\,{\sc ii}& *\\
*&3416.91\\
*&3417.69\\
Ne\,{\sc ii}&3542.84\\
%\hline
Ne\,{\sc ii}&*\\
*&3565.82\\
*&3568.50\\
*&3571.23\\
*&3574.61\\
Ne\,{\sc ii}& *\\
*&3643.93\\
*&3644.86\\
Ne\,{\sc ii}&3664.07\\
Ne\,{\sc ii}&3694.21\\
Ne\,{\sc ii}&3709.62\\
Ne\,{\sc ii}&3713.08\\
Ne\,{\sc ii}&3727.11\\
Ne\,{\sc ii}&3766.26\\
Ne\,{\sc ii}&3777.13\\
Ne\,{\sc ii}&4150.69\\
Ne\,{\sc ii}&* \\
*&4217.17\\
*&4219.37\\
*&4219.74\\
Ne\,{\sc ii}&4250.64\\ 
Ne\,{\sc ii}& * \\
*&4290.37\\
*&4290.60\\
Ne\,{\sc ii}&4379.55\\
Ne\,{\sc ii}&4397.99\\
Ne\,{\sc ii}&*\\
*&4409.30\\
*&4412.59\\
Ne\,{\sc ii}& *\\
*&4428.52\\
*&4428.63\\
*&4430.90\\
*&4431.81\\
\phantom{*}&\phantom{4431.81}\\
\hline

\end{tabular}
\end{center}
\end{minipage}
\end{table*}
\begin{table*}[h!]
Visual and UVA spectral lines used for the abundance determination (continued).%

\begin{minipage}{0.45\textwidth}
\begin{center}
\begin{tabular}{cc}
\hline\hline
\rule{0pt}{15pt}Element+Ion. stage&wavelength [\AA]\\
\hline
Mg\,{\sc ii}&4242.45\\
Mg\,{\sc ii}& *\\
*&4481.13\\
*&4481.33\\
Mg\,{\sc ii}& *\\
*&5916.43\\
%*&5917.40 \\ not in NIST or Kurucz 
*&5918.16\\
Mg\,{\sc ii}&5928.23\\
Mg\,{\sc ii}& *\\
*&6346.74\\
*&6346.96\\
%\hline
Al\,{\sc iii}&3601.63\\
Al\,{\sc iii}&4149.96\\
*&4159.91\\
*&4149.96\\
*&4150.17\\
%Al\,{\sc iii}&4479.88\\ %replace by doublet
Al\,{\sc iii}& *\\
*&4479.83\\
*&4479.97\\
Al\,{\sc iii}&4512.57\\
Al\,{\sc iii}&5722.73\\
%Al\,{\sc iii}&4512.57\\ % out of sequence
Al\,{\sc iii}&5696.60\\
Si\,{\sc iii}&3486.91\\
Si\,{\sc iii}&3791.41\\
Si\,{\sc iii}&3806.54\\
Si\,{\sc iii}&3924.47\\
Si\,{\sc iii}&4552.65\\
Si\,{\sc iii}&4567.87\\
Si\,{\sc iii}&4574.78\\
Si\,{\sc iii}&4716.66\\
\hline
\end{tabular}
\end{center}
\end{minipage}
\begin{minipage}{0.45\textwidth}
\begin{center}
\begin{tabular}{cc}
\hline\hline
\rule{0pt}{15pt}Element+Ion. stage&Wavelength [\AA]\\
\hline

%\hline

Si\,{\sc iii}&4813.39\\
Si\,{\sc iii}&4819.72\\
Si\,{\sc iii}&5739.73\\
%\hline
Si\,{\sc iv}&3773.15\\
Si\,{\sc iv}&4088.86\\
Si\,{\sc iv}&4116.10\\
Si\,{\sc iv}&4212.41\\
Si\,{\sc iv}&4314.10\\
Si\,{\sc iv}&4631.24\\
Si\,{\sc iv}&4654.32\\
Si\,{\sc iv}&4950.11\\
%\hline
P\,{\sc iv}&3347.74\\
P\,{\sc iv}&3364.47\\
P\,{\sc iv}&4249.66\\	
P\,{\sc iv}&4728.37\\	
P\,{\sc v}&5122.19\\
%\hline
S\,{\sc iii}&3632.02\\
S\,{\sc iii}&3662.01\\
S\,{\sc iii}&3717.77\\
S\,{\sc iii}&3747.85\\
S\,{\sc iii}&3794.68\\
S\,{\sc iii}&3837.79\\
S\,{\sc iii}&3928.61\\
S\,{\sc iii}&3961.51\\
S\,{\sc iii}&4099.15\\
S\,{\sc iii}&4253.59\\
S\,{\sc iii}&4284.99\\
S\,{\sc iii}&4361.47\\
%S\,{\sc iii}&4361.52\\ % probably same line as 4361.47, can not be resolved anyway
%\hline
S\,{\sc iv}&5488.25\\
S\,{\sc iv}&5497.75\\
\hline

\end{tabular}
\end{center}
\end{minipage}
\end{table*}

\begin{table*}[h!]
\caption{UV lines in the FUSE spectra used for the abundance determination.}\label{2xx}
\begin{minipage}{0.45\textwidth}
\begin{center}
\begin{tabular}{cc}
\hline\hline
\rule{0pt}{15pt}Element+Ion. stage&wavelength [\AA]\\
\hline
%He\,{\sc i}&2829.08\\
%He\,{\sc i}&2945.11\\
%\hline
C\,{\sc iii}&1051.15\\
%2xC\,{\sc iii}&1070.25\\
C\,{\sc iii}& * \\
*&1070.17\\
*&1070.33\\
%5xC\,{\sc iii}&1125.66\\
C\,{\sc iii}& *\\
*&1125.63\\
*&1125.64\\
*&1125.65\\
*&1125.67\\
*&1125.68\\
C\,{\sc iii}&1139.90\\
%5xC\,{\sc iii}&1148.93\\
C\,{\sc iii}& *\\
*&1148.89\\
*&1148.90\\
*&1148.91\\
*&1148.92\\
*&1148.94\\
%3xC\,{\sc iii}& 1165.65\\
C\,{\sc iii}& *\\
*&1165.62\\
*&1165.70\\
%6xC\,{\sc iii}&1175.5\\
C\,{\sc iii}& *\\
*&1174.93\\
*&1175.26\\
*&1175.59\\
*&1175.71\\
*&1175.99\\
*&1176.37\\
%C\,{\sc iii}&1240.28\\ %outside of spectral rang	
%C\,{\sc iii}&1247.00\\ 
%C\,{\sc iii}&1785.02\\
%C\,{\sc iii}&1880.60\\
%C\,{\sc iii}&1991.80\\
%\hline
C\,{\sc iv}&1107.98\\
%2xC\,{\sc iv}&1230.3\\	
%C\,{\sc iv}& *\\	
%*&1230.04\\ %outside of spectral range
%*&1230.52\\
%\hline
%2xN\,{\sc iii}+N\,{\sc iv}&1006\\
N\,{\sc iii}& *\\
*&1005.99\\
*&1006.04\\
%3xN\,{\sc iii}&1038.77\\
N\,{\sc iii}& *\\
*&1038.61\\
*&1038.98\\
*&1039.00\\
%2xN\,{\sc iii}&1055.96	\\
N\,{\sc iii}& *	\\	
*&1055.95\\
*&1055.97\\
N\,{\sc iii}&1103.05\\
%2xN\,{\sc iii}+N\,{\sc iv}&1104.32\\
N\,{\sc iii}+N\,{\sc iv}& *\\
*&1104.07\\
*&1104.12\\
*&1104.54\\
N\,{\sc iii}&1106.34\\
%2xN\,{\sc iii}&1120.8\\
N\,{\sc iii}& *\\
*&1120.60\\
*&1120.81\\
%2xN\,{\sc iii}&1140.08\\
N\,{\sc iii}& *\\
*&1140.05\\
*&1140.12\\
N\,{\sc iii}&\\
*&1182.97\\
*&1183.03\\
%2xN\,{\sc iii}&1184.55\\
N\,{\sc iii}& *\\
*&1184.51\\
*&1184.55\\
%\hline
N\,{\sc iv}&955.34\\
\hline
\end{tabular}
\end{center}
\end{minipage}
\begin{minipage}{0.45\textwidth}
\begin{center}
\begin{tabular}{cc}
\hline\hline
\rule{0pt}{15pt}Element+Ion. stage&wavelength [\AA]\\
%2xN\,{\sc iii}&1183\\	
\hline
%6xN\,{\sc iv}&1036.25\\
N\,{\sc iv}& *\\
*&1036.12\\
*&1036.15\\
*&1036.20\\
*&1036.23\\
*&1036.24\\
*&1036.33\\
N\,{\sc iv}&1117.93\\
%2xN\,{\sc iv}&1132.1\\
N\,{\sc iv}&*\\
*&1132.02\\
*&1132.23\\
%3xN\,{\sc iv}&1132.9\\
N\,{\sc iv}& *\\
*&1132.68\\
*&1132.94\\
*&1133.12\\
N\,{\sc iv}&1135.25\\
N\,{\sc iv}&1188.01\\
%\hline
%3xO\,{\sc iii}&1008\\
O\,{\sc iii}& *\\
*&1007.88\\
*&1008.10\\
*&1008.40\\
O\,{\sc iii}&1033.15\\
O\,{\sc iii}&1040.32\\
O\,{\sc iii}&1138.54\\
%2xO\,{\sc iii}&1150.9\\
O\,{\sc iii}& *\\
*&1150.88\\
*&1150.99\\
%2xO\,{\sc iii}&1153.1\\
O\,{\sc iii}&*\\
*&1153.02\\
*&1153.21\\
O\,{\sc iii}&1153.77\\
O\,{\sc iii}&1157.64\\
%\hline
O\,{\sc iv}&1106.93\\
O\,{\sc iv}&1164.55\\	
%\hline
Ne\,{\sc ii}&1133.92\\
%2xNe\,{\sc ii}&1139.3\\	
Ne\,{\sc ii}& * \\
*&1139.11\\
*&1139.36\\
Ne\,{\sc ii}&1143.89\\	
%\hline
%2xNe\,{\sc iii}&1044.9\\
Ne\,{\sc iii}& *\\
*&1044.94\\
*&1044.96\\
%3xNe\,{\sc iii}&1076.87\\
Ne\,{\sc iii}& *\\
*&1076.71\\
*&1076.79\\
*&1077.02\\
%\hline
Al\,{\sc iii}&1071.74\\
%2xAl\,{\sc iii}&1162.6\\
Al\,{\sc iii}&*\\
*&1162.59\\
*&1162.62\\
%\hline
Si\,{\sc iii}&997.39\\
%3xSi\,{\sc iii}&1113.2\\
Si\,{\sc iii}& *\\
*&1113.17\\
*&1113.20\\
*&1113.23\\
Si\,{\sc iii}&1140.50\\
Si\,{\sc iii}&1144.31\\

\hline
\end{tabular}
\end{center}
\end{minipage}
\end{table*}

\begin{table*}[h!]
UV lines in the FUSE spectra used for the abundance determination (continued).

\begin{minipage}{0.45\textwidth}
\begin{center}
\begin{tabular}{cc}
\hline\hline
\rule{0pt}{15pt}Element+Ion. stage&wavelength [\AA]\\
\hline
Si\,{\sc iii}&1161.58\\
Si\,{\sc iv}&996.87\\	
%2xSi\,{\sc iv}&1154.62\\
Si\,{\sc iv}& *\\
*&1154.62\\
*&1154.62\\
%\hline
%2xP\,{\sc iv}&1030.50\\
P\,{\sc iv}& *\\
*&1030.51\\
*&1030.52\\
P\,{\sc iv}&1033.10\\ %corrected from 1034.1
P\,{\sc iv}&1064.61\\
P\,{\sc iv}&1118.55\\
P\,{\sc iv}&1187.57\\
%\hline
P\,{\sc v}&997.60\\
P\,{\sc v}&1117.98\\
P\,{\sc v}&1128.01\\	
%\hline
S\,{\sc iv}&1072.97\\
S\,{\sc iv}&1073.52\\
S\,{\sc iv}&1098.36\\
S\,{\sc iv}&1098.92\\
%\hline
S\,{\sc v}&1122.04\\
%\hline
S\,{\sc vi}&933.38\\
S\,{\sc vi}&944.53\\
%\hline
Fe\,{\sc iii}&1141.27\\
Fe\,{\sc iii}&1142.96\\
%\hline
Fe\,{\sc iv}&1156.53\\
Fe\,{\sc iv}+2Fe\,{\sc v}&1112.18\\
*&1111.89\\
*&1112.17\\
*&1112.21\\
%2xFe\,{\sc v}&1116.11\\
Fe\,{\sc v}& *\\
*&1116.05\\
*&1116.16\\
%Fe\,{\sc iv}+Fe\,{\sc v}&1124.85\\
Fe\,{\sc iv}+Fe\,{\sc v}& *\\
*&1124.83\\
*&1124.89\\
\hline
\end{tabular}
\end{center}
\end{minipage}
\begin{minipage}{0.45\textwidth}
\begin{center}
\begin{tabular}{cc}
\hline\hline
\rule{0pt}{15pt}Element+Ion. stage&wavelength [\AA]\\
\hline

Fe\,{\sc v}&1118.38\\
Fe\,{\sc v}&1133.57\\
Fe\,{\sc v}&1149.12\\
%Fe\,{\sc v}&1149.12\\ %listed twice
Fe\,{\sc v}&1153.80\\
Fe\,{\sc v}&1161.92\\
Fe\,{\sc v}&1165.71\\
Ni\,{\sc iv}&1126.33\\
Ni\,{\sc iv}+Ni\,{\sc v}&1152.68\\
*&1152.67\\
*&1152.68\\
%Ni\,{\sc iv}+4xNi\,{\sc v}&1159.02\\
Ni\,{\sc iv}+4xNi\,{\sc v}& *\\
*&1158.86\\
*&1158.98\\
*&1158.02\\
*&1159.04\\
*&1159.09\\
%Ni\,{\sc iv}+2xNi\,{\sc v}&1187.8\\
Ni\,{\sc iv}+2xNi\,{\sc v}& * \\
*&1187.84\\
*&1187.67\\
*&1187.79\\
%6xNi\,{\sc v}&1178.9\\	
Ni\,{\sc v}& *\\
*&1178.80\\
*&1178.81\\
*&1178.82\\
*&1178.92\\
*&1178.94\\
*&1179.16\\
%3xNi\,{\sc v}&1182.62\\
Ni\,{\sc v}& *\\
*&1182.54\\
*&1182.62\\
*&1182.71\\
\phantom{*}& \phantom{1182.71}\\
\phantom{*}& \phantom{1182.71}\\
\phantom{*}& \phantom{1182.71}\\
\phantom{*}& \phantom{1182.71}\\
\hline
\end{tabular}
\end{center}
\end{minipage}
\vspace*{10\baselineskip}
\end{table*}

%%%%%%%%

\end{appendix}

\end{document}

%% file: beta_all.tex
   H     &   -3.22 $\pm$   0.20 & -2.79   $\pm$    0.25 &  -2.62   $\pm$    0.30 &  {-3.34}  $\pm$    0.25   &   -0.13    \\
   He    &   -0.0035   & -0.0138    &  -0.0142  &   -0.0121     &   -0.60 $\pm$  0.01\\
   C     &   -4.15  $\pm$   0.16 & -1.72   $\pm$    0.19 &  -1.65   $\pm$    0.13 &   -2.12   $\pm$    0.13   &   -2.63 $\pm$ 0.05\\
   N     &   -2.38  $\pm$   0.11 & -2.75   $\pm$    0.22 &  -3.05   $\pm$    0.19 &   -2.33   $\pm$    0.27   &   -3.16 $\pm$ 0.04\\
   O     &   -4.22  $\pm$   0.13 & -3.41   $\pm$    0.24 &  -3.47   $\pm$    0.22 &   -2.90   $\pm$    0.15   &   -2.24 $\pm$ 0.05\\
   Ne    &   -2.74  $\pm$   0.12 & -2.19   $\pm$    0.21 &  -2.28   $\pm$    0.16 &   -1.95   $\pm$    0.21   &   -2.90 $\pm$ 0.10\\
   Mg    &   -3.39  $\pm$   0.10 & -3.36   $\pm$    0.29 &  -3.64   $\pm$    0.16 &   -3.17   $\pm$    0.10   &   -3.15 $\pm$ 0.05\\
   Al    &   -4.47  $\pm$   0.21 & -3.45   $\pm$    0.30 &  -4.24   $\pm$    0.23 &   -4.05   $\pm$    0.10   &   -4.26 $\pm$ 0.04\\
   Si    &   -3.47  $\pm$   0.16 & -3.50   $\pm$    0.19 &  -3.70   $\pm$    0.22 &   -3.02   $\pm$    0.16   &   -3.18 $\pm$ 0.03\\
   P     &   -5.61  $\pm$   0.15 & -5.10   $\pm$    0.15 &  -5.25   $\pm$    0.22 &   -4.88   $\pm$    0.21   &   -5.24 $\pm$ 0.03\\
   S     &   -3.67  $\pm$   0.19 & -3.75   $\pm$    0.19 &  -4.29   $\pm$    0.20 &   -3.89   $\pm$    0.21   &   -3.51 $\pm$ 0.03\\
   Fe    &   {-3.48}  $\pm$   0.27 & {-3.46}   $\pm$   {0.18} &  {-3.55}   $\pm$    0.25 &   {-3.49}   $\pm$    {0.16}   &   -2.89 $\pm$ 0.04\\
   Ni    &   -3.94  $\pm$   0.23 & {-3.64}   $\pm$   {0.29} & {-3.96}   $\pm$    {0.26} &   {-3.78}   $\pm$    {0.31}   &   -4.15 $\pm$ 0.04\\

%% file: He_sdo_revised_accepted_arxiv.bbl
\begin{thebibliography}{94}
\expandafter\ifx\csname natexlab\endcsname\relax\def\natexlab#1{#1}\fi

\bibitem[{{Ahmad} \& {Jeffery}(2003)}]{2003A&A...402..335A}
{Ahmad}, A. \& {Jeffery}, C.~S. 2003, \aap, 402, 335

\bibitem[{{Asplund} {et~al.}(2009){Asplund}, {Grevesse}, {Sauval}, \&
  {Scott}}]{asplund2009}
{Asplund}, M., {Grevesse}, N., {Sauval}, A.~J., \& {Scott}, P. 2009, \araa, 47,
  481

\bibitem[{{Battich} {et~al.}(2018){Battich}, {Bertolami}, {C{\'o}rsico}, \&
  {Althaus}}]{battich2018}
{Battich}, T., {Bertolami}, M.~M.~M., {C{\'o}rsico}, A.~H., \& {Althaus}, L.~G.
  2018, \aap, 614, A136

\bibitem[{{Bauer} \& {Husfeld}(1995)}]{bauer1995}
{Bauer}, F. \& {Husfeld}, D. 1995, \aap, 300, 481

\bibitem[{{Berger} \& {Fringant}(1980)}]{1980A&A....85..367B}
{Berger}, J. \& {Fringant}, A.-M. 1980, \aap, 85, 367

\bibitem[{{Blair} {et~al.}(2009){Blair}, {Sankrit}, {Torres}, {Chayer}, \&
  {Danforth}}]{2009ApJ...692..335B}
{Blair}, W.~P., {Sankrit}, R., {Torres}, S.~I., {Chayer}, P., \& {Danforth},
  C.~W. 2009, \apj, 692, 335

\bibitem[{{Blanchette} {et~al.}(2008){Blanchette}, {Chayer}, {Wesemael},
  {Fontaine}, {Fontaine}, {Dupuis}, {Kruk}, \& {Green}}]{2008ApJ...678.1329B}
{Blanchette}, J.-P., {Chayer}, P., {Wesemael}, F., {et~al.} 2008, \apj, 678,
  1329

\bibitem[{{Brown} {et~al.}(1997){Brown}, {Ferguson}, {Davidsen}, \&
  {Dorman}}]{Brown1997}
{Brown}, T.~M., {Ferguson}, H.~C., {Davidsen}, A.~F., \& {Dorman}, B. 1997,
  \apj, 482, 685

\bibitem[{{Brown} {et~al.}(2001){Brown}, {Sweigart}, {Lanz}, {Landsman}, \&
  {Hubeny}}]{2001ApJ...562..368B}
{Brown}, T.~M., {Sweigart}, A.~V., {Lanz}, T., {Landsman}, W.~B., \& {Hubeny},
  I. 2001, \apj, 562, 368

\bibitem[{{Byrne} {et~al.}(2018){Byrne}, {Jeffery}, {Tout}, \&
  {Hu}}]{2018MNRAS.475.4728B}
{Byrne}, C.~M., {Jeffery}, C.~S., {Tout}, C.~A., \& {Hu}, H. 2018, \mnras, 475,
  4728

\bibitem[{{Cassisi} {et~al.}(2003){Cassisi}, {Schlattl}, {Salaris}, \&
  {Weiss}}]{2003ApJ...582L..43C}
{Cassisi}, S., {Schlattl}, H., {Salaris}, M., \& {Weiss}, A. 2003, \apjl, 582,
  L43

\bibitem[{{Castellani} \& {Castellani}(1993)}]{Castellani1993}
{Castellani}, M. \& {Castellani}, V. 1993, \apj, 407, 649

\bibitem[{{Chayer} {et~al.}(2006){Chayer}, {Fontaine}, {Fontaine}, {Wesemael},
  \& {Dupuis}}]{2006BaltA..15..131C}
{Chayer}, P., {Fontaine}, M., {Fontaine}, G., {Wesemael}, F., \& {Dupuis}, J.
  2006, Baltic Astronomy, 15, 131

\bibitem[{{Cutri} \& {et al.}(2013)}]{2013yCat.2328....0C}
{Cutri}, R.~M. \& {et al.} 2013, VizieR Online Data Catalog, 2328

\bibitem[{{Cutri} {et~al.}(2003){Cutri}, {Skrutskie}, {van Dyk}, {Beichman},
  {Carpenter}, {Chester}, {Cambresy}, {Evans}, {Fowler}, {Gizis}, {Howard},
  {Huchra}, {Jarrett}, {Kopan}, {Kirkpatrick}, {Light}, {Marsh}, {McCallon},
  {Schneider}, {Stiening}, {Sykes}, {Weinberg}, {Wheaton}, {Wheelock}, \&
  {Zacarias}}]{cutri2003}
{Cutri}, R.~M., {Skrutskie}, M.~F., {van Dyk}, S., {et~al.} 2003, VizieR Online
  Data Catalog, 2246

\bibitem[{{Dekker} {et~al.}(2000){Dekker}, {D'Odorico}, {Kaufer}, {Delabre}, \&
  {Kotzlowski}}]{2000SPIE.4008..534D}
{Dekker}, H., {D'Odorico}, S., {Kaufer}, A., {Delabre}, B., \& {Kotzlowski}, H.
  2000, in \procspie, Vol. 4008, Optical and IR Telescope Instrumentation and
  Detectors, ed. M.~{Iye} \& A.~F. {Moorwood}, 534--545

\bibitem[{{Dixon} {et~al.}(2007){Dixon}, {Sahnow}, {Barrett}, {Civeit},
  {Dupuis}, {Fullerton}, {Godard}, {Hsu}, {Kaiser}, {Kruk}, {Lacour},
  {Lindler}, {Massa}, {Robinson}, {Romelfanger}, \&
  {Sonnentrucker}}]{dixon2007}
{Dixon}, W.~V., {Sahnow}, D.~J., {Barrett}, P.~E., {et~al.} 2007, \pasp, 119,
  527

\bibitem[{{Dixon} {et~al.}(2006){Dixon}, {Sankrit}, \&
  {Otte}}]{2006ApJ...647..328D}
{Dixon}, W.~V.~D., {Sankrit}, R., \& {Otte}, B. 2006, \apj, 647, 328

\bibitem[{{Dorman} {et~al.}(1993){Dorman}, {Rood}, \&
  {O'Connell}}]{1993ApJ...419..596D}
{Dorman}, B., {Rood}, R.~T., \& {O'Connell}, R.~W. 1993, \apj, 419, 596

\bibitem[{{Dreizler}(1993)}]{Dreizler1993}
{Dreizler}, S. 1993, \aap, 273, 212

\bibitem[{{Dreizler} {et~al.}(1990){Dreizler}, {Heber}, {Werner}, {Moehler}, \&
  {de Boer}}]{1990A&A...235..234D}
{Dreizler}, S., {Heber}, U., {Werner}, K., {Moehler}, S., \& {de Boer}, K.~S.
  1990, \aap, 235, 234

\bibitem[{{Drilling} \& {Heber}(1987)}]{drilling1987}
{Drilling}, J.~S. \& {Heber}, U. 1987, in IAU Colloq. 95: Second Conference on
  Faint Blue Stars, ed. A.~G.~D. {Philip}, D.~S. {Hayes}, \& J.~W. {Liebert},
  603--606

\bibitem[{{Edelmann} {et~al.}(2003){Edelmann}, {Heber}, {Hagen}, {Lemke},
  {Dreizler}, {Napiwotzki}, \& {Engels}}]{2003A&A...400..939E}
{Edelmann}, H., {Heber}, U., {Hagen}, H.-J., {et~al.} 2003, \aap, 400, 939

\bibitem[{{Ferraro} {et~al.}(1997){Ferraro}, {Paltrinieri}, {Fusi Pecci},
  {Cacciari}, {Dorman}, \& {Rood}}]{Ferraro1997}
{Ferraro}, F.~R., {Paltrinieri}, B., {Fusi Pecci}, F., {et~al.} 1997, \apj,
  484, L145

\bibitem[{{Fitzpatrick}(1999)}]{1999PASP..111...63F}
{Fitzpatrick}, E.~L. 1999, \pasp, 111, 63

\bibitem[{{Fontaine} {et~al.}(2014){Fontaine}, {Green}, {Brassard}, {Latour},
  \& {Chayer}}]{2014ASPC..481...83F}
{Fontaine}, G., {Green}, E., {Brassard}, P., {Latour}, M., \& {Chayer}, P.
  2014, in Astronomical Society of the Pacific Conference Series, Vol. 481, 6th
  Meeting on Hot Subdwarf Stars and Related Objects, ed. V.~{van Grootel},
  E.~{Green}, G.~{Fontaine}, \& S.~{Charpinet}, 83

\bibitem[{{Fontaine} {et~al.}(2006){Fontaine}, {Chayer}, {Wesemael},
  {Fontaine}, \& {Lamontagne}}]{2006BaltA..15...99F}
{Fontaine}, M., {Chayer}, P., {Wesemael}, F., {Fontaine}, G., \& {Lamontagne},
  R. 2006, Baltic Astronomy, 15, 99

\bibitem[{{Friedman} {et~al.}(2002){Friedman}, {Howk}, {Chayer}, {Tripp},
  {H{\'e}brard}, {Andr{\'e}}, {Oliveira}, {Jenkins}, {Moos}, {Oegerle},
  {Sonneborn}, {Lamontagne}, {Sembach}, \&
  {Vidal-Madjar}}]{2002ApJS..140...37F}
{Friedman}, S.~D., {Howk}, J.~C., {Chayer}, P., {et~al.} 2002, \apjs, 140, 37

\bibitem[{{Gaia Collaboration} {et~al.}(2018){Gaia Collaboration}, {Brown},
  {Vallenari}, {Prusti}, {de Bruijne}, {Babusiaux}, {Bailer-Jones}, {Biermann},
  {Evans}, {Eyer}, \& et~al.}]{2018A&A...616A...1G}
{Gaia Collaboration}, {Brown}, A.~G.~A., {Vallenari}, A., {et~al.} 2018, \aap,
  616, A1

\bibitem[{{Garrison} \& {Hiltner}(1973)}]{1973ApJ...179L.117G}
{Garrison}, R.~F. \& {Hiltner}, W.~A. 1973, \apjl, 179, L117

\bibitem[{{Geier}(2013)}]{2013A&A...549A.110G}
{Geier}, S. 2013, \aap, 549, A110

\bibitem[{{Giddings}(1981)}]{giddings1981}
{Giddings}, J.~R. 1981, PhD thesis, , University of London, (1981)

\bibitem[{{Green} {et~al.}(2018){Green}, {Schlafly}, {Finkbeiner}, {Rix},
  {Martin}, {Burgett}, {Draper}, {Flewelling}, {Hodapp}, {Kaiser}, {Kudritzki},
  {Magnier}, {Metcalfe}, {Tonry}, {Wainscoat}, \&
  {Waters}}]{2018MNRAS.478..651G}
{Green}, G.~M., {Schlafly}, E.~F., {Finkbeiner}, D., {et~al.} 2018, \mnras,
  478, 651

\bibitem[{{Han} {et~al.}(2007){Han}, {Podsiadlowski}, \&
  {Lynas-Gray}}]{han2007}
{Han}, Z., {Podsiadlowski}, P., \& {Lynas-Gray}, A.~E. 2007, \mnras, 380, 1098

\bibitem[{{Heber}(2009)}]{heber2009}
{Heber}, U. 2009, Annual Review of Astronomy \& Astrophysics, 47, 211

\bibitem[{{Heber}(2016)}]{heber2016}
{Heber}, U. 2016, \pasp, 128, 082001

\bibitem[{{Heber}(2017)}]{2017ASPC..509...85H}
{Heber}, U. 2017, in Astronomical Society of the Pacific Conference Series,
  Vol. 509, 20th European White Dwarf Workshop, ed. P.-E. {Tremblay},
  B.~{Gaensicke}, \& T.~{Marsh}, 85

\bibitem[{{Heber} {et~al.}(2018){Heber}, {Irrgang}, \&
  {Schaffenroth}}]{2018OAst...27...35H}
{Heber}, U., {Irrgang}, A., \& {Schaffenroth}, J. 2018, Open Astronomy, 27, 35

\bibitem[{{H{\'e}brard} \& {Moos}(2003)}]{2003ApJ...599..297H}
{H{\'e}brard}, G. \& {Moos}, H.~W. 2003, \apj, 599, 297

\bibitem[{{Henden} {et~al.}(2016){Henden}, {Templeton}, {Terrell}, {Smith},
  {Levine}, \& {Welch}}]{2016yCat.2336....0H}
{Henden}, A.~A., {Templeton}, M., {Terrell}, D., {et~al.} 2016, VizieR Online
  Data Catalog, 2336

\bibitem[{{Hirsch}(2009)}]{Hirsch2009}
{Hirsch}, H.~A. 2009, PhD thesis, Friedrich-Alexander University
  Erlangen-N{\"u}rnberg

\bibitem[{{H{\o}g} {et~al.}(2000){H{\o}g}, {Fabricius}, {Makarov}, {Urban},
  {Corbin}, {Wycoff}, {Bastian}, {Schwekendiek}, \&
  {Wicenec}}]{2000A&A...355L..27H}
{H{\o}g}, E., {Fabricius}, C., {Makarov}, V.~V., {et~al.} 2000, \aap, 355, L27

\bibitem[{{Hubeny} \& {Lanz}(1995)}]{Hubeny1995}
{Hubeny}, I. \& {Lanz}, T. 1995, \apj, 439, 875

\bibitem[{{Hubeny} \& {Lanz}(2017{\natexlab{a}})}]{2017arXiv170601859H}
{Hubeny}, I. \& {Lanz}, T. 2017{\natexlab{a}}, ArXiv e-prints
  [\eprint[arXiv]{1706.01859}]

\bibitem[{{Hubeny} \& {Lanz}(2017{\natexlab{b}})}]{2017arXiv170601935H}
{Hubeny}, I. \& {Lanz}, T. 2017{\natexlab{b}}, ArXiv e-prints
  [\eprint[arXiv]{1706.01935}]

\bibitem[{{Hubeny} \& {Lanz}(2017{\natexlab{c}})}]{2017arXiv170601937H}
{Hubeny}, I. \& {Lanz}, T. 2017{\natexlab{c}}, ArXiv e-prints
  [\eprint[arXiv]{1706.01937}]

\bibitem[{Hubeny {et~al.}(1994)Hubeny, Lanz, \& Jeffery}]{synspec94}
Hubeny, I., Lanz, T., \& Jeffery, C. 1994, Newsletter on Analysis of
  Astronomical Spectra 20, p.30, ed. C.S. Jeffery,, Report, St. Andrews
  University

\bibitem[{{Husfeld} {et~al.}(1989){Husfeld}, {Butler}, {Heber}, \&
  {Drilling}}]{1989A&A...222..150H}
{Husfeld}, D., {Butler}, K., {Heber}, U., \& {Drilling}, J.~S. 1989, \aap, 222,
  150

\bibitem[{{Jenkins}(2013)}]{2013ApJ...764...25J}
{Jenkins}, E.~B. 2013, \apj, 764, 25

\bibitem[{{Kaufer} {et~al.}(1999){Kaufer}, {Stahl}, {Tubbesing},
  {N{\o}rregaard}, {Avila}, {Francois}, {Pasquini}, \&
  {Pizzella}}]{1999Msngr..95....8K}
{Kaufer}, A., {Stahl}, O., {Tubbesing}, S., {et~al.} 1999, The Messenger, 95, 8

\bibitem[{{Kilkenny} \& {Muller}(1989)}]{kilkenny1989}
{Kilkenny}, D. \& {Muller}, S. 1989, South African Astronomical Observatory
  Circular, 13, 69

\bibitem[{Kramida {et~al.}(2018)Kramida, Ralchenko, Reader, \&
  NIST~ASD~Team}]{nist18}
Kramida, A., Ralchenko, Y., Reader, J., \& NIST~ASD~Team, p. 2018, NIST Atomic
  Spectra Database (version 5.5.3), [Online]. Available:
  https://physics.nist.gov/asd [Tue Mar 20 2018], Report, National Institute of
  Standards and Technology, Gaithersburg, MD.

\bibitem[{{Lallement} {et~al.}(2008){Lallement}, {H{\'e}brard}, \&
  {Welsh}}]{2008A&A...481..381L}
{Lallement}, R., {H{\'e}brard}, G., \& {Welsh}, B.~Y. 2008, \aap, 481, 381

\bibitem[{{Landolt}(2007)}]{2007AJ....133.2502L}
{Landolt}, A.~U. 2007, \aj, 133, 2502

\bibitem[{{Lanz} {et~al.}(2004){Lanz}, {Brown}, {Sweigart}, {Hubeny}, \&
  {Landsman}}]{2004ApJ...602..342L}
{Lanz}, T., {Brown}, T.~M., {Sweigart}, A.~V., {Hubeny}, I., \& {Landsman},
  W.~B. 2004, \apj, 602, 342

\bibitem[{{Lanz} {et~al.}(1997){Lanz}, {Hubeny}, \& {Heap}}]{lanz1997}
{Lanz}, T., {Hubeny}, I., \& {Heap}, S.~R. 1997, \apj, 485, 843

\bibitem[{{Latour} {et~al.}(2018){Latour}, {Chayer}, {Green}, {Irrgang}, \&
  {Fontaine}}]{2018A&A...609A..89L}
{Latour}, M., {Chayer}, P., {Green}, E.~M., {Irrgang}, A., \& {Fontaine}, G.
  2018, \aap, 609, A89

\bibitem[{{Latour} {et~al.}(2013){Latour}, {Fontaine}, {Chayer}, \&
  {Brassard}}]{2013ApJ...773...84L}
{Latour}, M., {Fontaine}, G., {Chayer}, P., \& {Brassard}, P. 2013, \apj, 773,
  84

\bibitem[{{Lawrence} {et~al.}(2007){Lawrence}, {Warren}, {Almaini}, {Edge},
  {Hambly}, {Jameson}, {Lucas}, {Casali}, {Adamson}, {Dye}, {Emerson},
  {Foucaud}, {Hewett}, {Hirst}, {Hodgkin}, {Irwin}, {Lodieu}, {McMahon},
  {Simpson}, {Smail}, {Mortlock}, \& {Folger}}]{2007MNRAS.379.1599L}
{Lawrence}, A., {Warren}, S.~J., {Almaini}, O., {et~al.} 2007, \mnras, 379,
  1599

\bibitem[{{Lisker} {et~al.}(2005){Lisker}, {Heber}, {Napiwotzki}, {Christlieb},
  {Han}, {Homeier}, \& {Reimers}}]{2005A&A...430..223L}
{Lisker}, T., {Heber}, U., {Napiwotzki}, R., {et~al.} 2005, \aap, 430, 223

\bibitem[{{Mermilliod} {et~al.}(1997){Mermilliod}, {Mermilliod}, \&
  {Hauck}}]{1997A&AS..124..349M}
{Mermilliod}, J.-C., {Mermilliod}, M., \& {Hauck}, B. 1997, \aaps, 124

\bibitem[{{Miller Bertolami}(2016)}]{2016A&A...588A..25M}
{Miller Bertolami}, M.~M. 2016, \aap, 588, A25

\bibitem[{{Miller Bertolami} {et~al.}(2008){Miller Bertolami}, {Althaus},
  {Unglaub}, \& {Weiss}}]{Miller2008}
{Miller Bertolami}, M.~M., {Althaus}, L.~G., {Unglaub}, K., \& {Weiss}, A.
  2008, A\&A, 491, 253

\bibitem[{{Moos} {et~al.}(2000){Moos}, {Cash}, {Cowie}, {Davidsen}, {Dupree},
  {Feldman}, {Friedman}, {Green}, {Green}, {Gry}, {Hutchings}, {Jenkins},
  {Linsky}, {Malina}, {Michalitsianos}, {Savage}, {Shull}, {Siegmund}, {Snow},
  {Sonneborn}, {Vidal-Madjar}, {Willis}, {Woodgate}, {York}, {Ake},
  {Andersson}, {Andrews}, {Barkhouser}, {Bianchi}, {Blair}, {Brownsberger},
  {Cha}, {Chayer}, {Conard}, {Fullerton}, {Gaines}, {Grange}, {Gummin},
  {Hebrard}, {Kriss}, {Kruk}, {Mark}, {McCarthy}, {Morbey}, {Murowinski},
  {Murphy}, {Oegerle}, {Ohl}, {Oliveira}, {Osterman}, {Sahnow}, {Saisse},
  {Sembach}, {Weaver}, {Welsh}, {Wilkinson}, \& {Zheng}}]{2000ApJ...538L...1M}
{Moos}, H.~W., {Cash}, W.~C., {Cowie}, L.~L., {et~al.} 2000, \apjl, 538, L1

\bibitem[{{Morton}(2003)}]{2003ApJS..149..205M}
{Morton}, D.~C. 2003, \apjs, 149, 205

\bibitem[{{Morton} \& {Dinerstein}(1976)}]{1976ApJ...204....1M}
{Morton}, D.~C. \& {Dinerstein}, H.~L. 1976, \apj, 204, 1

\bibitem[{{Napiwotzki}(1999)}]{napiwotzki1999}
{Napiwotzki}, R. 1999, \aap, 350, 101

\bibitem[{{Napiwotzki}(2008)}]{napiwotzki2008}
{Napiwotzki}, R. 2008, in Astronomical Society of the Pacific Conference
  Series, Vol. 392, Hot Subdwarf Stars and Related Objects, ed. U.~{Heber},
  C.~S. {Jeffery}, \& R.~{Napiwotzki}, 139

\bibitem[{{Napiwotzki} {et~al.}(2004){Napiwotzki}, {Yungelson}, {Nelemans},
  {Marsh}, {Leibundgut}, {Renzini}, {Homeier}, {Koester}, {Moehler},
  {Christlieb}, {Reimers}, {Drechsel}, {Heber}, {Karl}, \&
  {Pauli}}]{2004ASPC..318..402N}
{Napiwotzki}, R., {Yungelson}, L., {Nelemans}, G., {et~al.} 2004, in
  Astronomical Society of the Pacific Conference Series, Vol. 318,
  Spectroscopically and Spatially Resolving the Components of the Close Binary
  Stars, ed. R.~W. {Hilditch}, H.~{Hensberge}, \& K.~{Pavlovski}, 402--410

\bibitem[{{Naslim} {et~al.}(2010){Naslim}, {Jeffery}, {Ahmad}, {Behara}, \&
  {{\c S}ah{\`i}n}}]{2010MNRAS.409..582N}
{Naslim}, N., {Jeffery}, C.~S., {Ahmad}, A., {Behara}, N.~T., \& {{\c
  S}ah{\`i}n}, T. 2010, \mnras, 409, 582

\bibitem[{{N{\'e}meth} {et~al.}(2012){N{\'e}meth}, {Kawka}, \&
  {Vennes}}]{peter2012}
{N{\'e}meth}, P., {Kawka}, A., \& {Vennes}, S. 2012, \mnras, 427, 2180

\bibitem[{{N\'{e}meth} {et~al.}(2012){N\'{e}meth}, {Kawka}, \&
  {Vennes}}]{nemeth2012}
{N\'{e}meth}, P., {Kawka}, A., \& {Vennes}, S. 2012, in Astronomical Society of
  the Pacific Conference Series, Vol. 452, Fifth Meeting on Hot Subdwarf Stars
  and Related Objects, ed. D.~{Kilkenny}, C.~S. {Jeffery}, \& C.~{Koen}, 33

\bibitem[{{Ohl} {et~al.}(2000){Ohl}, {Chayer}, \& {Moos}}]{2000ApJ...538L..95O}
{Ohl}, R.~G., {Chayer}, P., \& {Moos}, H.~W. 2000, \apjl, 538, L95

\bibitem[{{Paczy{\'n}ski}(1971)}]{1971AcA....21....1P}
{Paczy{\'n}ski}, B. 1971, \actaa, 21, 1

\bibitem[{{Paunzen}(2015)}]{2015A&A...580A..23P}
{Paunzen}, E. 2015, \aap, 580, A23

\bibitem[{{Rauch} {et~al.}(2014){Rauch}, {Rudkowski}, {Kampka}, {Werner},
  {Kruk}, \& {Moehler}}]{2014A&A...566A...3R}
{Rauch}, T., {Rudkowski}, A., {Kampka}, D., {et~al.} 2014, \aap, 566, A3

\bibitem[{{Sahnow} {et~al.}(2000){Sahnow}, {Moos}, {Ake}, {Andersen},
  {Andersson}, {Andre}, {Artis}, {Berman}, {Blair}, {Brownsberger}, {Calvani},
  {Chayer}, {Conard}, {Feldman}, {Friedman}, {Fullerton}, {Gaines}, {Gawne},
  {Green}, {Gummin}, {Jennings}, {Joyce}, {Kaiser}, {Kruk}, {Lindler}, {Massa},
  {Murphy}, {Oegerle}, {Ohl}, {Roberts}, {Romelfanger}, {Roth}, {Sankrit},
  {Sembach}, {Shelton}, {Siegmund}, {Silva}, {Sonneborn}, {Vaclavik}, {Weaver},
  \& {Wilkinson}}]{2000ApJ...538L...7S}
{Sahnow}, D.~J., {Moos}, H.~W., {Ake}, T.~B., {et~al.} 2000, \apjl, 538, L7

\bibitem[{{Schaffenroth} {et~al.}(2014){Schaffenroth}, {Classen}, {Nagel},
  {Geier}, {Koen}, {Heber}, \& {Edelmann}}]{2014A&A...570A..70S}
{Schaffenroth}, V., {Classen}, L., {Nagel}, K., {et~al.} 2014, \aap, 570, A70

\bibitem[{{Schlafly} \& {Finkbeiner}(2011{\natexlab{a}})}]{schlafly}
{Schlafly}, E.~F. \& {Finkbeiner}, D.~P. 2011{\natexlab{a}}, \apj, 737, 103

\bibitem[{{Schlafly} \& {Finkbeiner}(2011{\natexlab{b}})}]{2011ApJ...737..103S}
{Schlafly}, E.~F. \& {Finkbeiner}, D.~P. 2011{\natexlab{b}}, \apj, 737, 103

\bibitem[{{Schwab}(2018)}]{2018MNRAS.476.5303S}
{Schwab}, J. 2018, \mnras, 476, 5303

\bibitem[{{Skrutskie} {et~al.}(2006){Skrutskie}, {Cutri}, {Stiening},
  {Weinberg}, {Schneider}, {Carpenter}, {Beichman}, {Capps}, {Chester},
  {Elias}, {Huchra}, {Liebert}, {Lonsdale}, {Monet}, {Price}, {Seitzer},
  {Jarrett}, {Kirkpatrick}, {Gizis}, {Howard}, {Evans}, {Fowler}, {Fullmer},
  {Hurt}, {Light}, {Kopan}, {Marsh}, {McCallon}, {Tam}, {Van Dyk}, \&
  {Wheelock}}]{2006AJ....131.1163S}
{Skrutskie}, M.~F., {Cutri}, R.~M., {Stiening}, R., {et~al.} 2006, \aj, 131,
  1163

\bibitem[{{Soker}(1998)}]{1998AJ....116.1308S}
{Soker}, N. 1998, \aj, 116, 1308

\bibitem[{{Sonneborn} {et~al.}(2002){Sonneborn}, {Andr{\'e}}, {Oliveira},
  {H{\'e}brard}, {Howk}, {Tripp}, {Chayer}, {Friedman}, {Kruk}, {Jenkins},
  {Lemoine}, {Moos}, {Oegerle}, {Sembach}, \&
  {Vidal-Madjar}}]{2002ApJS..140...51S}
{Sonneborn}, G., {Andr{\'e}}, M., {Oliveira}, C., {et~al.} 2002, \apjs, 140, 51

\bibitem[{{Stroeer} {et~al.}(2007){Stroeer}, {Heber}, {Lisker}, {Napiwotzki},
  {Dreizler}, {Christlieb}, \& {Reimers}}]{Stroer2007}
{Stroeer}, A., {Heber}, U., {Lisker}, T., {et~al.} 2007, \aap, 462, 269

\bibitem[{{Sweigart}(1997)}]{1997fbs..conf....3S}
{Sweigart}, A.~V. 1997, in The Third Conference on Faint Blue Stars, ed.
  A.~G.~D. {Philip}, J.~{Liebert}, R.~{Saffer}, \& D.~S. {Hayes}, 3

\bibitem[{{Tailo} {et~al.}(2015){Tailo}, {D'Antona}, {Vesperini}, {di
  Criscienzo}, {Ventura}, {Milone}, {Bellini}, {Dotter}, {Decressin},
  {D'Ercole}, {Caloi}, \& {Capuzzo-Dolcetta}}]{2015Natur.523..318T}
{Tailo}, M., {D'Antona}, F., {Vesperini}, E., {et~al.} 2015, \nat, 523, 318

\bibitem[{{Ulla} \& {Thejll}(1998)}]{ulla1998}
{Ulla}, A. \& {Thejll}, P. 1998, \aaps, 132, 1

\bibitem[{{Walker}(1981)}]{1981MNRAS.197..241W}
{Walker}, A.~R. 1981, \mnras, 197, 241

\bibitem[{{Webbink}(1984)}]{webbink1984}
{Webbink}, R.~F. 1984, The Astrophysical Journal, 277, 355

\bibitem[{{Werner} \& {Dreizler}(1999)}]{1999JCoAM.109...65W}
{Werner}, K. \& {Dreizler}, S. 1999, Journal of Computational and Applied
  Mathematics, 109, 65

\bibitem[{{Wolf} {et~al.}(2018){Wolf}, {Onken}, {Luvaul}, {Schmidt}, {Bessell},
  {Chang}, {Da Costa}, {Mackey}, {Martin-Jones}, {Murphy}, {Preston}, {Scalzo},
  {Shao}, {Smillie}, {Tisserand}, {White}, \& {Yuan}}]{2018PASA...35...10W}
{Wolf}, C., {Onken}, C.~A., {Luvaul}, L.~C., {et~al.} 2018, \pasa, 35, e010

\bibitem[{{Wood} {et~al.}(2004){Wood}, {Linsky}, {H{\'e}brard}, {Williger},
  {Moos}, \& {Blair}}]{2004ApJ...609..838W}
{Wood}, B.~E., {Linsky}, J.~L., {H{\'e}brard}, G., {et~al.} 2004, \apj, 609,
  838

\bibitem[{{Zhang} \& {Jeffery}(2012)}]{zhang2012}
{Zhang}, X. \& {Jeffery}, C.~S. 2012, MNRAS, 419, 452

\end{thebibliography}
